\documentclass[10pt]{amsart}
\usepackage{epsfig,a4,graphics,a4wide,amsmath,amsthm,amssymb, amssymb}

\newcommand{\beq}{\begin{equation}}  % start likning
\newcommand{\eeq}{\end{equation}}    % slutt likning
\newcommand{\beqa}{\begin{eqnarray}} % start rad av likninger
\newcommand{\eeqa}{\end{eqnarray}}   % slutt rad av likninger
\newcommand{\bca}{\begin{cases}}
\newcommand{\eca}{\end{cases}}
\newcommand{\nn}{\nonumber}

\newcommand{\n}{\perp}
\renewcommand{\(}{\left(}
\renewcommand{\)}{\right)}

\newcommand{\half}{\frac{1}{2}}

\numberwithin{figure}{section}
\numberwithin{equation}{section}
\numberwithin{theorem}{section}

\begin{document}

\title{Numerical comparison of Riemann solvers for astrophysical hydrodynamics}
\author{
Christian Klingenberg$^1$,
Wolfram Schmidt$^2$,
Knut Waagan$^3$
}
\thanks{
$^1$Departement of Mathematics, W\"urzburg University, Am Hubland 97074
W\"urzburg, Germany
}
\thanks{
$^2$Departement of Theoretical Physics, W\"urzburg University, Am Hubland 97074
W\"urzburg, Germany
}
\thanks{
$^3$Center of Mathematics for Applications,
P.O. Box 1053 Blindern,
NO-0316 Oslo,
Norway. Visitor to W{\"u}rzburg University, Germany.
\bf knut.waagan@cma.uio.no
}

\date{\today}

\begin{abstract}
The idea of this work is to compare a new positive and entropy stable approximate Riemann
solver by Francois Bouchut with a state-of the-art algorithm for
astrophysical fluid dynamics. We implemented
the new Riemann solver into an astrophysical PPM-code, the Prometheus code, and also made a version with a
different, more theoretically grounded higher order algorithm than PPM. We present
shock tube tests, two-dimensional instability tests and forced turbulence
simulations in three dimensions. We find subtle differences
between the codes in the shock tube tests, and in the statistics of the
turbulence simulations. The new Riemann solver increases the computational
speed without significant loss of accuracy.
\end{abstract}

\maketitle

\section{Introduction}

In modern astrophysics the interplay between observations and numerical
experiments plays a central role. Typically hydrodynamical
flows with high Reynolds numbers and Mach numbers are studied, and they are modelled by the Euler equations
\begin{align}
\rho_t + \text{div}(\rho {\bf u})=0\nn\\
(\rho u)_t +  \text{div}(\rho u{\bf u}) + p_x=\rho f_1\nn\\
(\rho v)_t +  \text{div}(\rho v{\bf u}) + p_y=\rho f_2\nn\\
(\rho w)_t +  \text{div}(\rho w{\bf u}) + p_z=\rho f_3\nn\\
E_t + \text{div}((E+p){\bf u})=\rho {\bf f }\cdot{\bf u}.
\label{euler}
\end{align}
Here $\rho$ is the mass density, ${\bf u}=(u,v,w)$ is the velocity field, $p$
is the pressure, and $E$ is the energy density $E=\half\rho {\bf u}^2 + \rho
e$ with $e$ the specific internal energy. External forces are given in units of acceleration by ${\bf f }=(f_1,f_2,f_3)$. The system is closed by the equation
of state that relates $p$ to $\rho$ and $e$. In this work we consider ideal
gases where $\rho e=p/(\gamma-1)$ for some $\gamma>1$, and isothermal gases.
An isothermal gas has constant temperature $T$, which implies
$p=a^2\rho$ for $a=\frac{RT}{\mu}$, with $R$ the gas constant and $\mu$ the
mean molecular weight. In a real astrophysical flow additional physical phenomena such as magnetic fields, gravitational forces and electromagnetic radiation may be important, but this paper is only concerned with the hydrodynamics.
The specific physical entropy $s$ is defined by the relation
\beq
{\text d}e + p {\text d}\frac{1}{\rho}=T{\text d}s
\eeq
with $T=T(\rho,e)>0$ the temperature. The second law of thermodynamics implies that
\beq
(\rho \phi(s))_t + \text{div}(\rho {\bf u} \phi(s))\leq 0
\label{entropyinequality}
\eeq
for any smooth, nonincreasing and convex $\phi$. In high Mach number flows this condition is needed to ensure the dissipativity of shocks, since the viscous forces are ignored in \eqref{euler}. 

To numerically solve \eqref{euler}, shock-capturing finite volume schemes are widely used.
In astrophysics it is often done with the PPM algorithm described in \cite{ppmpaper}, often with an iterative method
to approximate the exact midpoint value of the Riemann fan. This was
implemented in the Prometheus code in 1989, using the iterative Riemann solver
of \cite{ppmriemann} with a fixed number of iterations, see
\cite{prometheus}. An efficient, parallelised version was then implemented in
2001, see \cite{Reinecke1} and \cite{Reinecke2}. Stochastic forcing for turbulence simulations was added later, see \cite{schmidtphd}, \cite{schmidtnum}. Results produced by the Prometheus code have been presented in many astrophysical
publications, for example in \cite{Schmidtsupernova2}, \cite{Reinecke1},
\cite{Reinecke2} and \cite{Schmidtsupernova}. We used this code from 2001 as
the basis for this work, and implemented our changes into it.

First, we switched the Riemann solver to an HLLC solver with the signal speeds
of Bouchut (\cite{BouchutBGK}, \cite{Bouchutbook}). As far as we know, this is the first implementation of this advancement into an astrophysics code. It is well known that approximate Riemann solvers like HLLC are computationally efficient and easy to implement. In addition, the Riemann solver of Bouchut has
two good properties: a) It automatically ensures that a discrete version of
the entropy inequality \eqref{entropyinequality} holds. b) The density and
pressure stay positive. Often, finite volume codes need to check for each cell update whether the new data are physically reasonable, but with these two a priori estimates, no checks are necessary apart from underflow treatment. The two estimates are true in the first order case.

When using this Riemann solver in a higher order scheme, these two properties are not automatically inherited. Hence,
we introduced a piecewise linear reconstruction, and
replaced the characteristic backtracing with Runge-Kutta time
integration. This was done in such a way that positivity is preserved at one half
the CFL-number required in the first order case. Such a second-order scheme
which also satisfies the entropy inequality is however impractical, see
\cite{secordentropic}, but entropy stability for first order schemes has so far seemed to be
a good condition in practice. A different notion of stability comes from
scalar conservation laws, which have solutions with nonincreasing total
variation. This notion is also important in the design of higher order methods
for systems. The reconstruction algorithm and the Runge-Kutta integration we use
ensure a nonincreasing total variation when applied to scalar equations.

This gave rise to four codes as summarized in Figure \ref{codestable}.
\begin{figure}
%\\[6pt]
\begin{tabular}[c]{|l||c|c|}
\hline
& \multicolumn{2}{|c|} {}\\[-8pt]
& \multicolumn{2}{|c|}{Riemann solver:}\\[3pt]
\cline{2-3}
& &\\[-8pt]
Higher order algorithm:& Iterative of Prometheus& HLLC-Bouchut\\[3pt]
\hline
\hline
&&\\
PPM reconstruction and back-tracing& PPM & PPM-HLLC\\[6pt]
\hline
&&\\
Piecewise linear in space, Runge-Kutta in time&RK-exact&RK-HLLC\\[6pt]
\hline
\end{tabular}
\caption{The table summarises the four codes we tested. Along the horizontal
axis the Riemann solver changes, while vertically the higher order algorithm varies}
\label{codestable}
\end{figure}
The codes RK-HLLC and PPM-HLLC both run about 20\% faster than PPM on the same
data with the same resolution. However, the difference between the algorithms might be larger, since only the original PPM-code was optimised.
The RK-exact code is the slowest, but it was of
fundamental interest in this project to compare its accuracy to the RK-HLLC code's.

In the remainder of this introduction we will first sum up the main ideas
of the underlying PPM algorithm in the Prometheus code. Then we will describe
the new Bouchut-HLLC solver and its theoretical advantages. The second order
algorithm is outlined next, and we show how it preserves positivity. In
sections 2,3 and 4 the one-, two- and three-dimensional test comparisons are
presented. At the end there is a conclusion. 

\subsection{PPM and the Prometheus code}
The basis of the Prometheus code is the one-dimensional PPM-method of \cite{ppmpaper}
with the iterative Riemann solver from \cite{ppmriemann}. Strang splitting is then used
to handle multidimensions. The crucial point of PPM is the so-called
characteristic back-tracing. This technique produces a second order
approximation to the states at the cell interfaces at the half time
step, allowing the use of the midpoint method in time. These approximate states are then used as input to the Riemann
solver. Although the overall accuracy is second order, the spatial
reconstruction is piecewise parabolic, which is reported to give better
resolution than piecewise linear reconstruction. Furthermore, the accuracy at
contact discontinuities is improved by an algorithm that detects them, and then
steepens the reconstructed density. There is also an algorithm that adds
artificial diffusion in order to avoid oscillations behind strong, slow-moving
shocks without smearing out the solution much. The reconstruction is required to be monotocity preserving. This means that the order of the scheme may
drop locally at extremal points of a reconstructed quantity, which means a primitive quantity in the case of PPM as in \cite{ppmpaper}. This drop in order is also a feature of the second order reconstruction we use with the Runge-Kutta time integration.

In order to resolve shocks and shock interactions a Riemann problem is solved with the data from the back-tracing operation as input. For the Euler equations there is no general explicit formula for the solution of the Riemann problem, so in Prometheus an iterative procedure provides instead approximate values of the fluxes at cell interfaces. For efficiency the number of iterations is limited to a fixed number.

\subsection{The HLLC solver of Bouchut and stability}
The notion of an approximate Riemann solver goes back to the Roe solver, see \cite{roescheme}, which is based on a local linearization of the fluxes at the cell interface. We refer
to \cite{Torobook} and \cite{Bouchutbook} for a modern presentation of the basic
ideas. The basic idea is to replace the exact Riemann fans in Godunov's
method with something simpler that still gives a numerical flux that is
consistent and conservative. In addition entropy consistency and
positivity of density must be somehow ensured. For this linearized solvers
require additional treatment, a so-called entropy fix. 

The simplest
approximate Riemann solver is the HLL solver, where the Riemann fan is
replaced by a constant state separated by two discontinuities moving with
constant speeds $C_l$ and $C_r$. A sufficient condition for stability is that
the exact Riemann solution does not have waves with speed outside the interval
$[C_l,C_r]$. The main weaknesses of this approach is that material contact
waves are smeared out, and that the signal velocities $C_l$ and $C_r$ have to
be guessed. A solution to the first problem was hinted at already in \cite{HLL},
and was carried out in \cite{HLLC}, see also \cite{Torobook}, with the so called
HLLC-solver. The HLLC-solver consists of three discontinuities traveling with
speeds  $C_l$, $u^*$ and $C_r$, where velocity and pressure are held constant
across the middle wave, and $u^*$ is this intermediate value of the velocity.

In \cite{Bouchutbook} the HLLC-solver was improved by showing that it results from a relaxation
system, which established its entropy stability. Furthermore, sharp explicit
formulas for the signal speeds that ensure positivity and entropy
stability could be given. We refer to these as Bouchut speeds, and they are
given by formula (2.133) and Proposition 2.18 of \cite{Bouchutbook}.

\subsection{The new second order algorithm}
When going to higher order, requiring the reconstruction to be entropy
dissipative leads to impractical methods, but there is a way to preserve positivity in a
rigorous manner. In the rewritten version of Prometheus RK-HLLC and RK-exact we
used the following reconstruction, based on \cite{Bouchutbook}

As a limiter we use
\beq
\text{minmod}(a_l,a_r)=\bca 0, \quad\quad a_la_r\leq 0\\
\text{sign}(a_l)\min\(\alpha |a_l|,\,\, \half(|a_l|+|a_r|),\,\,\alpha |a_r|\), \quad a_l a_r>0
\eca
\eeq
with $\alpha$ set to 1.4. This is applied to produce the discrete differential
\beq
D\rho_i=\frac{1}{h}\text{minmod}(\rho_{i+1}-\rho_i, \rho_i-\rho_{i-1}),
\eeq
and $Du_i$, $D(u_\n)_i$, $D(\rho e)_i$ similarily. The
positivity of the reconstructed density is guaranteed since
$\rho_i-\frac{h}{2}|D\rho_i|>\min_k(\rho_k)$. Conservation of momentum
dictates that we take the reconstruction slope
\beq
D(\rho u)_i=\rho_i Du_i+u D\rho_i-\frac{h^2}{4}D\rho_i Du_i,
\eeq
and
similarily for $D(\rho u_\n)_i$. Energy is conserved by replacing $e_i$ with
\beq
\tilde{e}_i=e_i-\frac{h^2}{8}\(1-\frac{h^2}{4\rho_i^2}D\rho_i^2\)\(Du_i^2+(Du_\n)_i^2\)
\eeq
when computing the reconstructed internal energy.
The extra terms cancel out the conservation errors in kinetic energy caused by
the linear reconstruction. Hence positivity means that
$\rho_i\tilde{e}_i-\frac{h}{2}|D(\rho e)_i|>0$, or in other words
\beq
\frac{h^2}{8}\(Du_i^2+(Du_\n)_i^2\)<
\frac{1}{1-\frac{h^2}{4\rho_i^2}D\rho_i^2}\(\rho_i e_i-\frac{h}{2}|D(\rho e)_i|\)\hat{=}\Lambda_i^2
\label{posreconstr}
\eeq
To ensure \eqref{posreconstr} in a consistent way we
first restrict $|D(u_\n)_i|$ to less than or equal to $\Lambda_i$, and then set
$Du_i^2$ less than or equal to $\Lambda_i^2-|D(u_n)_i|^2$. Note that in
practice we multiply $\Lambda_i $ with a number slightly less than one to
ensure that the inequality \eqref{posreconstr} is strict.

We did not apply any special treatment of material contact waves in this code version,
and no artificial diffusion was added at shocks.

The numerical time integration is a second order Runge-Kutta method. That is,
one does two full time steps, and then averages the resulting cell average
with the initial one. This procedure preserves positivity, and is total
variation diminishing. Multidimensionality is taken care of by Strang splitting just as in the PPM-codes.

\section{One space dimension: Shock tube tests of Toro}
\label{1dtests}
A basic setup for testing these methods are one-dimensional Riemann problems,
or shock tube tests. In \cite{Torobook} five very useful such test problems are
given and subjected to several different Riemann solvers. The problems are
carefully devised to exhibit phenomena known to be hard to reproduce
numerically. 

As reference solutions we simulated all tests with $10^5$ grid cells using
the original PPM-code. In some cases this was not locally converged due to
spurious oscillations etc., and we point out these anomalies when they
occur. In all the runs the CFL-number was 0.4, and we considered $x\in(0,1)$
with a resolution of 100 grid cells.

\subsection{Test 1}
The first test is not the most severe, but it contains a transsonic rarefaction,
which nonentropic schemes have trouble with. The initial data are
\beq
(\rho,u,p)=\bca 
(1,0.75,1),\quad &x<0.5\\
(0.125,0,0.1),\quad &x>0.5.
\eca
\eeq
All schemes handle the transsonic rarefaction without any signs of a nonentropic glitch, but there are differences in the resolution at the rear end of
the rarefaction with the PPM doing the best job. However PPM gives
large oscillations behind the contact discontinuity compared to the other
codes. With the RK codes there is little difference between the Riemann
solvers. We note the undershoot in front of the contact, and the less sharp
resolution of the contact compared to PPM. These observations also hold true at increasing resolution, as illustrated by the $L^1$-errors in mass density in Figure \ref{toro1table}.
\begin{figure}\centering
\begin{tabular}{cc}
\includegraphics[scale=0.47]{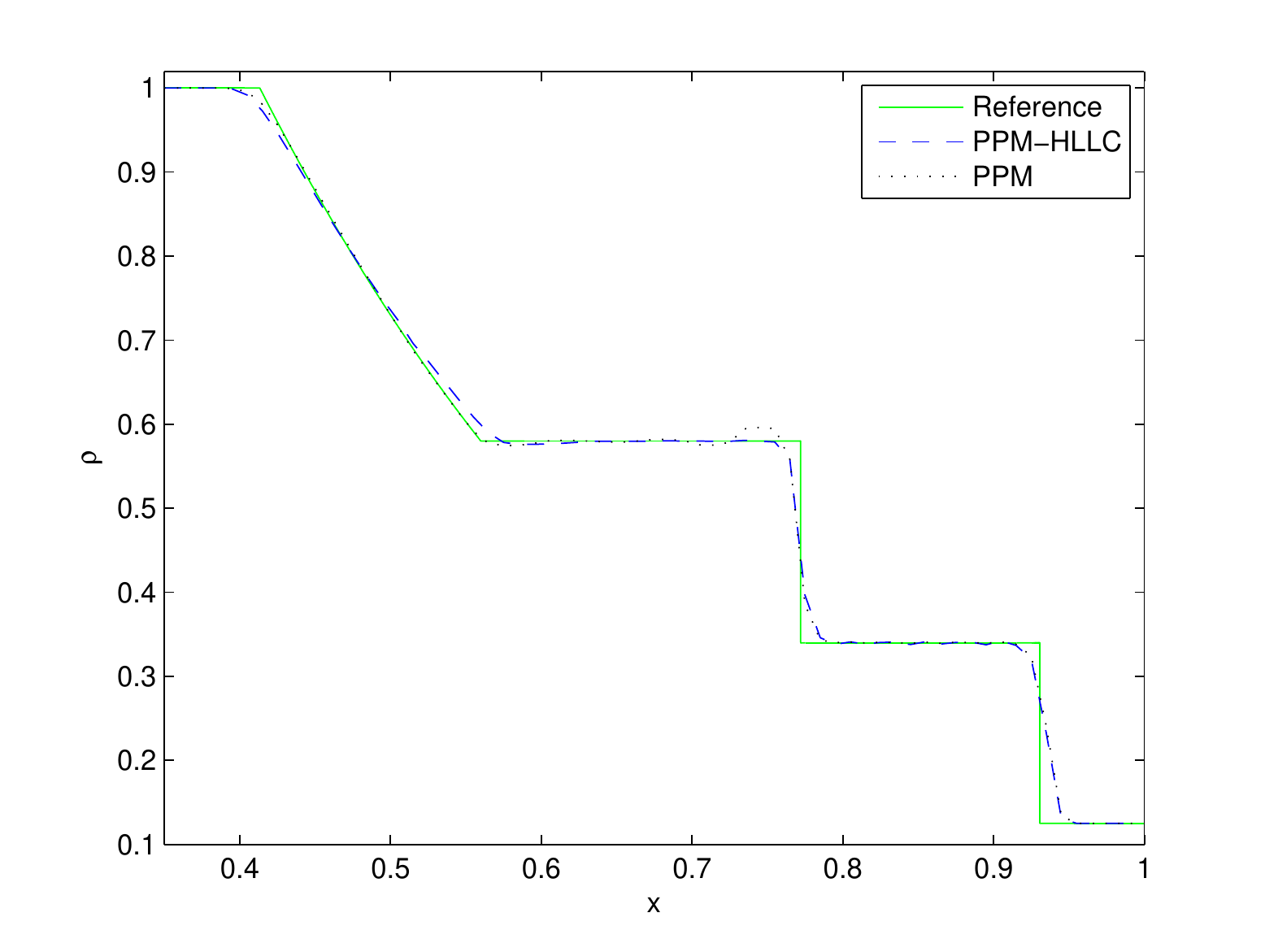}
&
\includegraphics[scale=0.47]{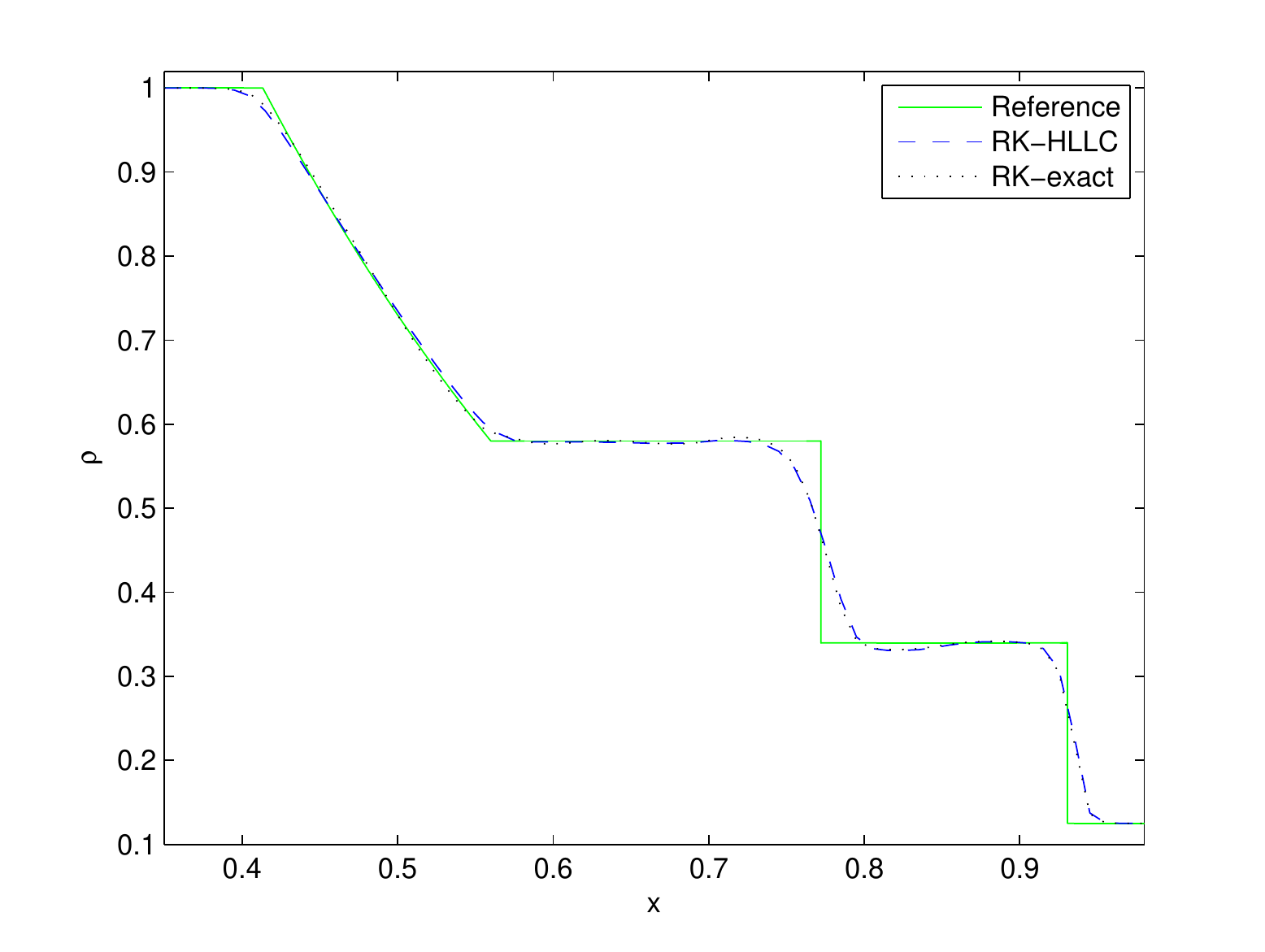}
\end{tabular}
\caption{Results for Toro test 1.}
\label{toro1rho}
\end{figure}
\begin{figure}
%\\[6pt]
\begin{tabular}[c]{|l||c|c|c|c|}
\hline
%& \multicolumn{2}{|c|} {}\\[-8pt]
%& \multicolumn{2}{|c|}{Riemann solver:}\\[3pt]
%\cline{2-3}
%& &\\[-8pt]
Resolution:&PPM & PPM-HLLC &RK-HLLC & RK-exact \\
\hline
\hline
$h=0.01$ & 3.75 & 3.11 & 5.07 & 5.60\\
\hline
$h=0.005$ & 1.48 & 1.61 & 2.63 & 2.92\\
\hline
$h=0.0025$ & 0.88 & 0.93 & 1.73 & 1.85\\
\hline
$h=0.00125$ & 0.42 & 0.39 & 0.95 & 0.98\\
\hline
\end{tabular}
\caption{The table shows the $L^1$-error in the computed mass density of Toro's test 1 for different codes and resolutions. The errors are given in units of $10^{-3}$.}
\label{toro1table}
\end{figure}

\subsection{Test 2}
Test 2 has two rarefactions going apart creating a low density region. The initial data are
\beq
(\rho,u,p)=\bca 
(1,-2,0.4),\quad x<0.5\\
(1,2,0.4),\quad x>0.5
\eca
\eeq
The solver should be able to handle this without giving negative density or
pressure. In particular, linearized solvers
have trouble with such cases.
\begin{figure}\centering
\begin{tabular}{cc}
\includegraphics[scale=0.47]{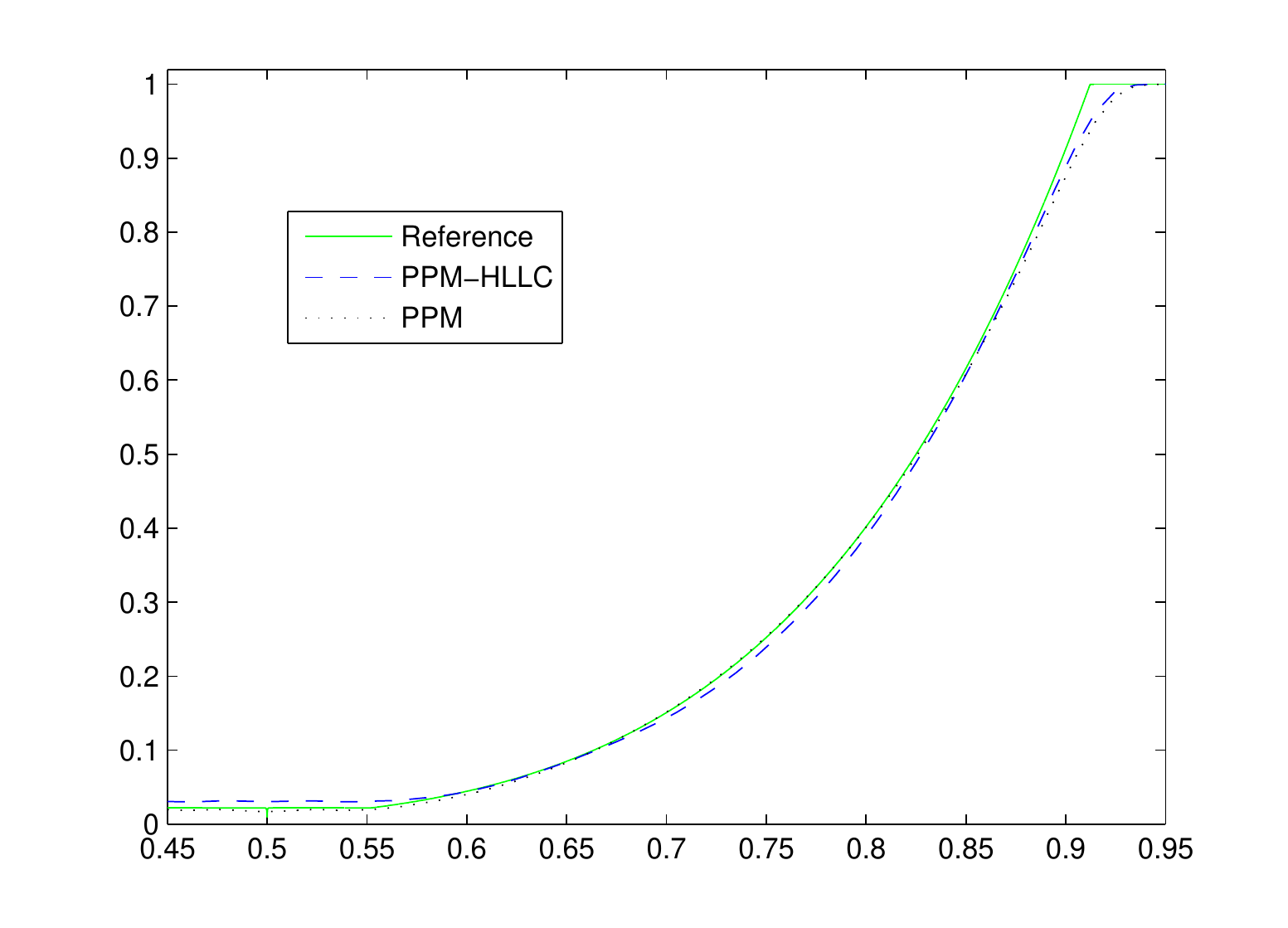}
&
\includegraphics[scale=0.47]{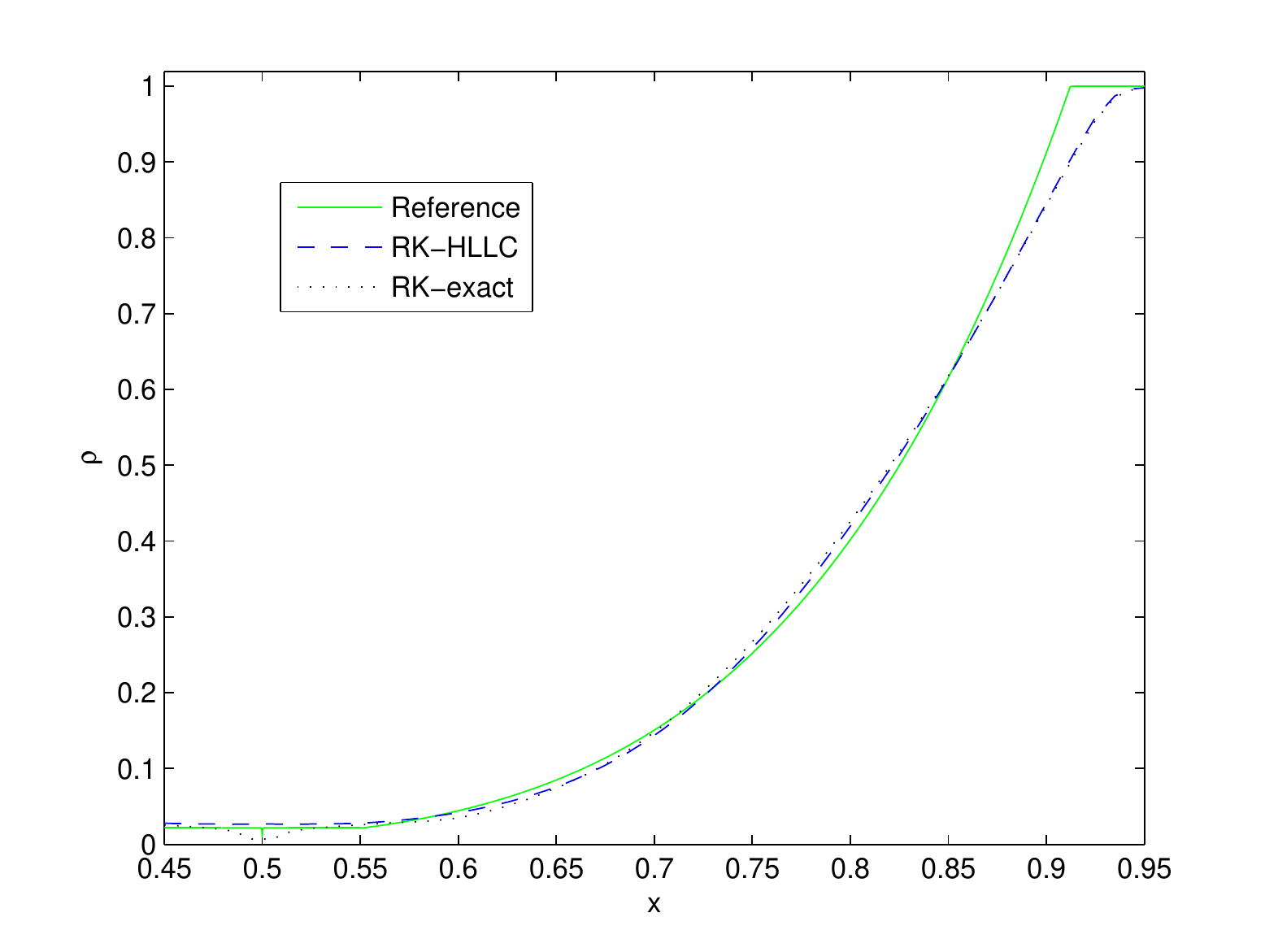}
\end{tabular}
\caption{Results for Toro test 2. The results are symmetric
  around $x=0.5$.}
\label{toro2rho}
\end{figure}
In the density plots, Figure \ref{toro2rho}, we note a bump in the density at
$x=0.5$ with the RK-exact code. We see similar tendencies for the PPM
simulation, and in the PPM reference solution there is a deep narrow bump.
\begin{figure}\centering
\begin{tabular}{cc}
\includegraphics[scale=0.47]{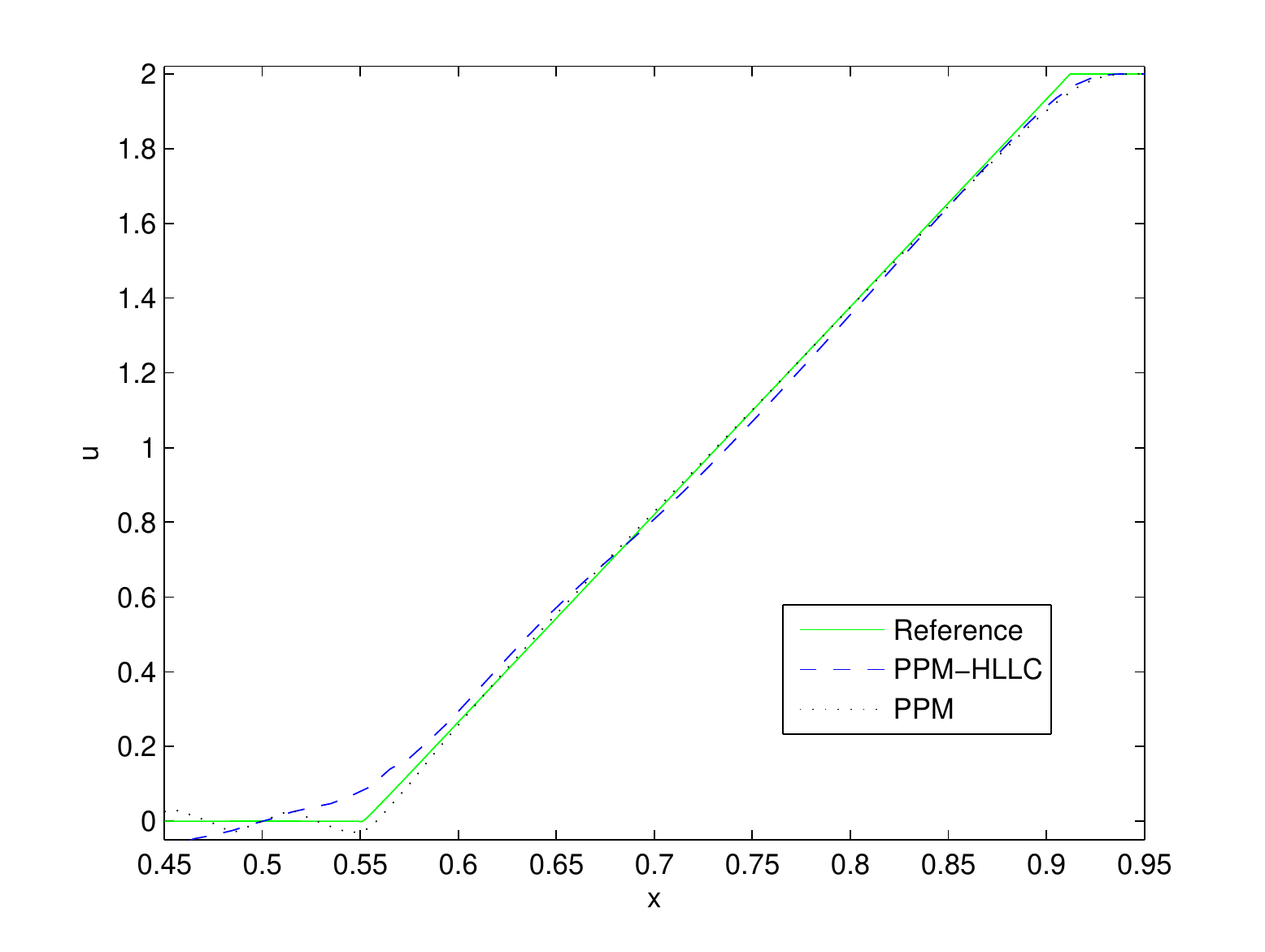}
&
\includegraphics[scale=0.47]{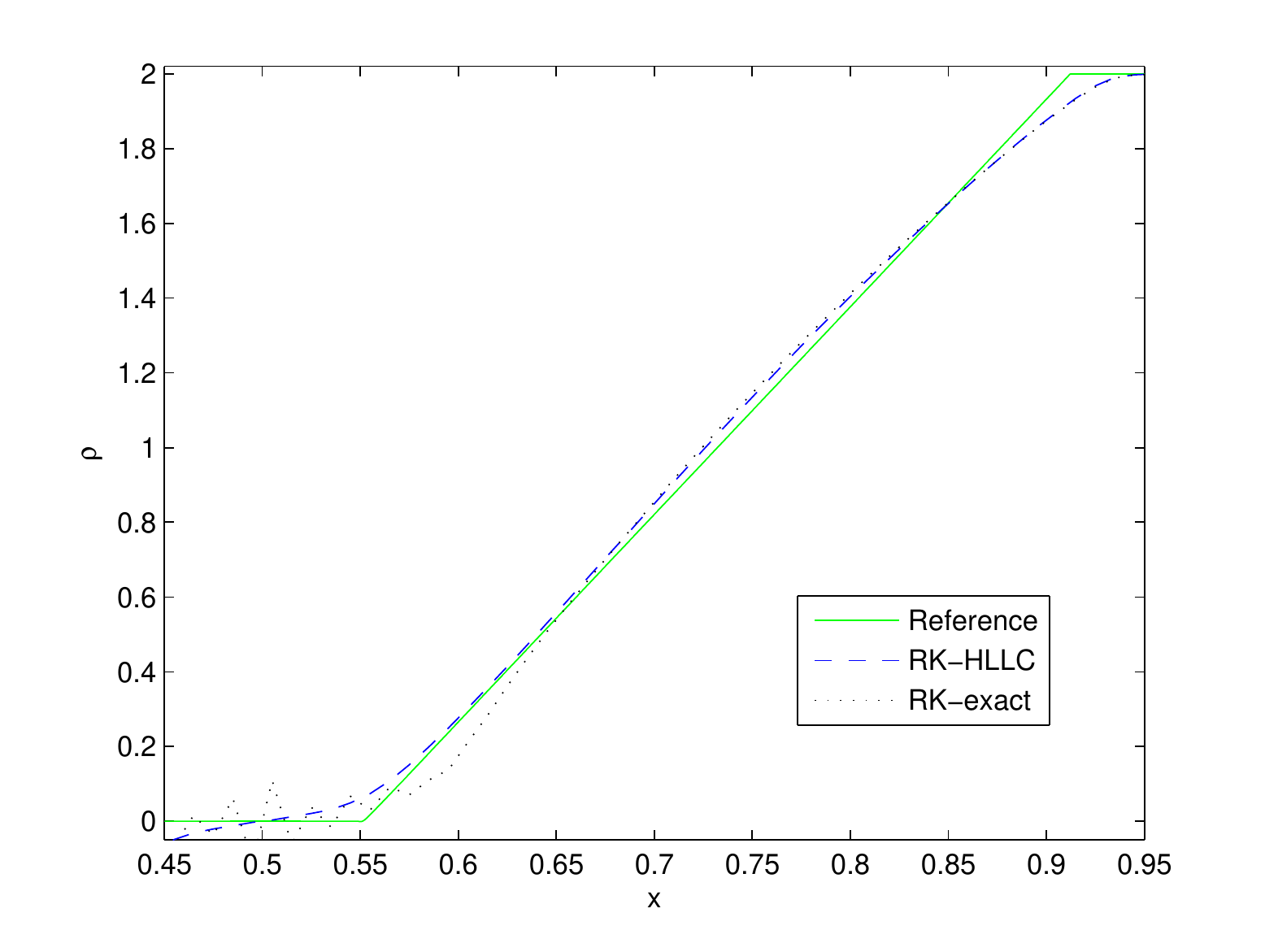}
\end{tabular}
\caption{Results for Toro test 2. The results are symmetric
  around $x=0.5$.}
\label{toro2u}
\end{figure}

For the RK-HLLC-code positivity was automatically
maintained, and it is interesting that we get a better approximation of the
density value in the middle compared to PPM-HLLC, and also of the velocity,
Figure \ref{toro2u}. The front of the rarefaction is however best resolved by
the PPM-codes.

Notice in Figure \ref{toro2u} that both PPM and RK-exact (which have the same
Riemann solver) has
oscillations in the velocity near $x=0.5$. The RK-exact code especially had
problems with this test, and positivity had to be artificially imposed for CFL-numbers larger than around 0.05. Theoretically a CFL-number less than 0.25
should ensure positivity with an exact Riemann solver, so this has to do with
the iterative procedure in the Riemann solver not automatically ensuring the positivity
property. With the iterative solver as part of PPM however, this seemed not to
cause serious problems.

\subsection{Test 3}
\begin{figure}\centering
\begin{tabular}{cc}
\includegraphics[scale=0.47]{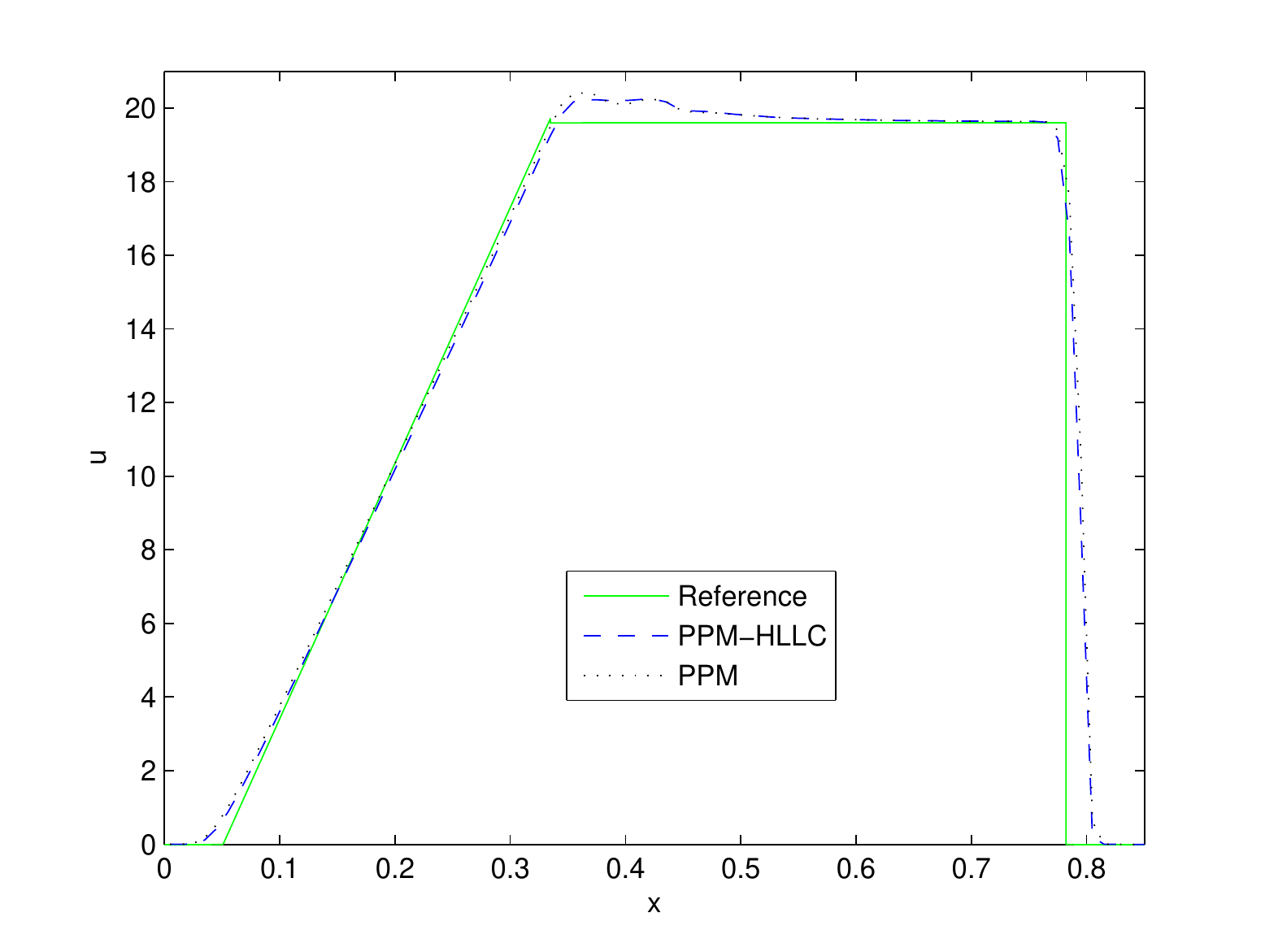}
&
\includegraphics[scale=0.47]{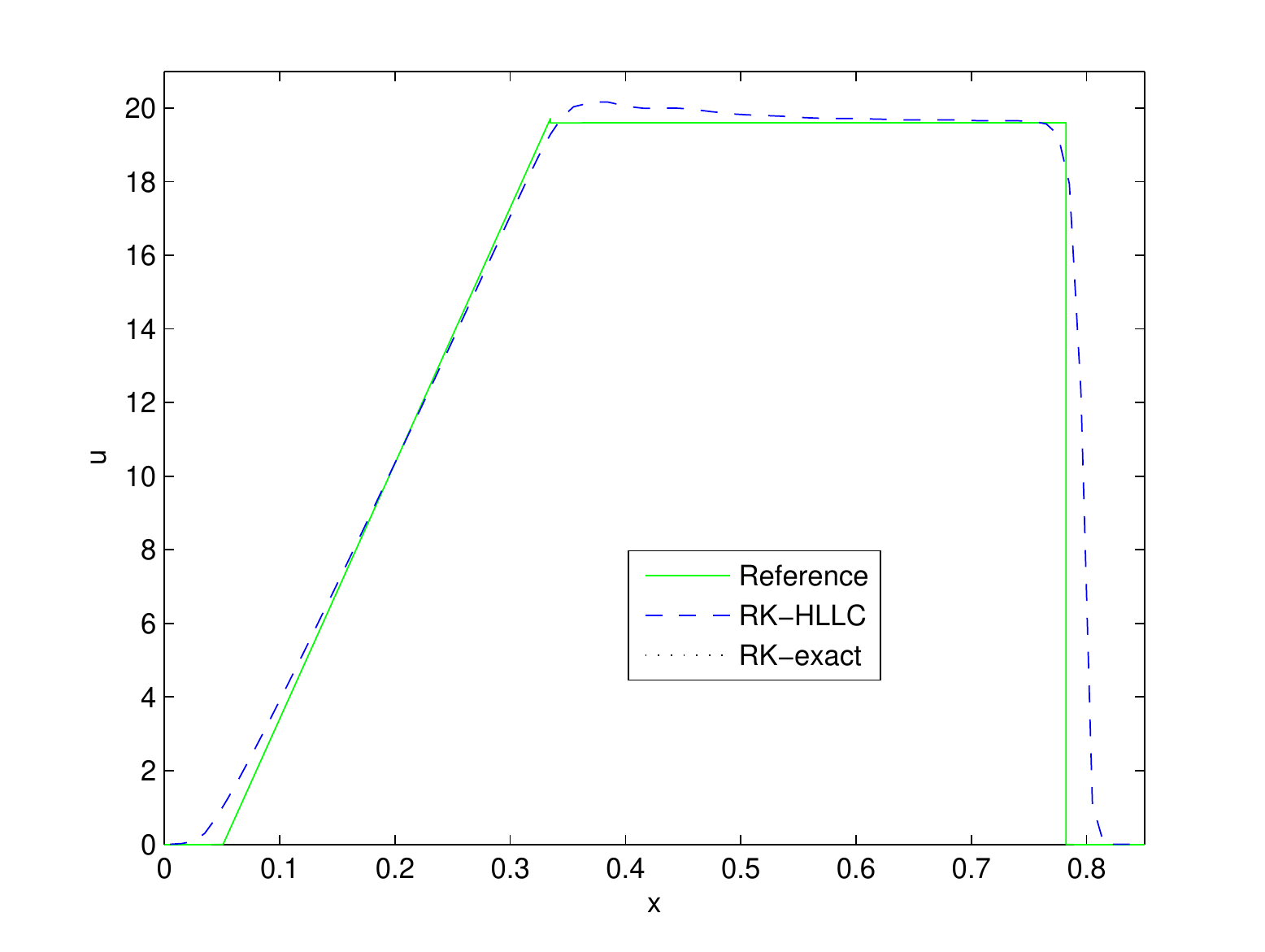}
\end{tabular}
\caption{Results for Toro test 3.}
\label{toro3u}
\end{figure}

Test 3 is a high Mach number shock tube with initial data
\beq
(\rho,u,p)=\bca 
(1,0,1000),\quad &x<0.5\\
(1,0,0.01),\quad &x>0.5.
\eca
\eeq
We found little difference between
the codes here apart from the expected slightly sharper resolution of the
PPM-codes, which was most prominent on the contact wave. All codes produced spurious effects behind the rarefaction, as seen
in the overshoot in the velocity plots, Figure \ref{toro3u}.

\begin{figure}\centering
\begin{tabular}{cc}
\includegraphics[scale=0.47]{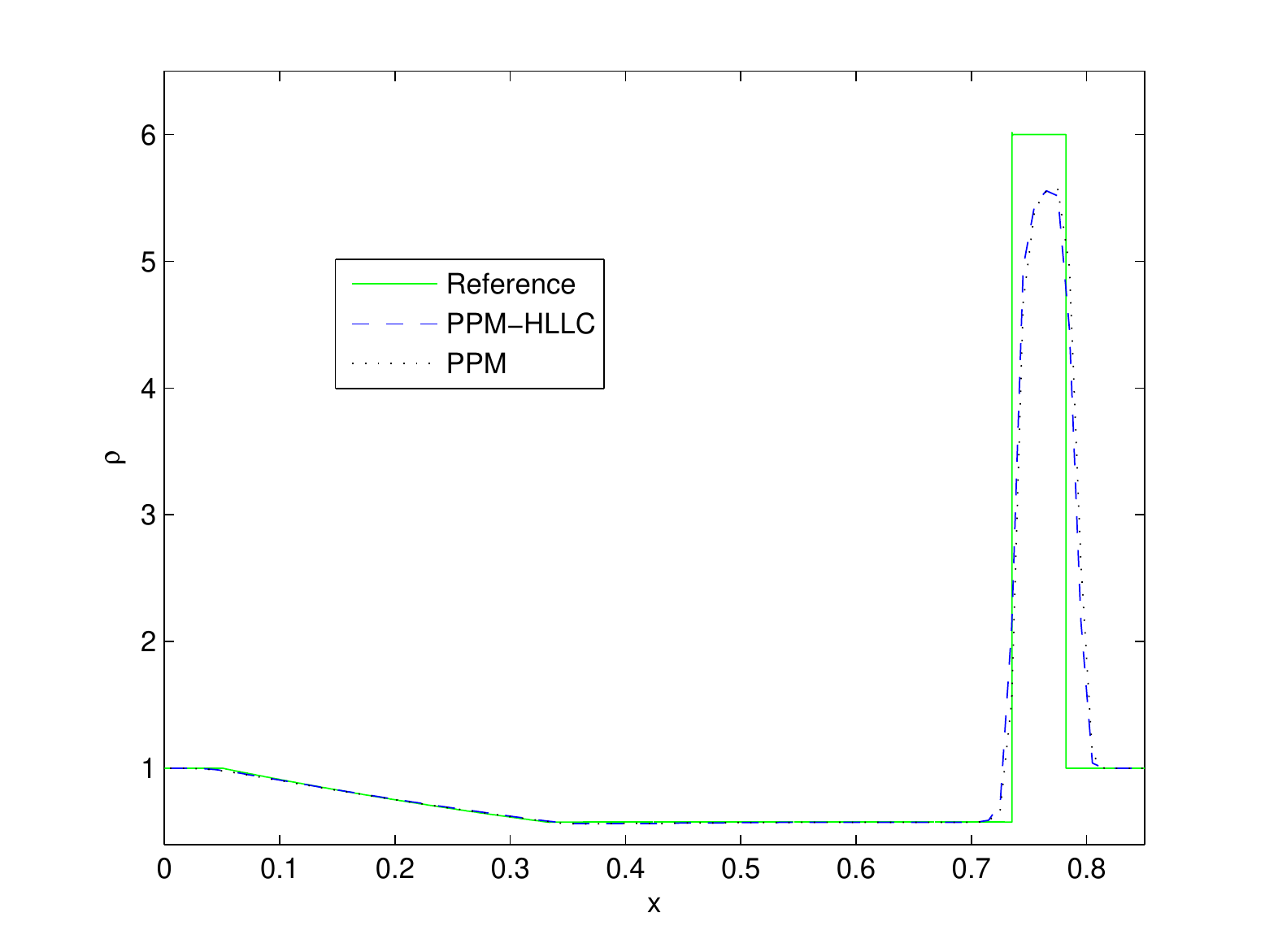}
&
\includegraphics[scale=0.47]{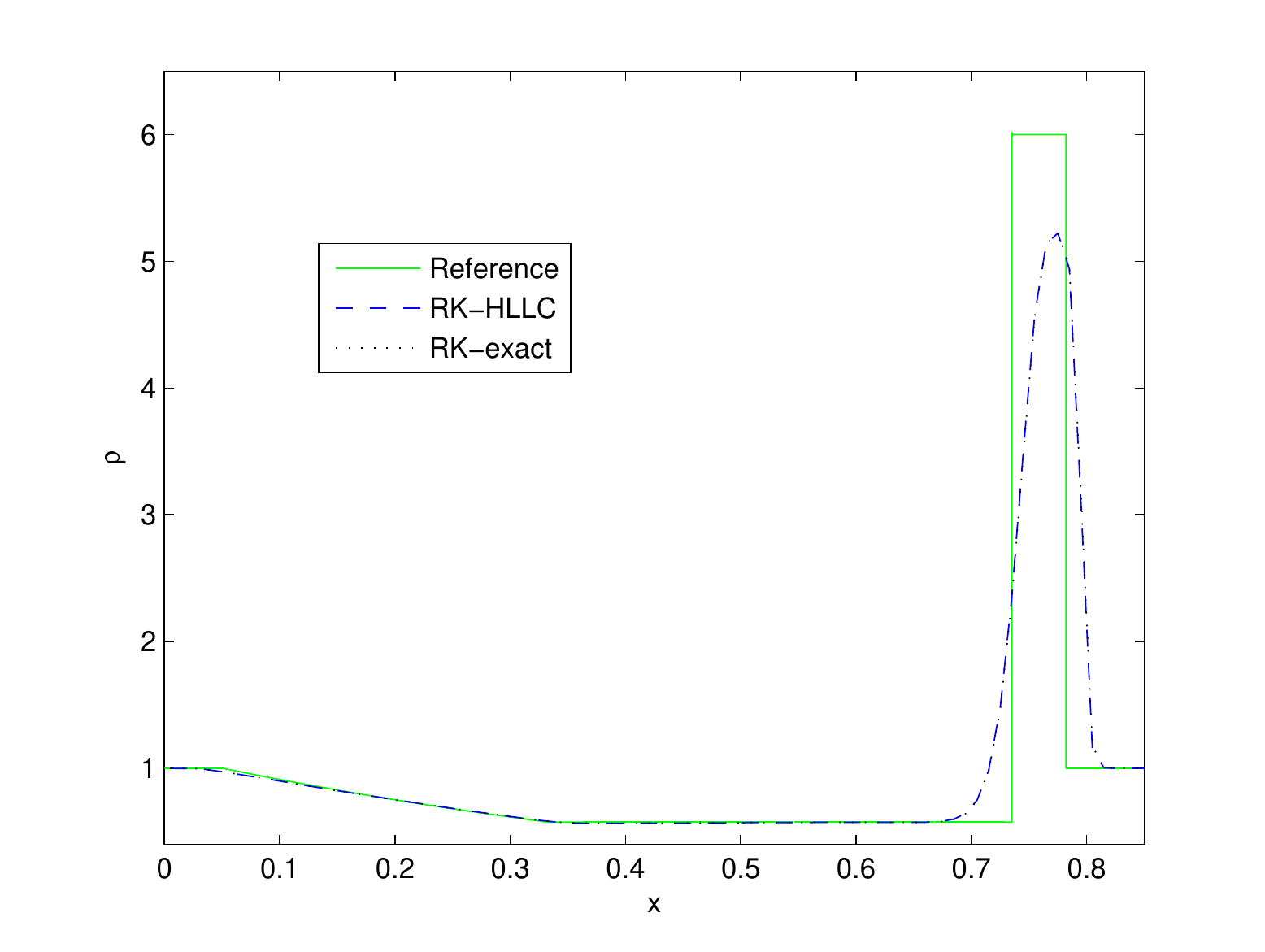}
\end{tabular}
\caption{Results for Toro test 3.}
\label{toro3rho}
\end{figure}

\subsection{Test 4}
The solution of test 4 has a near stationary shock, that is, the shock speed
is small compared to the characteristic speeds, hence if the numerical
diffusion applied to this shock is high, it will be particularly
pronounced. The initial data are
\beq
(\rho,u,p)=\bca 
(5.99924,19.5975,460.894),\quad &x<0.5\\
(5.99242,-6.19633,46.0950),\quad &x>0.5.
\eca
\eeq
Also, oscillations behind such shocks is a well known phenomenon, and in the
original PPM paper, a so called 'flattening' algorithm was introduced which
essentially adds diffusion. This algorithm was used in all the PPM simulations
here, but for this test we also tried to switch it off, resulting in
oscillations in the density of magnitude around 5 percent of the postshock
density both for PPM and PPM-HLLC. The RK-codes only show small oscillation
here, and no special treatment was necessary. In the reference PPM solution there are
pronounced oscillations both after the near stationary shock as well as
between the contact and the right moving shock.
\begin{figure}\centering
\begin{tabular}{cc}
\includegraphics[scale=0.47]{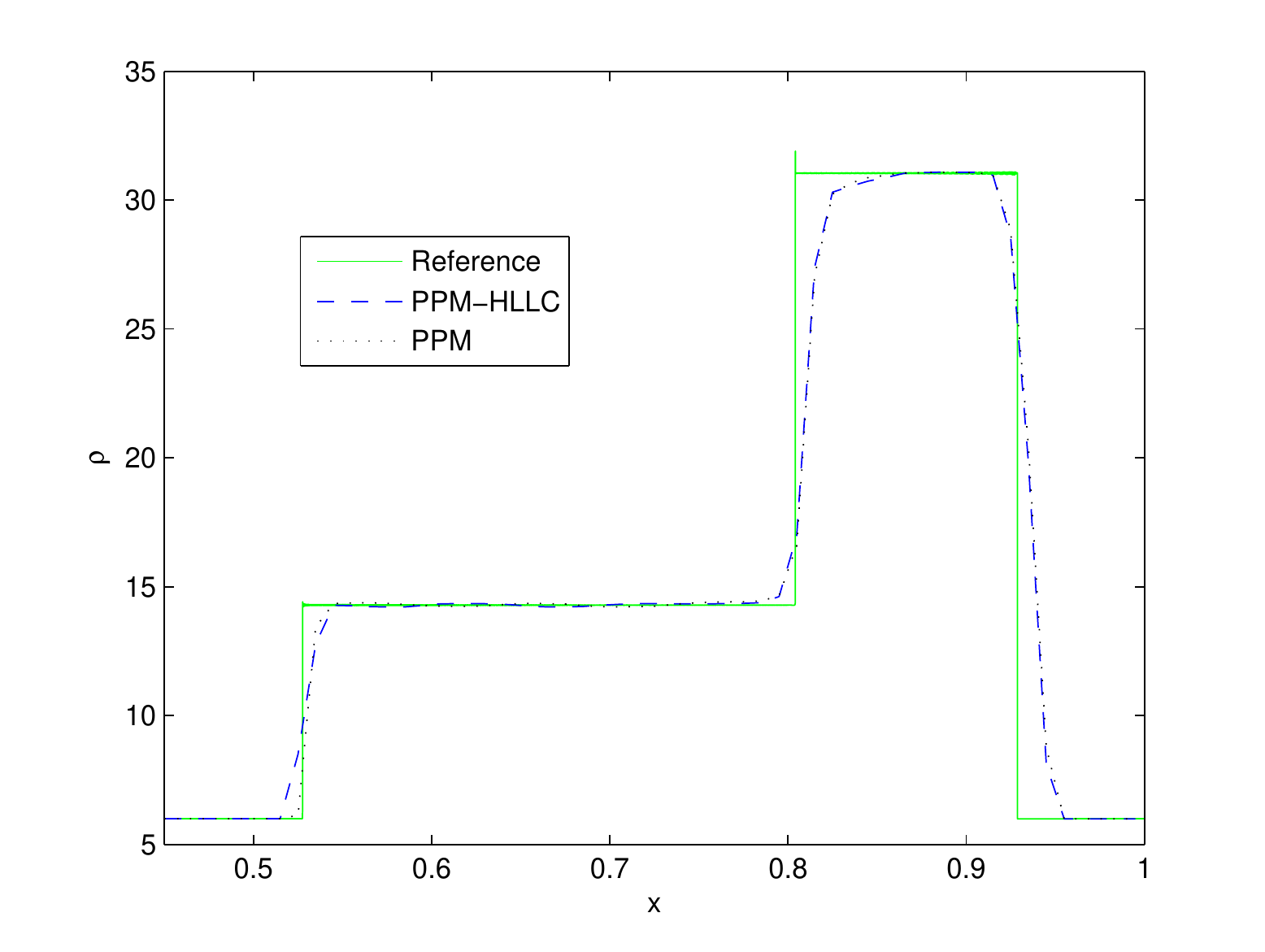}
&
\includegraphics[scale=0.47]{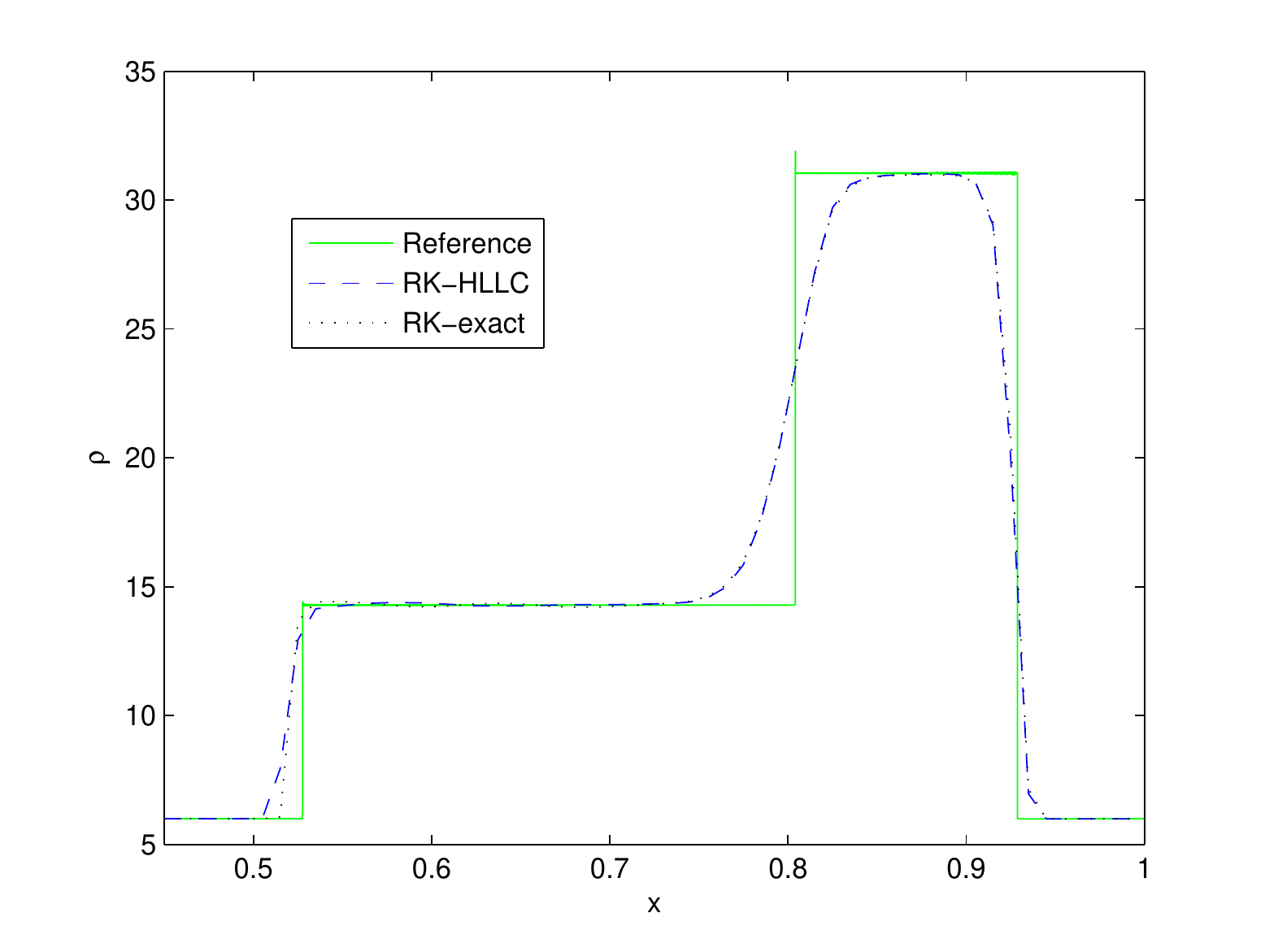}
\end{tabular}
\caption{Results for Toro test 4.}
\label{toro4rho}
\end{figure}
Both HLLC-codes smear the near stationary shock out with 1-2 grid cells
more in front of the shock, as seen Figure \ref{toro4rho} (We show the results
from PPM with flattening). The difference of 1-2 grid cells was maintained
when refining to 200 and 400 grid cells. It is probably caused by the signal
speeds of HLLC-Bouchut slightly overestimating the shock speeds. Otherwise we only note the lower resolution
of the contact wave with the RK-codes compared to HLLC.
\subsection{Test 5}
\begin{figure}\centering
\begin{tabular}{cc}
\includegraphics[scale=0.47]{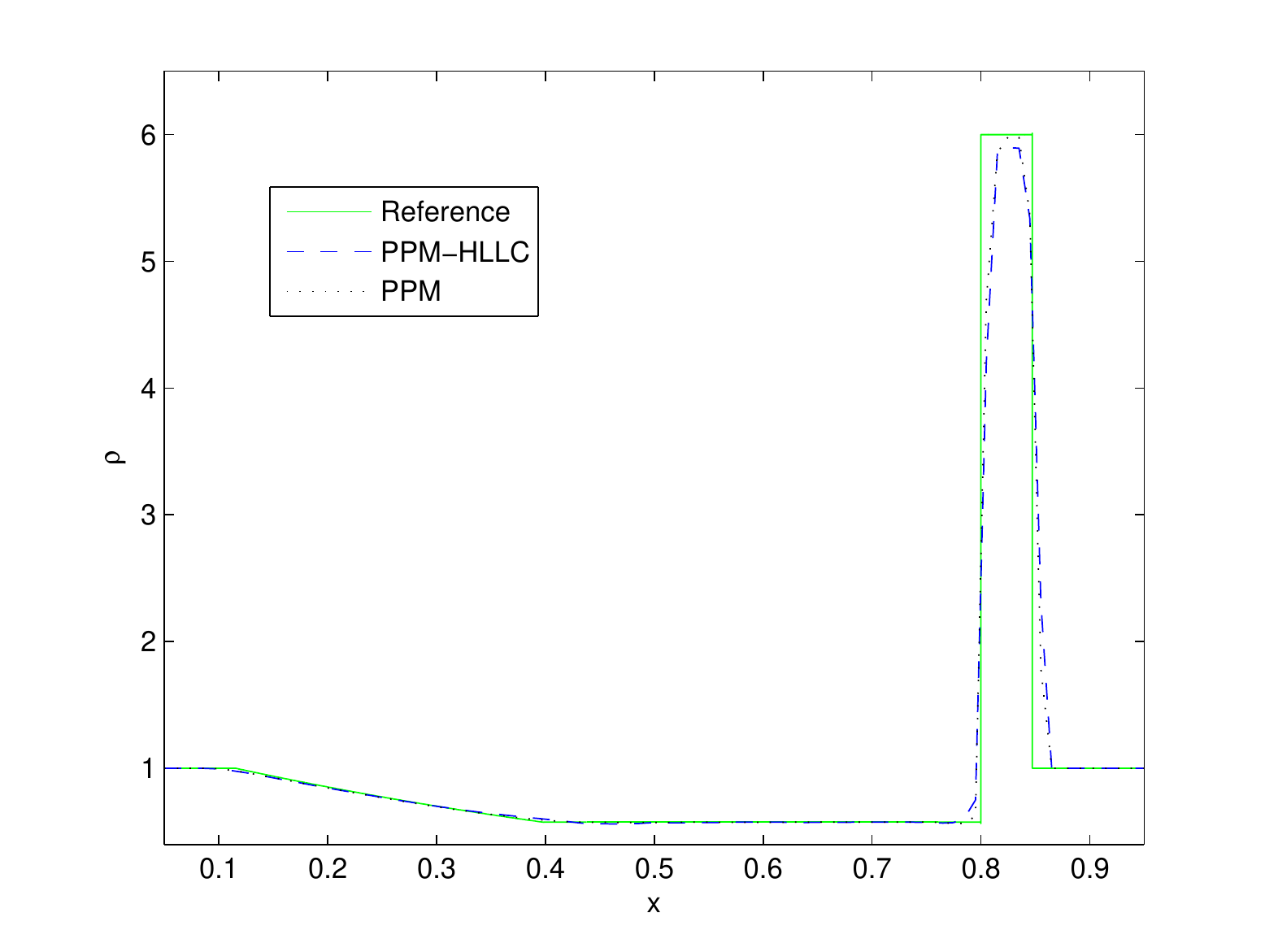}
&
\includegraphics[scale=0.47]{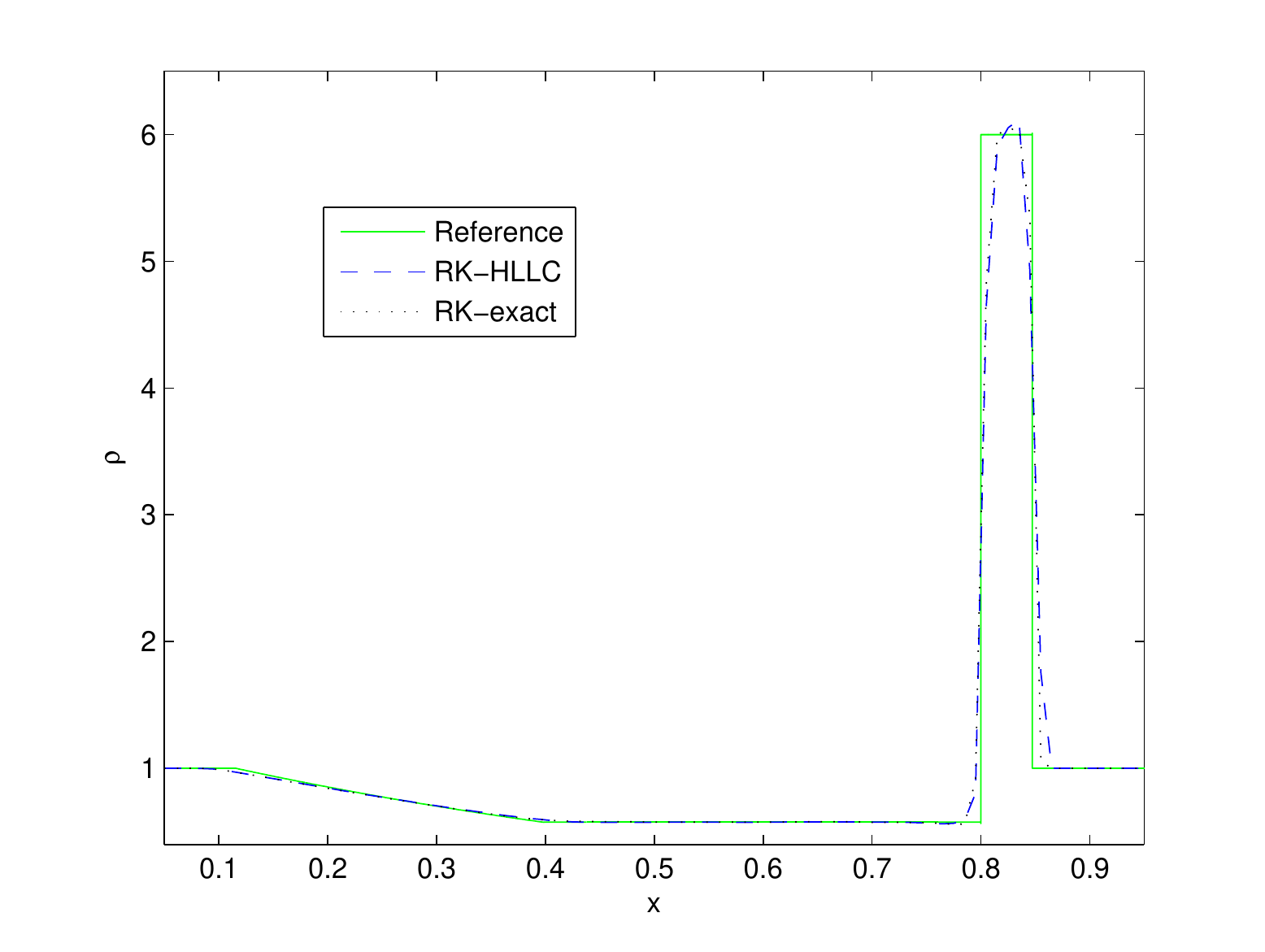}
\end{tabular}
\caption{Results for Toro test 5.}
\label{toro5rho}
\end{figure}

\begin{figure}\centering
\begin{tabular}{cc}
\includegraphics[scale=0.47]{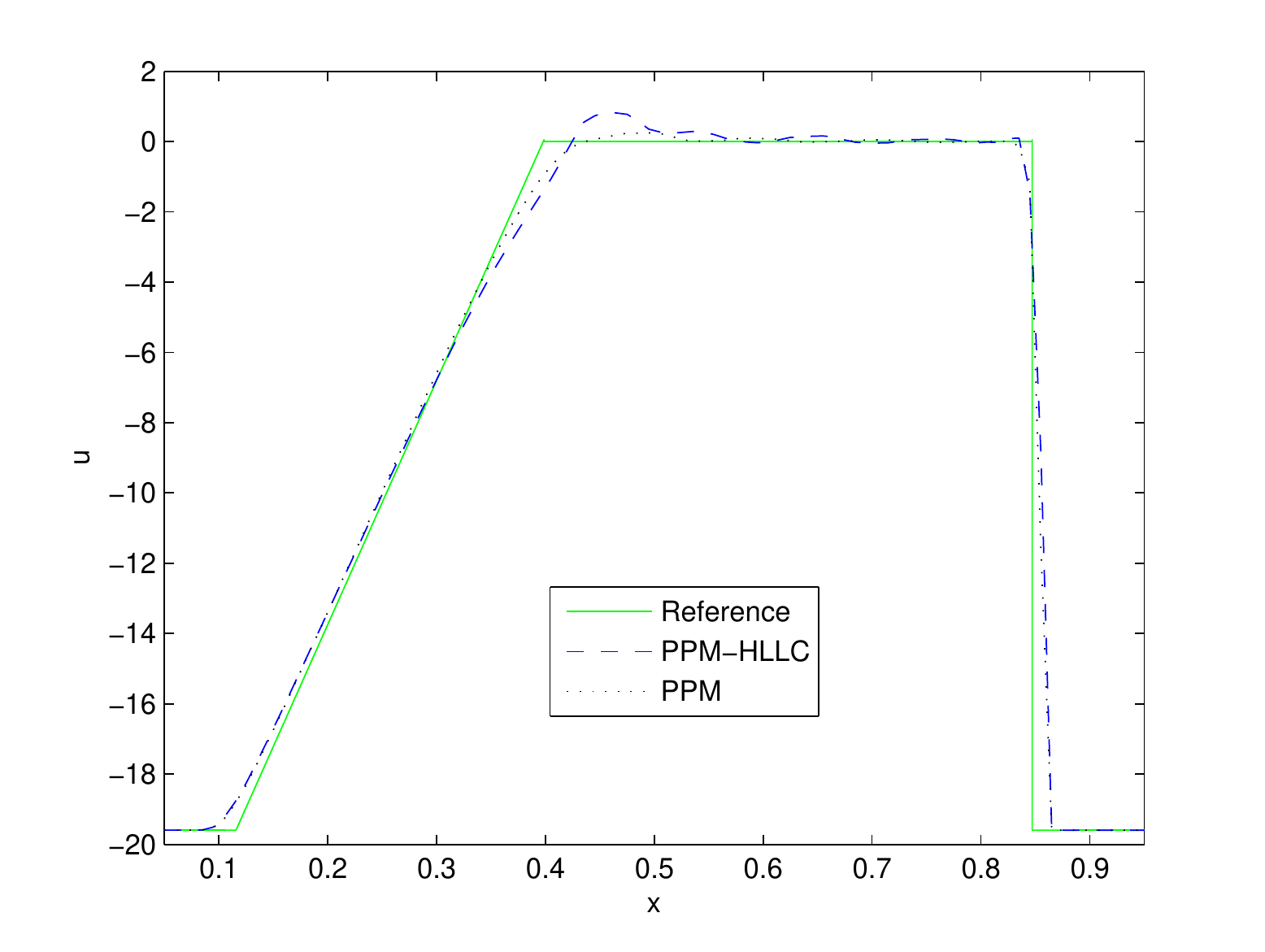}
&
\includegraphics[scale=0.47]{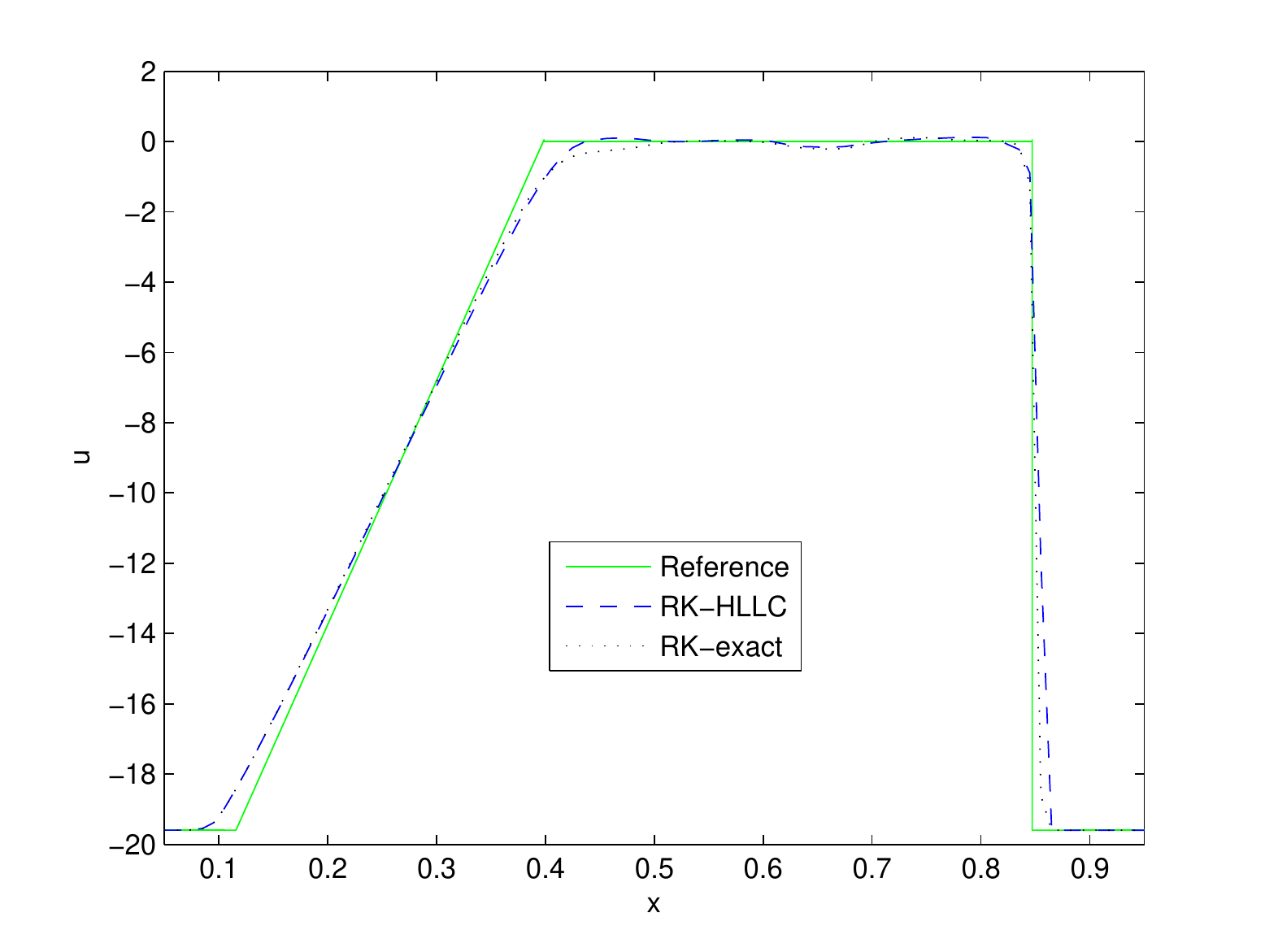}
\end{tabular}
\caption{Results for Toro test 5.}
\label{toro5u}
\end{figure}

Test 5 is like test 3 with a background velocity resulting in a near stationary
contact discontinuity. The initial data are
\beq
(\rho,u,p)=\bca 
(1,-19.59745,1000),\quad &x<0.8\\
(1,-19.59745,0.01),\quad &x>0.8.
\eca
\eeq
All codes handle this feature reasonably well, which was expected since
both Riemann solvers exactly resolve contact waves, see Figure \ref{toro5rho}. Note that in this case
the RK-codes have comparable resolution of the discontinuities to the
PPM-codes, but they overshoot the right intermediate density slightly.
The oscillations occuring behind the rarefaction in the velocity, see Figure
\ref{toro5u}, is not reported with the 1st
order schemes tested by Toro, so it must have something to do with the higher
order algorithms. It is especially pronounced with PPM-HLLC, but
visible in all simulations.
\section{Two space dimensions: Mixing layers}
We now look at transitions from laminar into unstable flows in two
dimensions. By the nature of the underlying unstable flows, although we observe differences in the output, we are not really
able to infer much about the quality of the respective codes. However, we observed that the instabilities needed to be highly
developed before any differences could be seen between the Riemann solvers. In
other words, at
the onset of instability the Riemann solvers seemed to give the same
results. We recall from the one-dimensional tests that changing the Riemann
solver had only a small effect on the numerical smearing. This indicates that
the 'numerical viscosity' varies little between the different schemes, and the
sensitivity of our instabilities to numerical viscosity seems to
be relatively small at their onset. 

\subsection{Kelvin-Helmholtz instability}
Two layers of fluid moving with different parallel velocities are always unstable in the
absence of viscosity and external forces. This is referred to as a Kelvin-Helmholtz instability, and seems to be an important source of turbulence in many applications. We consider a grid-aligned jump
in velocity here, and make a small periodic perturbation.
The initial data are $\rho=1$,  $\gamma=1.4$, and we let $p$ vary to
allow different Mach numbers. The velocity is in the $y$-direction with $v=0.5$ for $x<0.5$, and $v=-0.5$ for $x>0.5$, however we moved the velocity jump one grid cell to the left to break the symmetry. We perturb $v$ with $2\pi e^{2\pi x}cos(2\pi y)/100$ for $x<0.5$ and $-2\pi
e^{-2\pi x}cos(2\pi y)/100$ for $x>0.5$. This means we can compute $y\in
(0,1)$ with periodic boundary conditions, and in the $x$-direction we consider
$x\in(0.1)$ with reflecting boundary conditions. The CFL-number was 0.8 in all
simulations.

\begin{figure}\centering
\begin{tabular}{cc}
\includegraphics[scale=0.65]{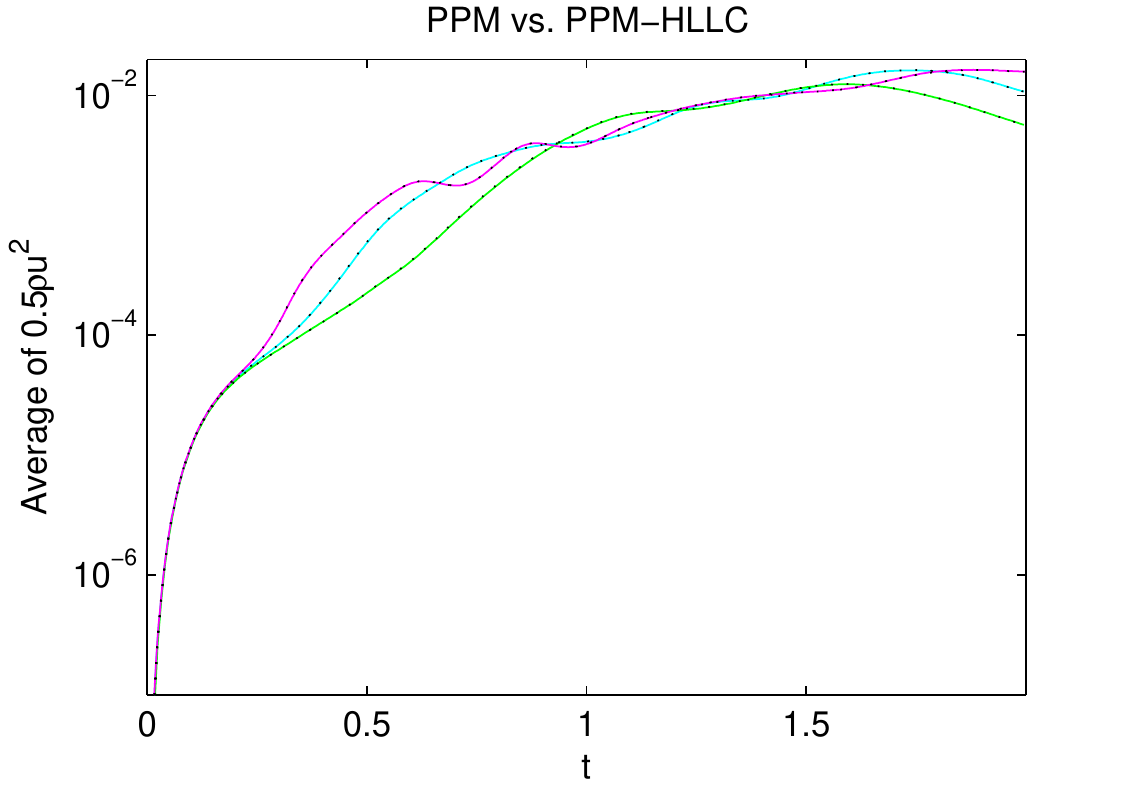}
&
\includegraphics[scale=0.65]{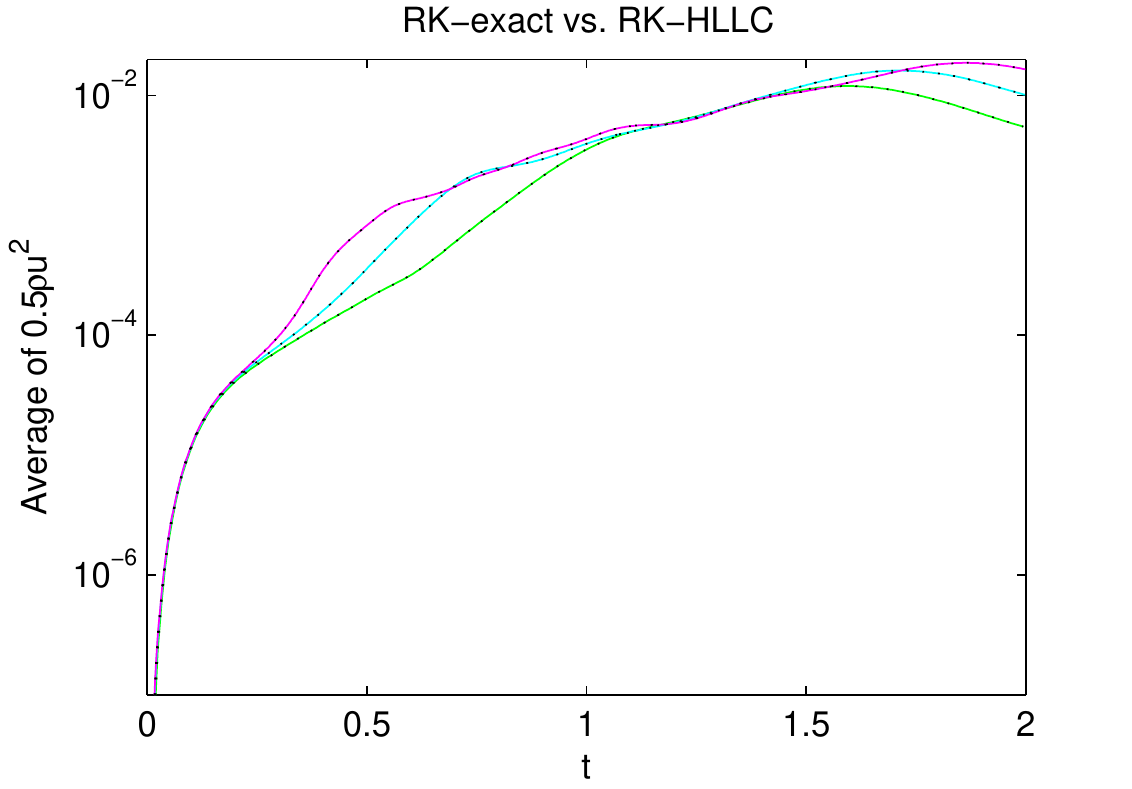}
\end{tabular}
\caption{Growth of transversal kinetic energy component. The codes with exact
  solver are represented with dotted lines, and the HLLC version with solid
  lines. We show data from runs with $100^2$ $200^2$ and $4002$ grid cell, the curve steepness increasing with resolution.}
\label{growthM1}
\end{figure}

First we take $p=1/\gamma$, which means that the relative velocity between the
layers equals the sound speed. We consider the time history of the
average of $\half \rho u^2$, as this quantity is often used as a measure of the growth of the instability. Figure \ref{growthM1} shows $\log \overline{\half \rho u^2}$
as a function of time for the different codes. We used three different resolutions, $100^2$,
$200^2$ and $400^2$ points, and we see here that the instability growth rate increases with resolution. This is plausible, since by linear instability theory, the growth rate is inverse proportional to the perturbation wavelength. The Riemann solver, however, seems to have no influence at all.
We measured the CPU-times for the different codes for this simulation with $200^2$ grid points until time $t=2.0$. We got the following results:
%\begin{figure}
%\\[6pt]
\begin{center}
\begin{tabular}[c]{|l||c|c|c|c|}
\hline
%& \multicolumn{2}{|c|} {}\\[-8pt]
%& \multicolumn{2}{|c|}{Riemann solver:}\\[3pt]
%\cline{2-3}
%& &\\[-8pt]
Code &PPM & PPM-HLLC &RK-HLLC & RK-exact \\
\hline
CPU-time & 1.00 & 0.81  & 0.79 & 1.19\\
\hline
\end{tabular}
\end{center}
These are the averages of two runs with each code. The numbers are normalised to the result from PPM.

\begin{figure}\centering
\begin{tabular}{cc}
\includegraphics[scale=0.47]{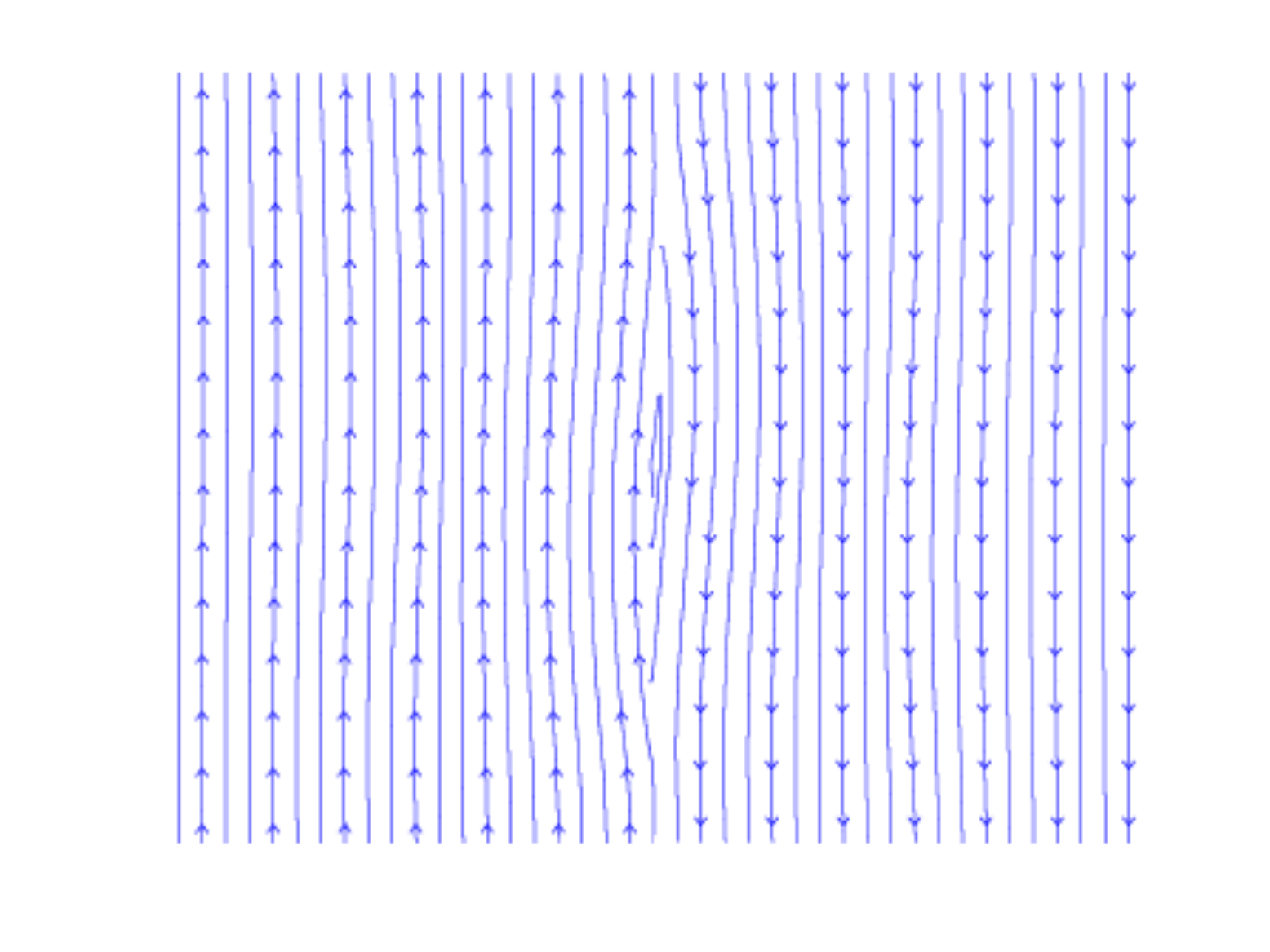}
&
\includegraphics[scale=0.47]{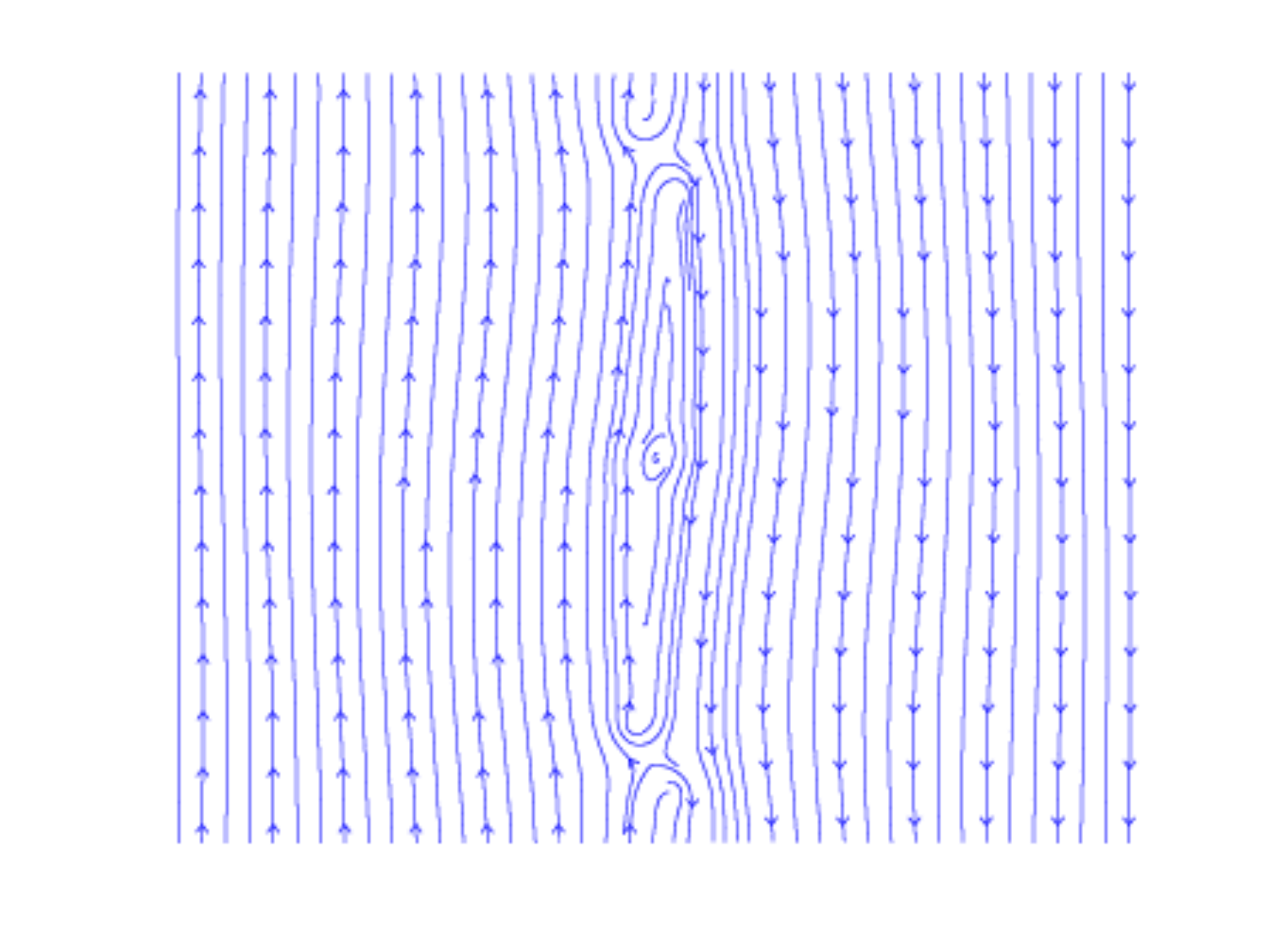}
\\
\includegraphics[scale=0.47]{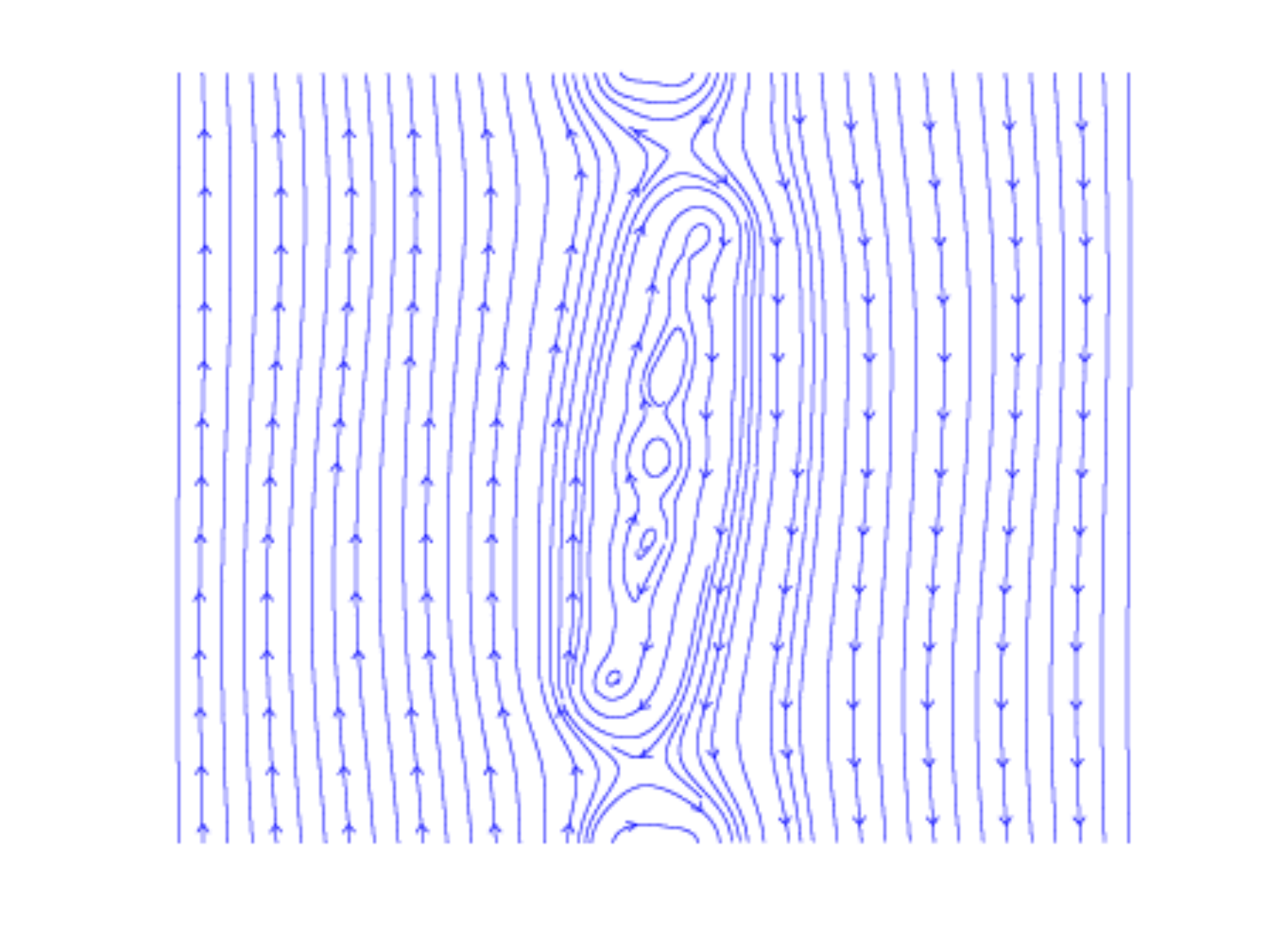}
&
\includegraphics[scale=0.47]{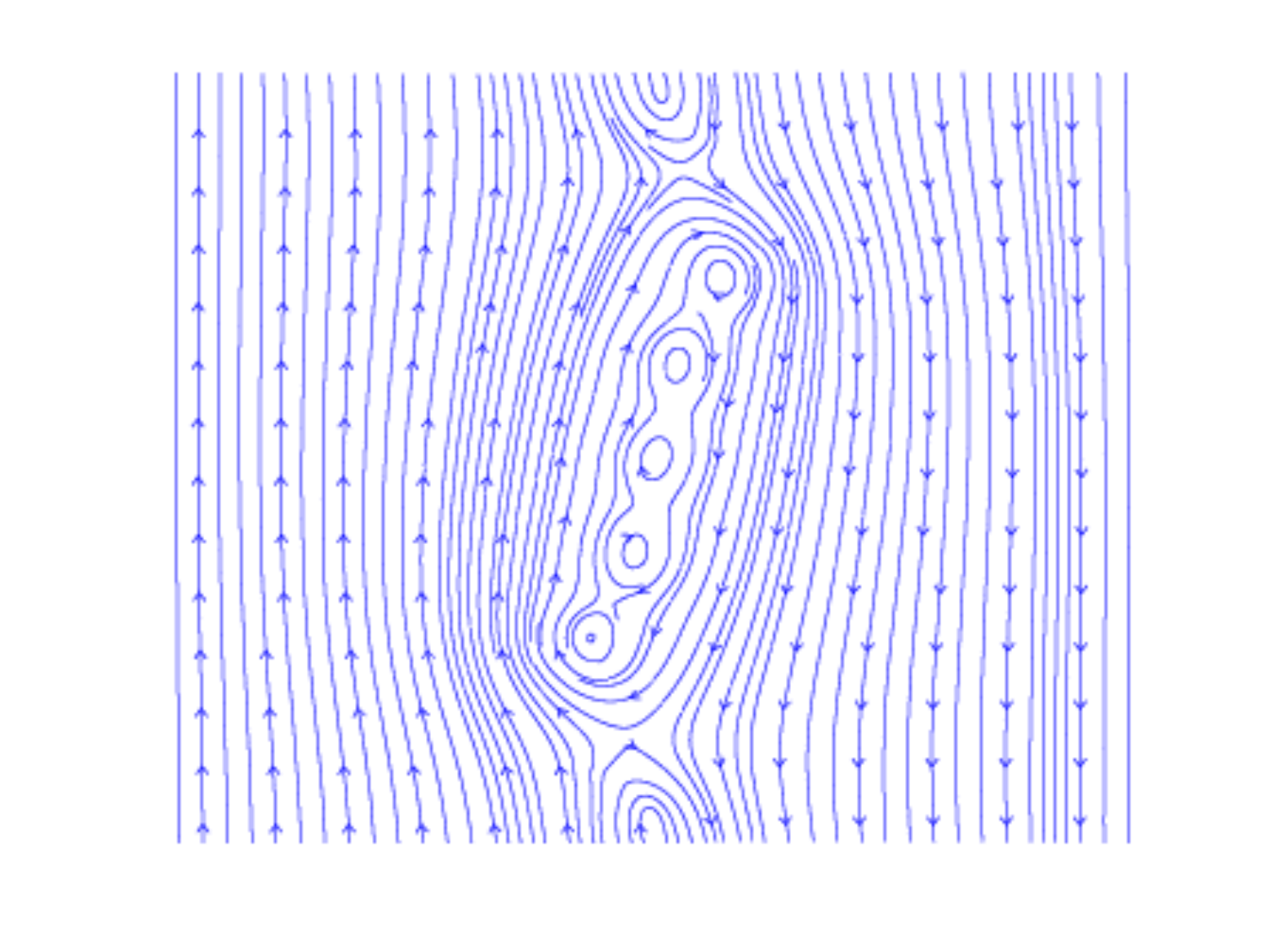}
\end{tabular}
\caption{The development of the Kelvin-Helmholtz instability is illustrated by
  the streamlines at times $0.25, 0.50$ and $1.0$ from PPM-HLLC with $200^2$ resolution.$^*$}
\label{ppmhllcearly}
\end{figure}

\begin{figure}\centering
\begin{tabular}{cc}
\includegraphics[scale=0.47]{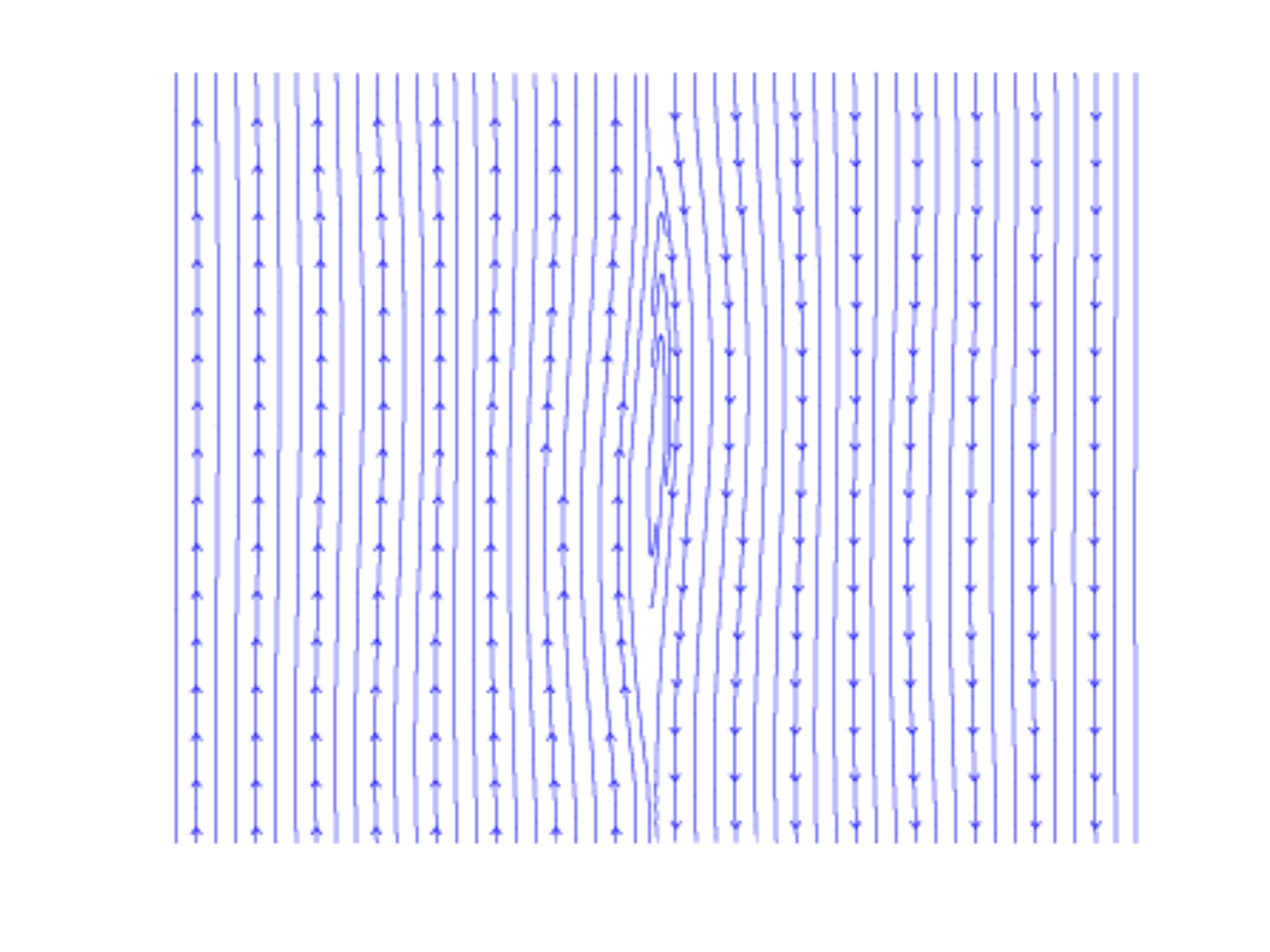}
&
\includegraphics[scale=0.47]{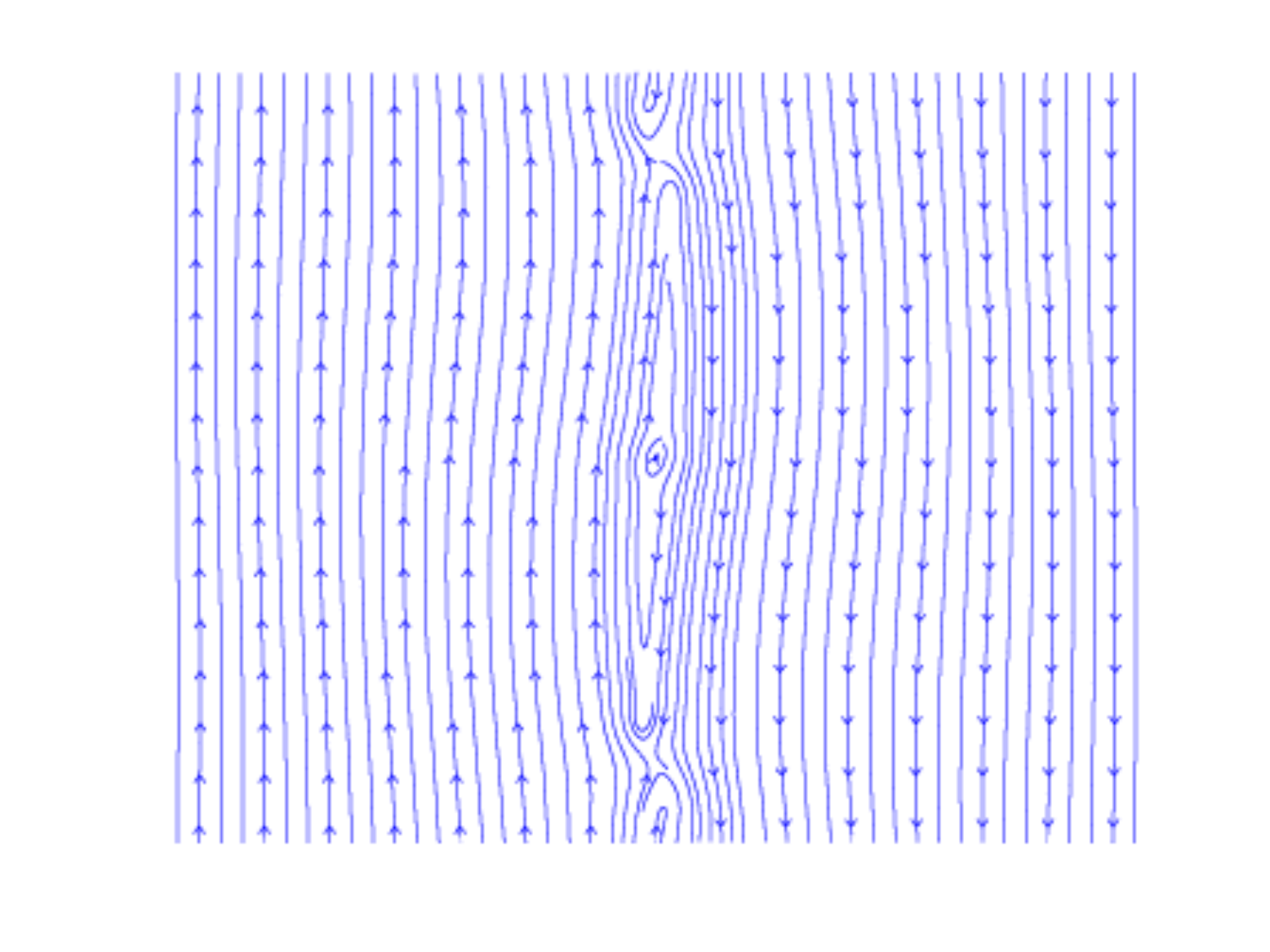}
\\
\includegraphics[scale=0.47]{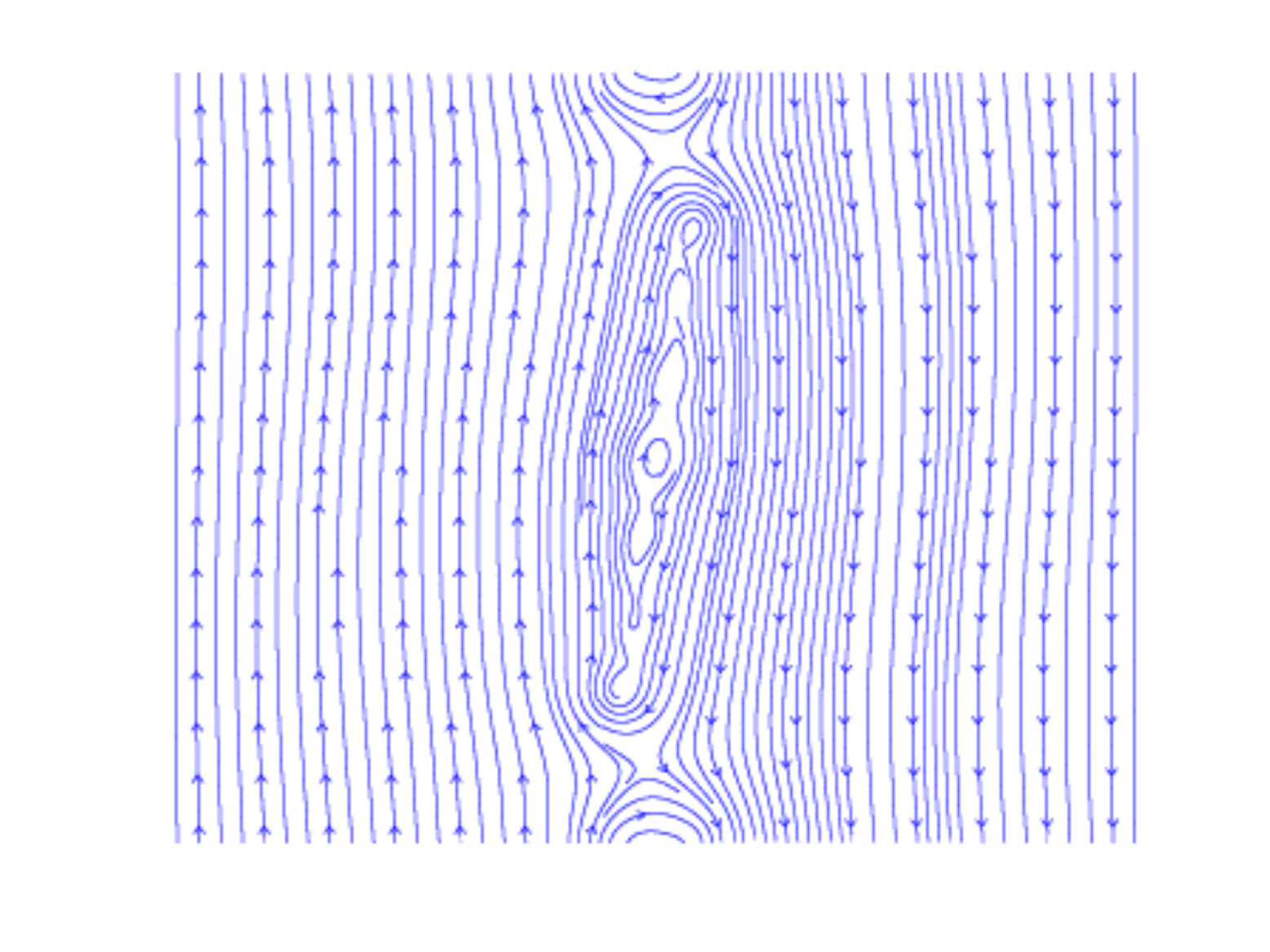}
&
\includegraphics[scale=0.47]{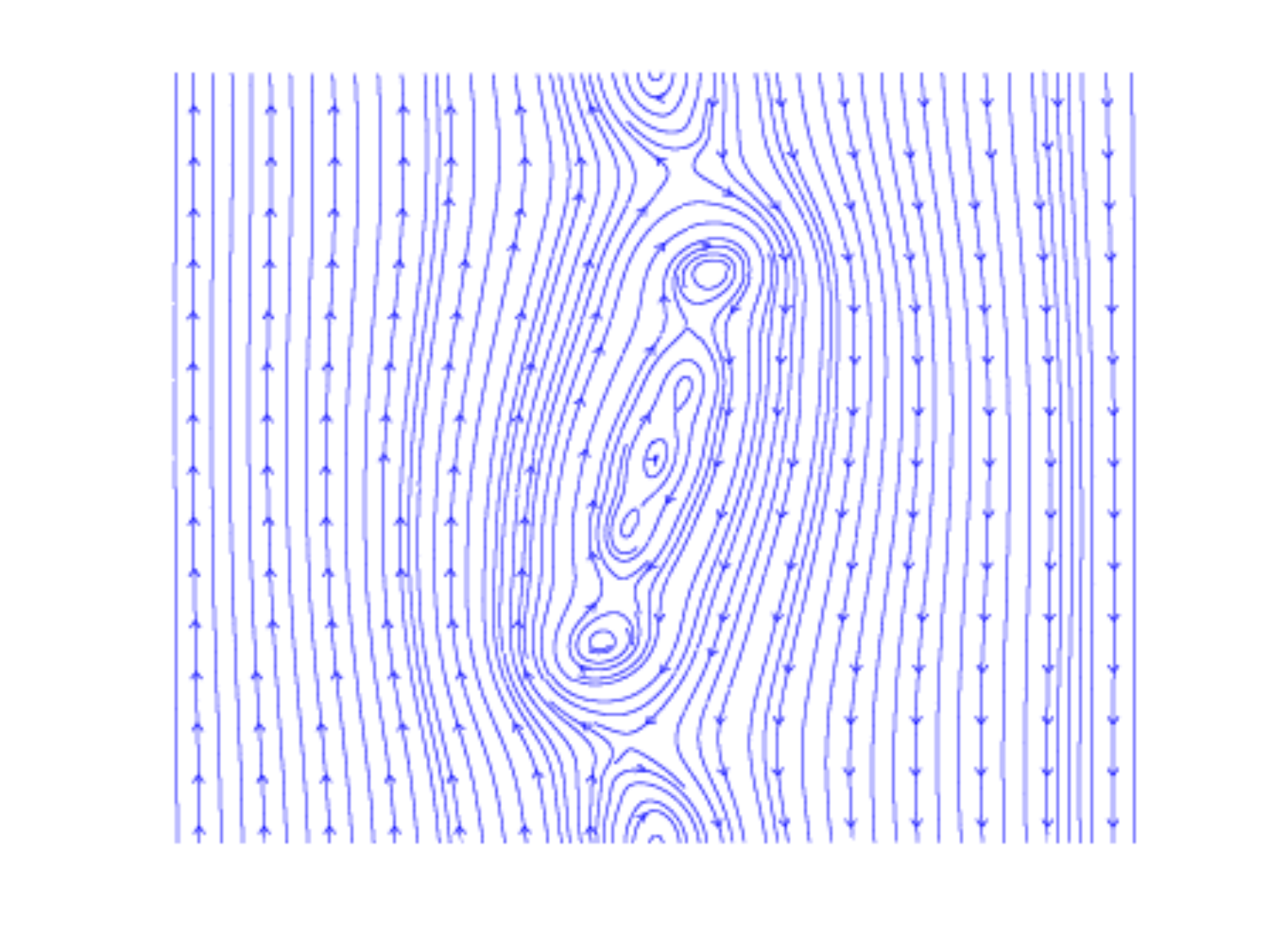}
\end{tabular}
\caption{The same as Figure \protect\ref{ppmhllcearly} with RK-HLLC.$^*$}
\label{rkhllcearly}
\end{figure}

Figure \ref{ppmhllcearly} shows the time evolution of the velocity field with
PPM-HLLC,
illustrated by streamlines at times $0.25, 0.50$ and $1.0$. The
streamlines were produced with the intrinsic Matlab routine
'streamslice'.
Note that the density of streamlines plotted does not accurately reflect the numerical resolution, but were chosen to give a clear representation of the observable topological flow features.
The plots from the original PPM looks very similar. Figure \ref{rkhllcearly}
shows the same but this time with the RK-HLLC code. Also with these codes we see no
significant differences between the two Riemann solvers.

\begin{figure}\centering
\begin{tabular}{cc}
\includegraphics[scale=0.47]{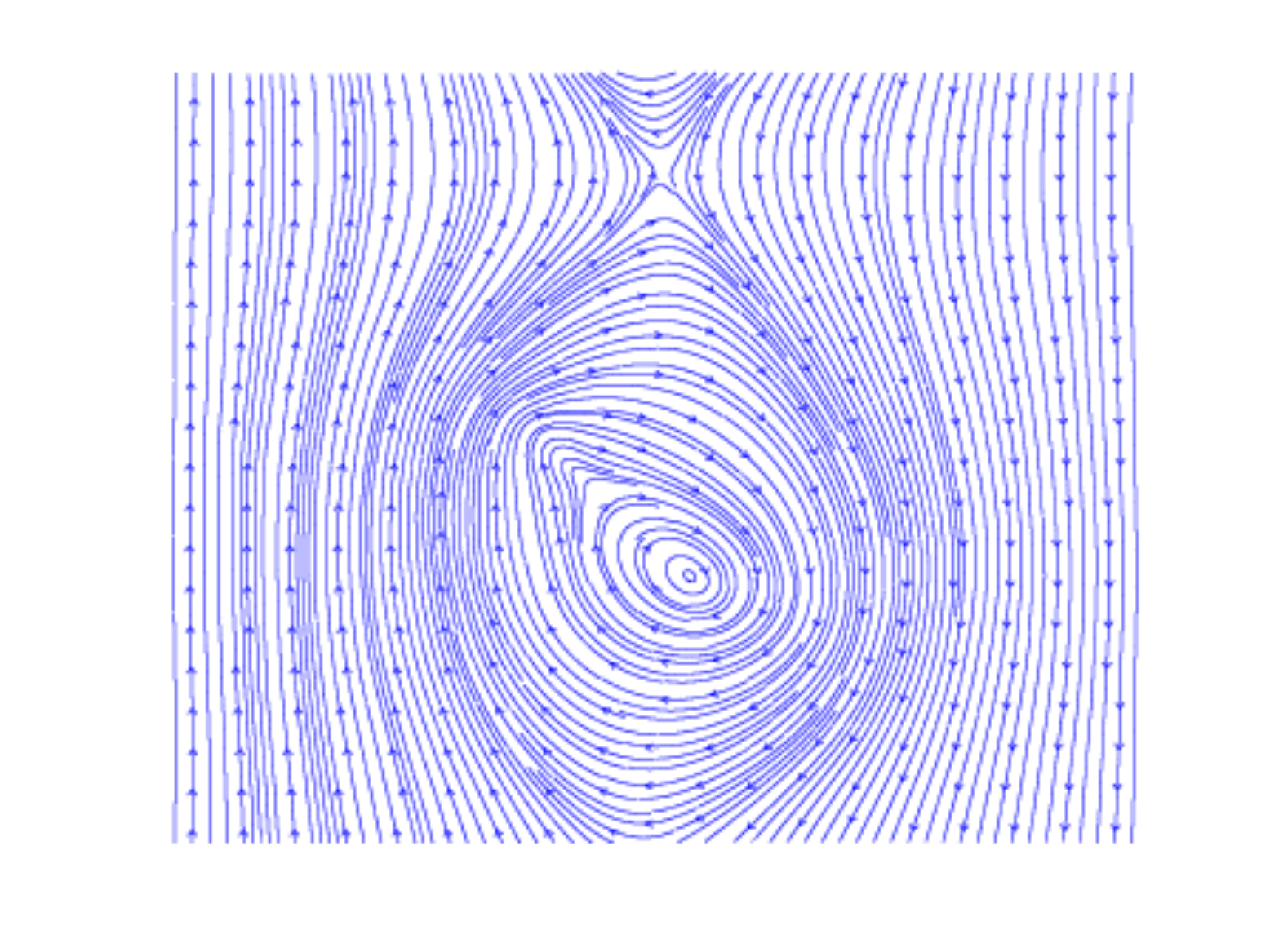}
&
\includegraphics[scale=0.47]{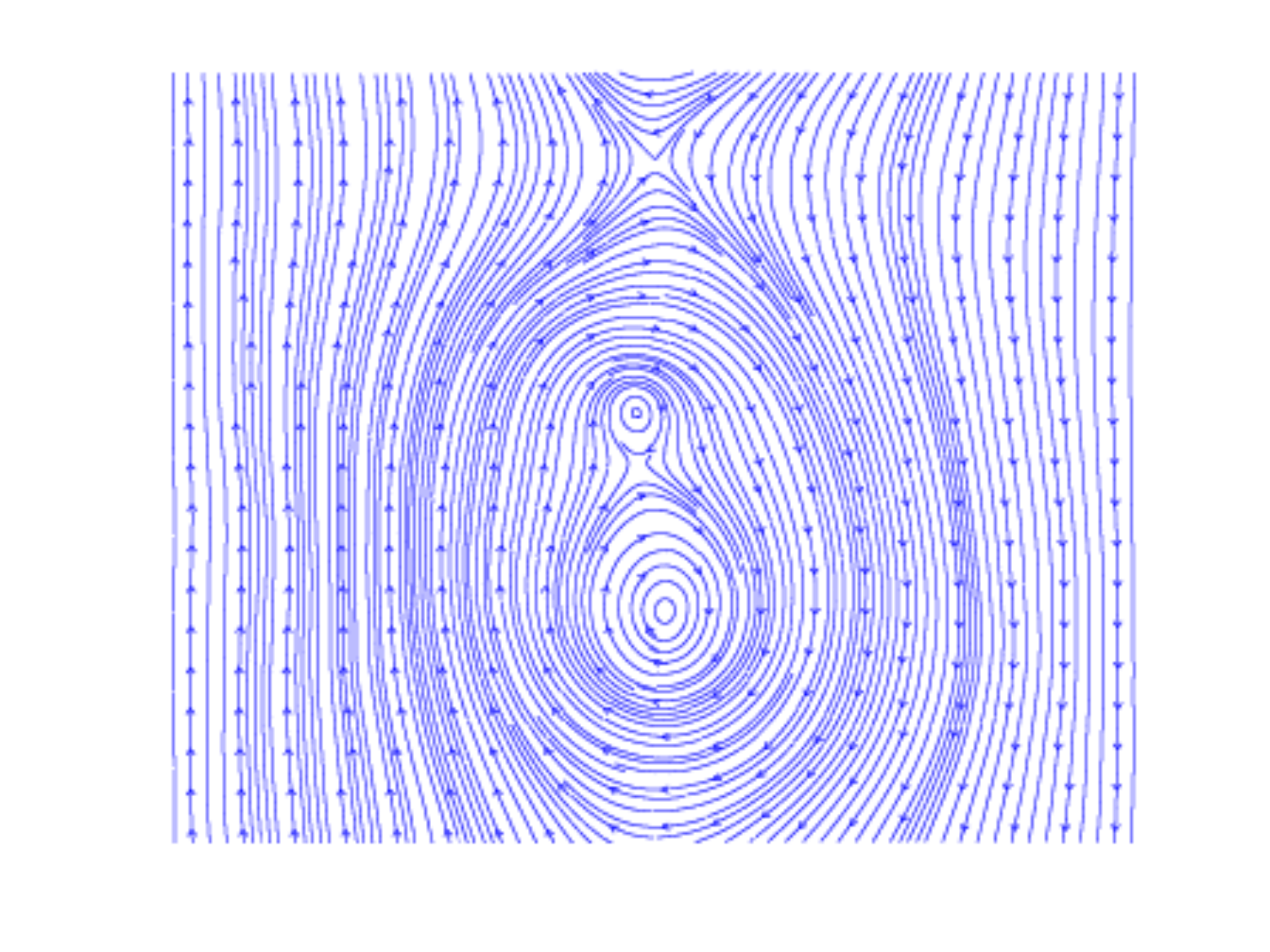}
\end{tabular}
\caption{Streamlines from simulations with $200^2$ cells at time
  $t=20$. PPM-HLLC is shown on the left, and PPM to the right.$^*$}
\label{ppm22late}
\end{figure}

\begin{figure}\centering
\begin{tabular}{cc}
\includegraphics[scale=0.47]{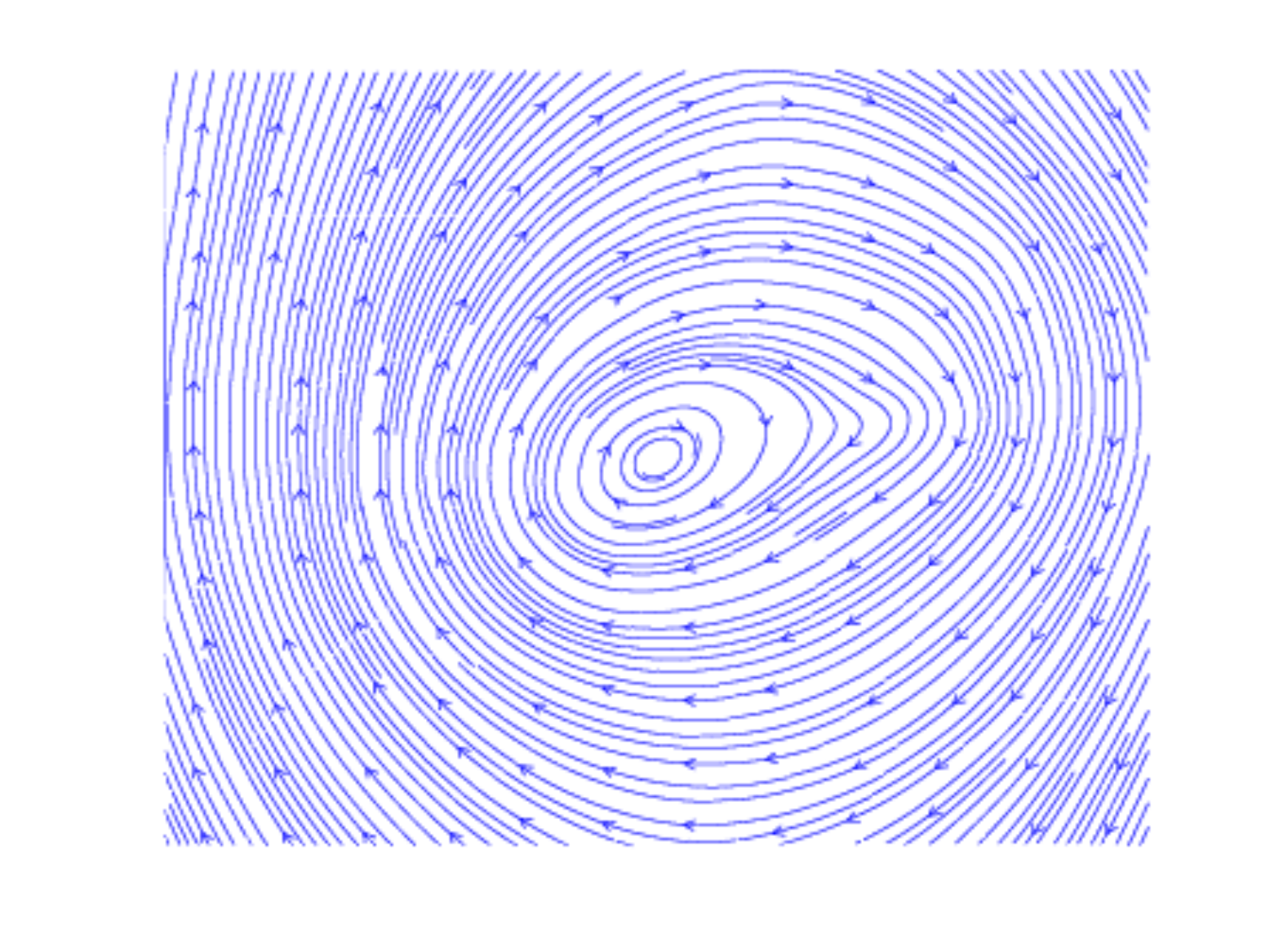}
&
\includegraphics[scale=0.47]{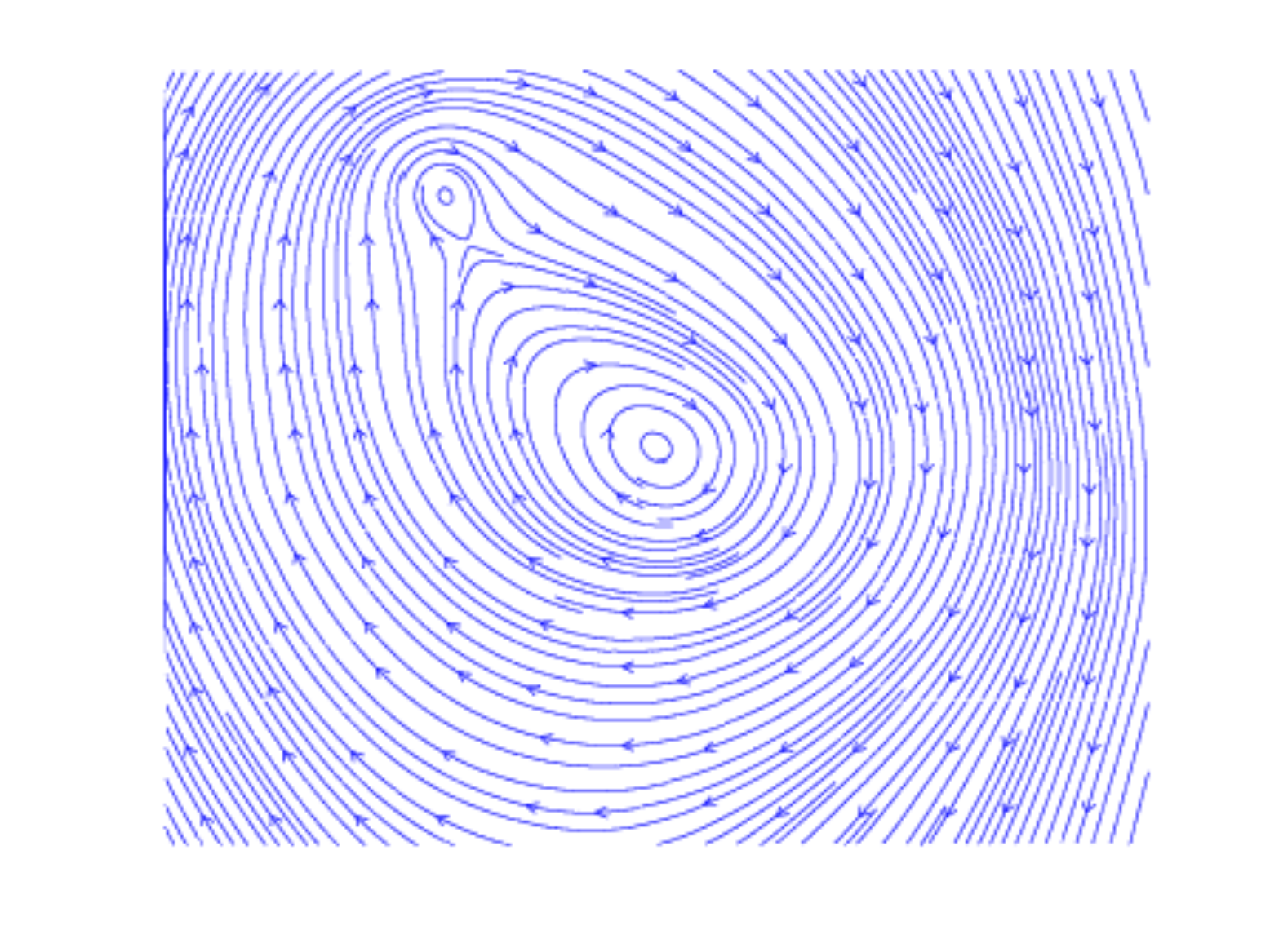}
\end{tabular}
\caption{Streamlines from simulations with  $400^2$ cells at $t=18$.  PPM-HLLC is shown on the left, and PPM to the right.$^*$}
\label{ppm44late}
\end{figure}

\begin{figure}\centering
\begin{tabular}{cc}
\includegraphics[scale=0.47]{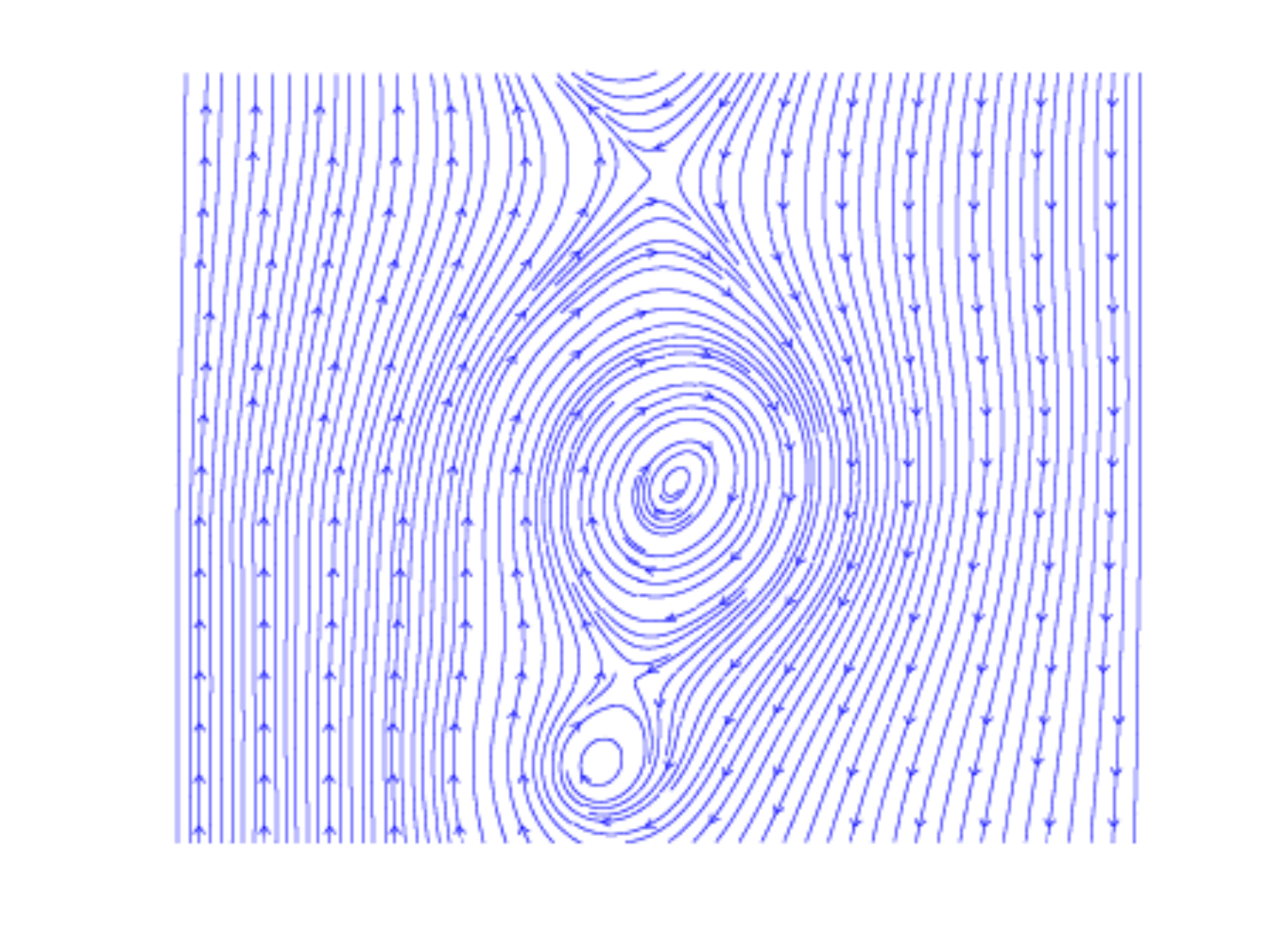}
&
\includegraphics[scale=0.47]{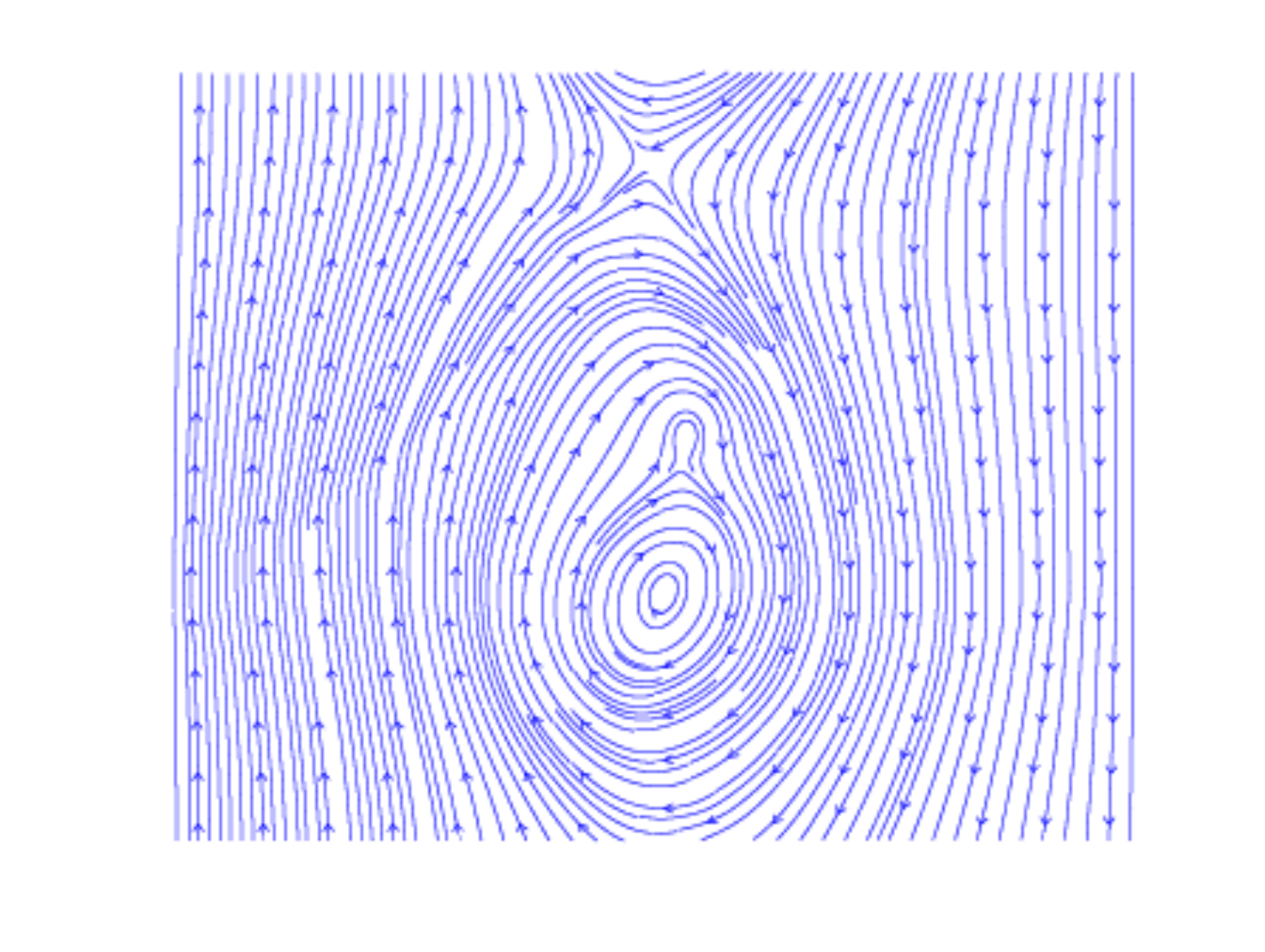}
\end{tabular}
\caption{Streamlines from simulations with$ 200^2$ cells at time
  $t=13$. RK-HLLC is shown on the left, and RK-exact on the right.$^*$}
\label{rk22late}
\end{figure}

\begin{figure}\centering
\begin{tabular}{cc}
\includegraphics[scale=0.47]{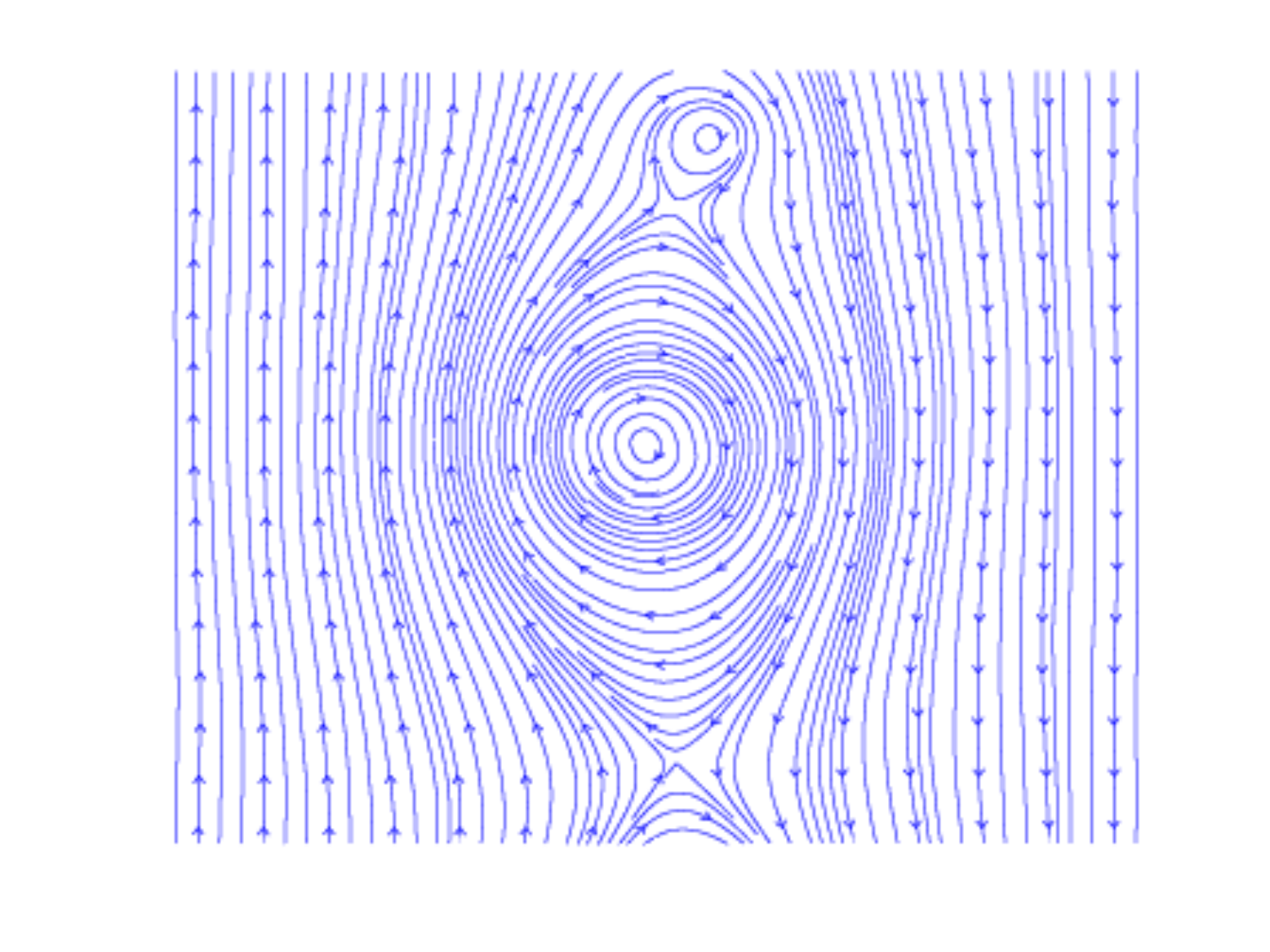}
&
\includegraphics[scale=0.47]{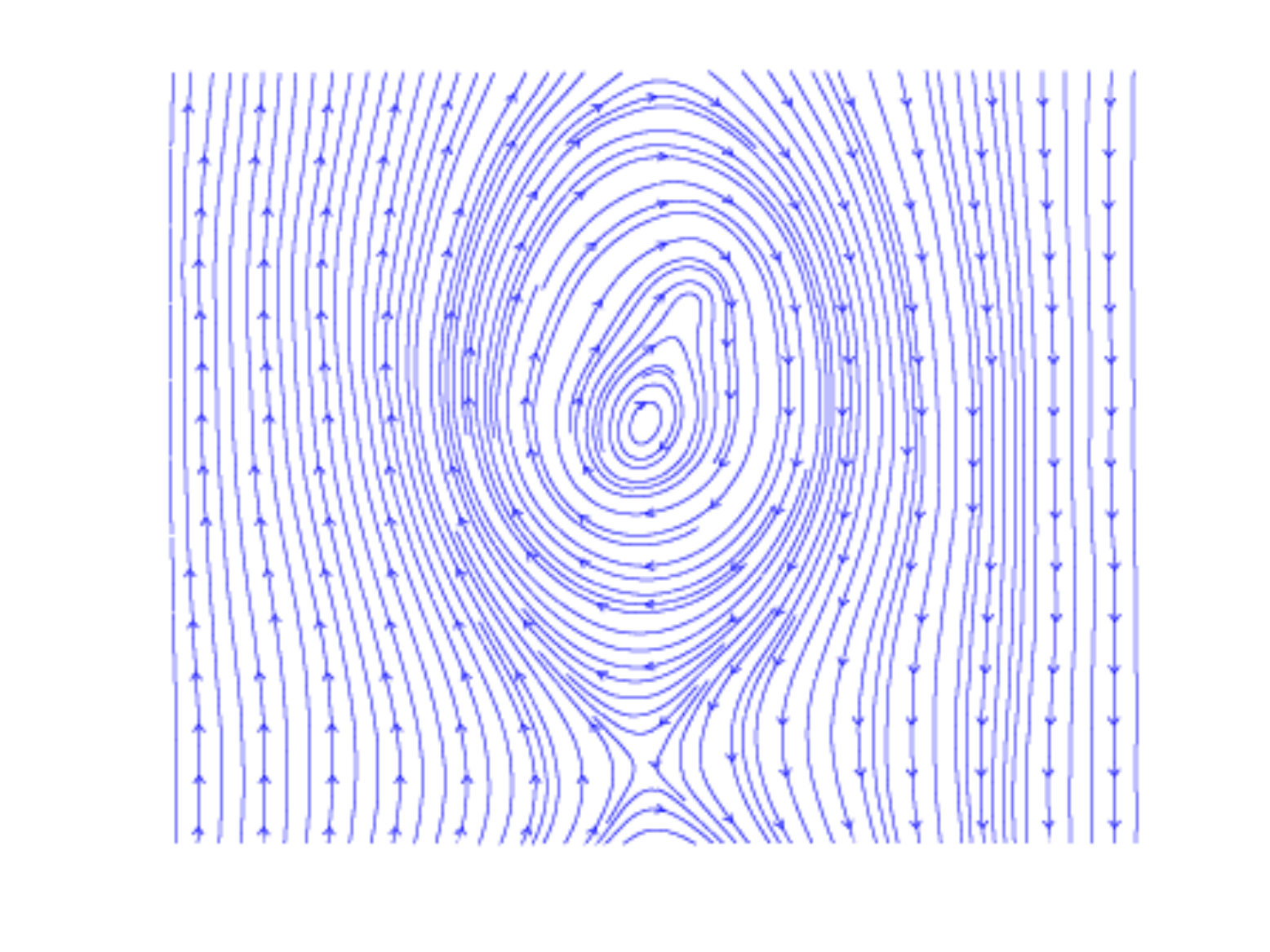}
\end{tabular}
\caption{Streamlines from simulations with $400^2$ cells at time
  $t=16$. RK-HLLC is shown on the left, and RK-exact right.$^*$}
\label{rk44late}
\end{figure}

Differences between the schemes only become apparent at later
times. 
Eventually all vortices are
swallowed by the domain-centered vortex, and we see some differences in at
which time $t_1$ this happens. We used the streamline plots to find
approximately when this change in flow
topology occurs.
For example in Figure \ref{ppm22late}, we plotted the flow at time $t=20$,
since the PPM-simulation then still had two distinct vortices, while with
PPM-HLLC we could only see one. In
Figures \ref{ppm44late}-\ref{rk44late} we do the same with
different resolutions and codes. Since the
resolution and the underlying code also influenced the time $t_1$, we chose
different plotting times in Figures \ref{ppm22late}-\ref{rk44late}. In all four cases considered it is clear that
the Riemann solver induces some difference in $t_1$.

\begin{figure}\centering
\begin{tabular}{cc}
\includegraphics[scale=0.47]{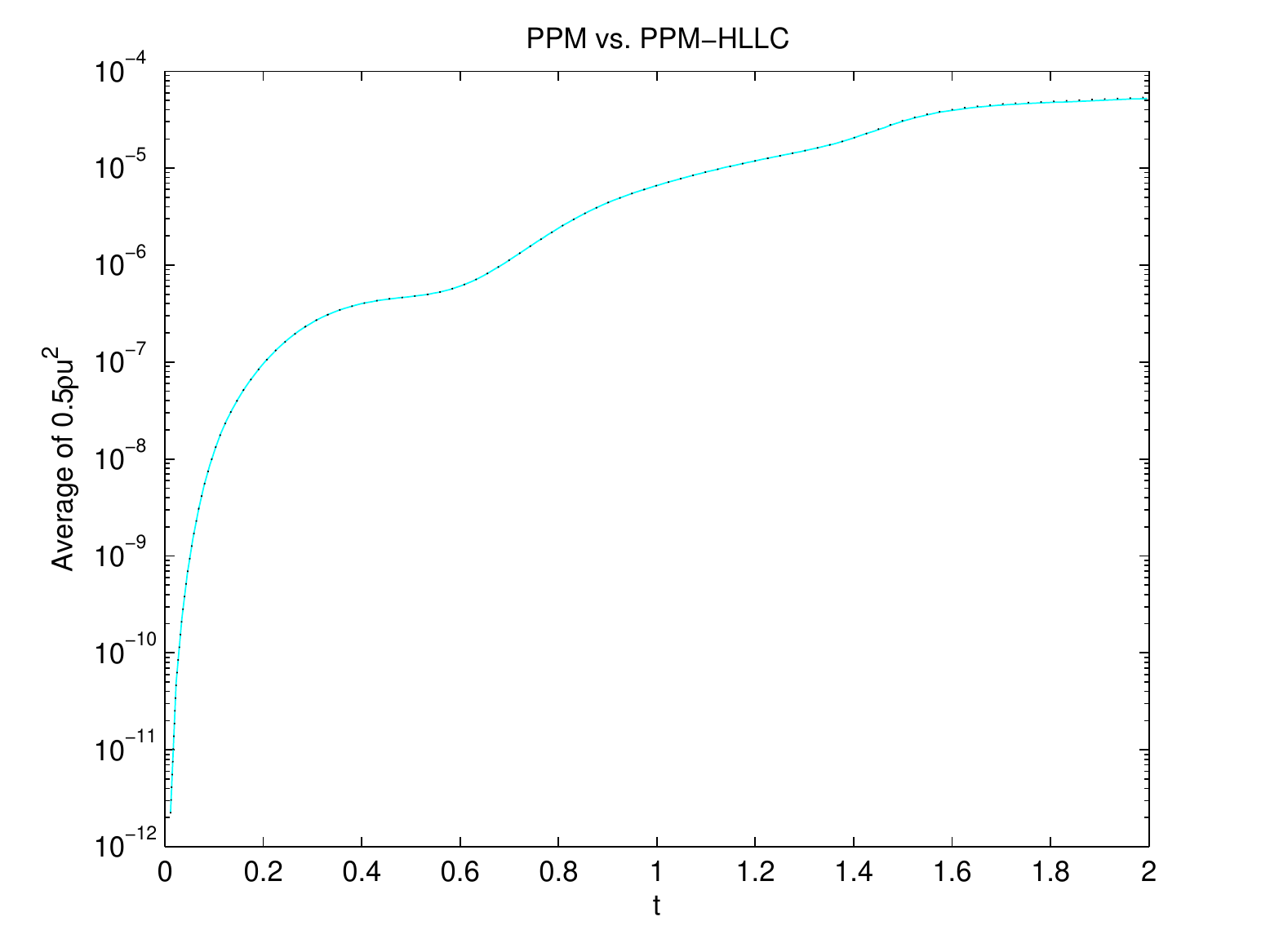}
&
\includegraphics[scale=0.47]{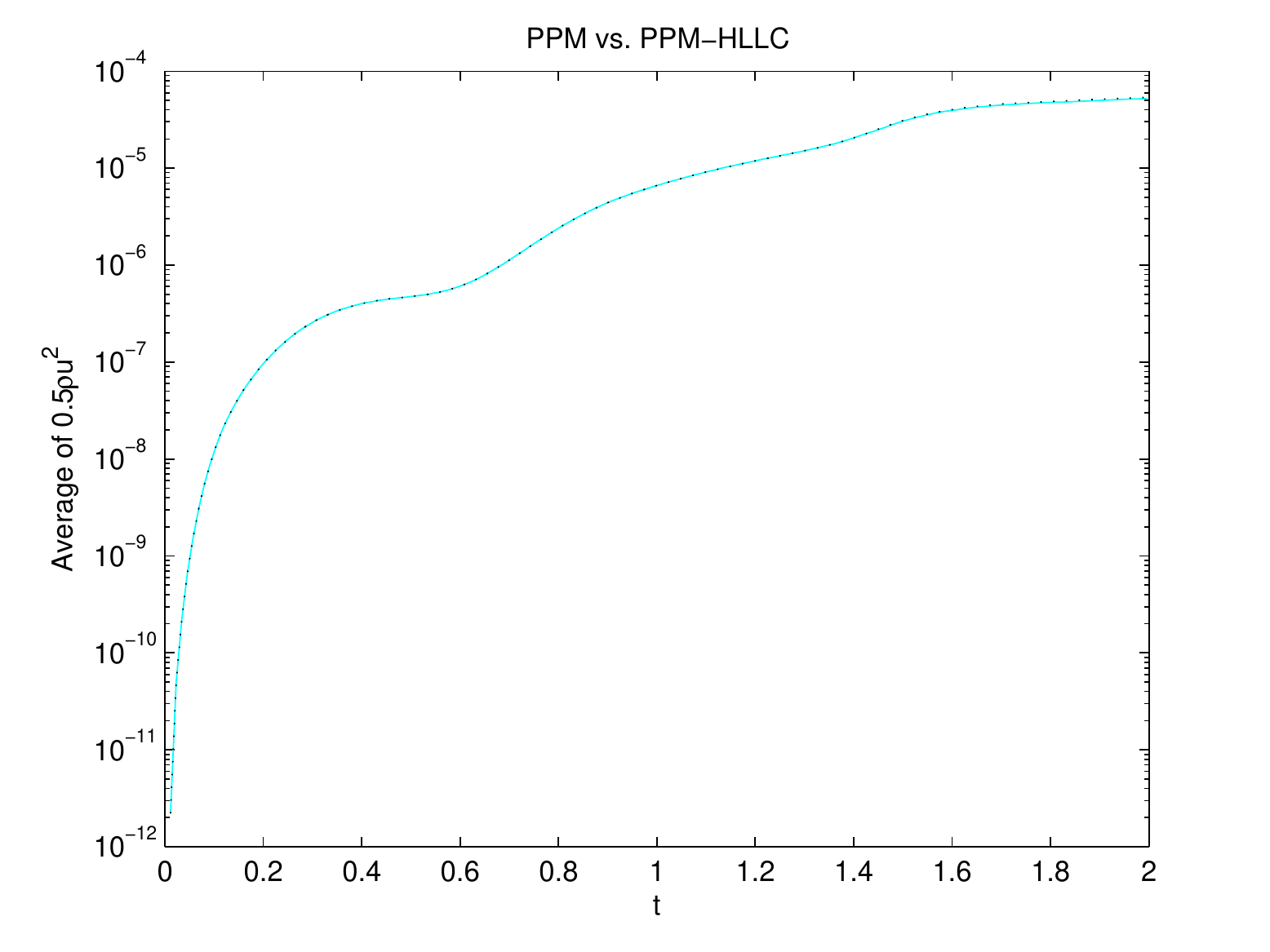}
\end{tabular}
\caption{Growth of transversal kinetic energy component with relative Mach number
  10. The codes with exact solver are represented with dotted lines, and the
      HLLC version with solid coloured lines. They look the same.}
\label{growthM10}
\end{figure}

\begin{figure}\centering
\begin{tabular}{cc}
\includegraphics[scale=0.47]{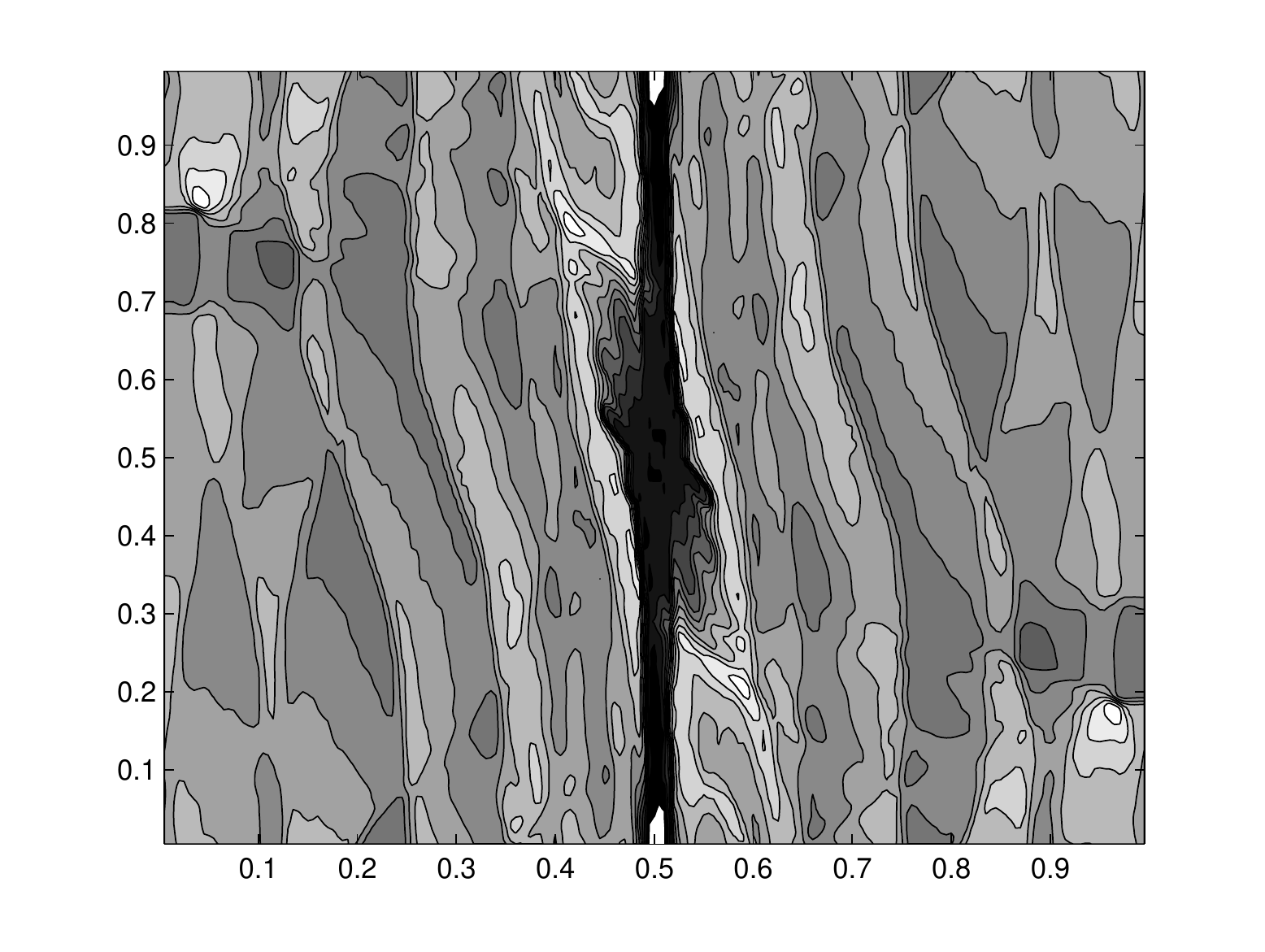}
&
\includegraphics[scale=0.47]{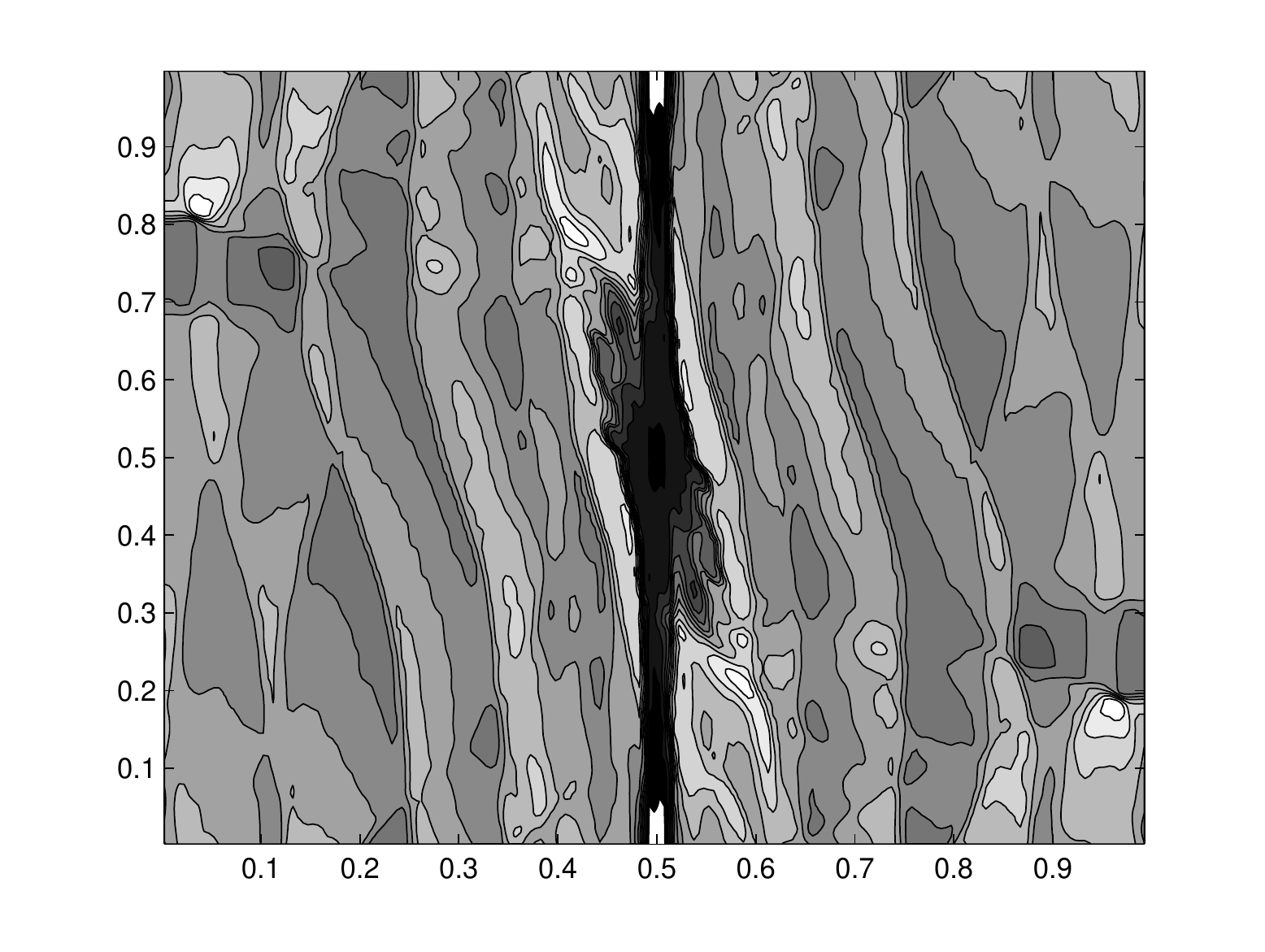}
\end{tabular}
\caption{Density contours from Kelvin-Helmholtz instability with Mach number
  10 at time $t=10$. The resolution was $200^2$ cells. PPM-HLLC is shown on the left,
  and PPM on the right. Density contours range from 0.4 to 1.4.$^*$}
\label{ppm22lateM10}
\end{figure}

\begin{figure}\centering
\begin{tabular}{cc}
\includegraphics[scale=0.47]{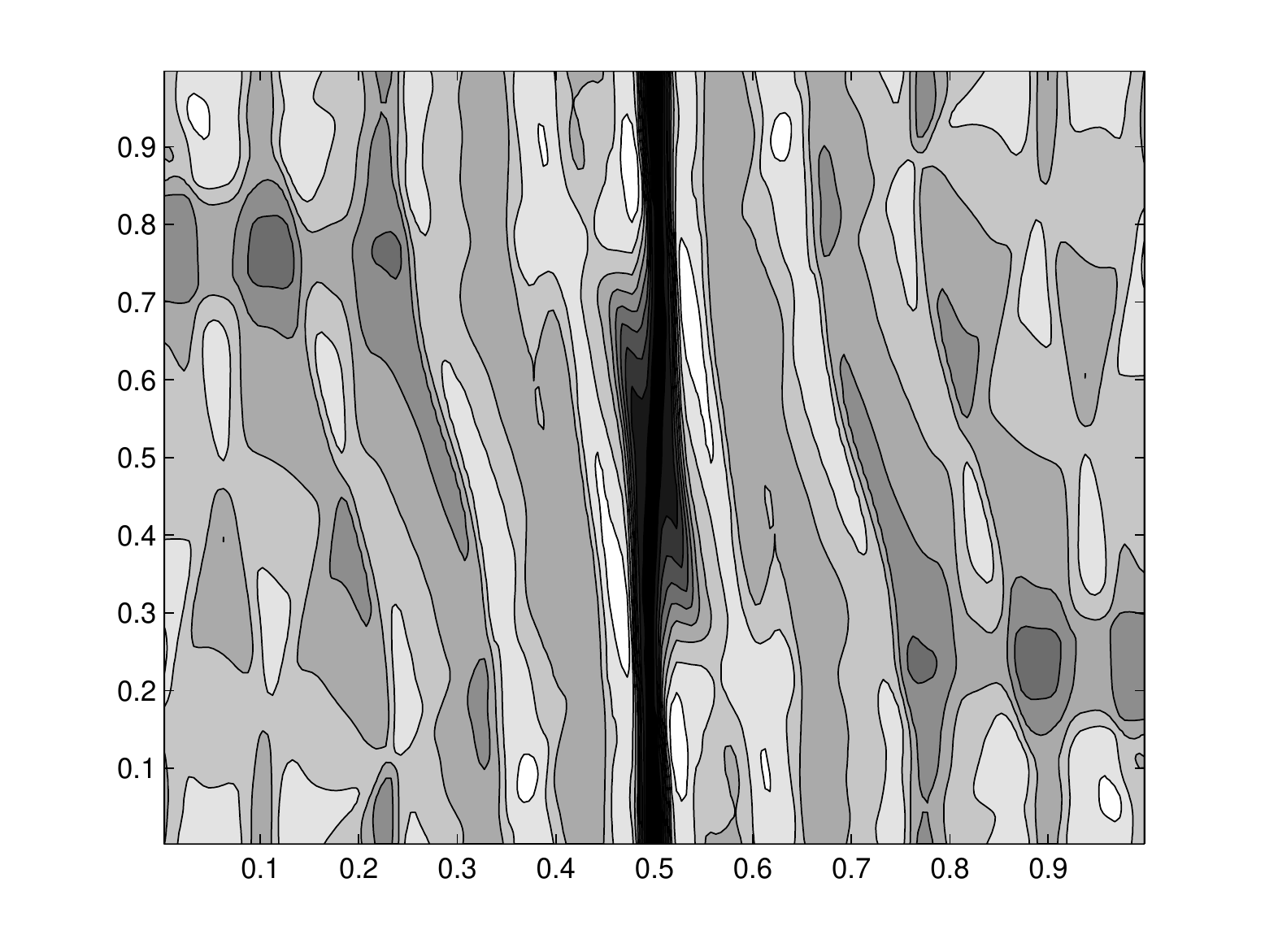}
&
\includegraphics[scale=0.47]{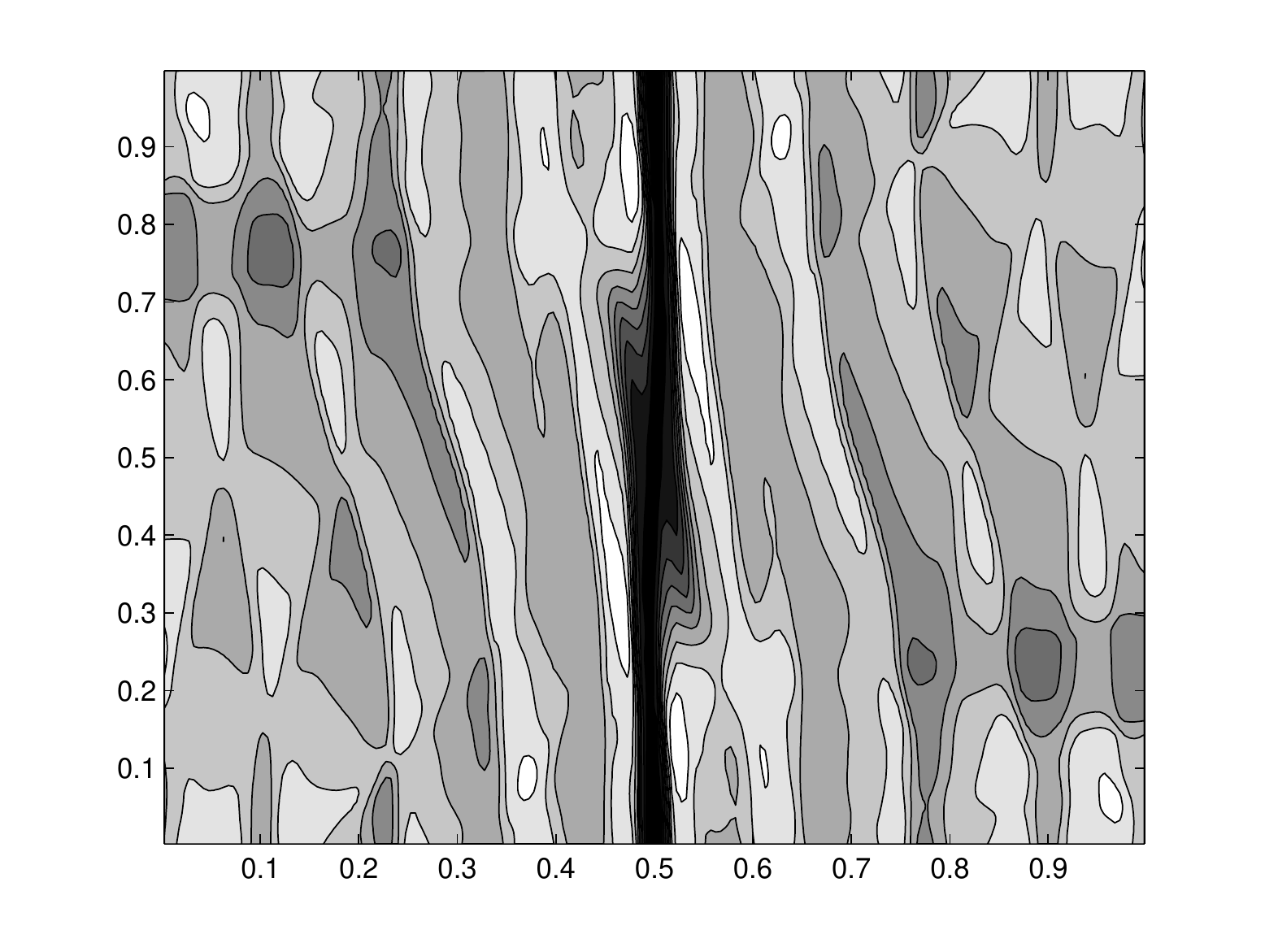}
\end{tabular}
\caption{The same as in Figure \protect\ref{ppm22lateM10}, but with RK-HLLC on the
  left and  RK-exact on the right.$^*$}
\label{rk22lateM10}
\end{figure}

\begin{figure}\centering
\begin{tabular}{cc}
\includegraphics[scale=0.47]{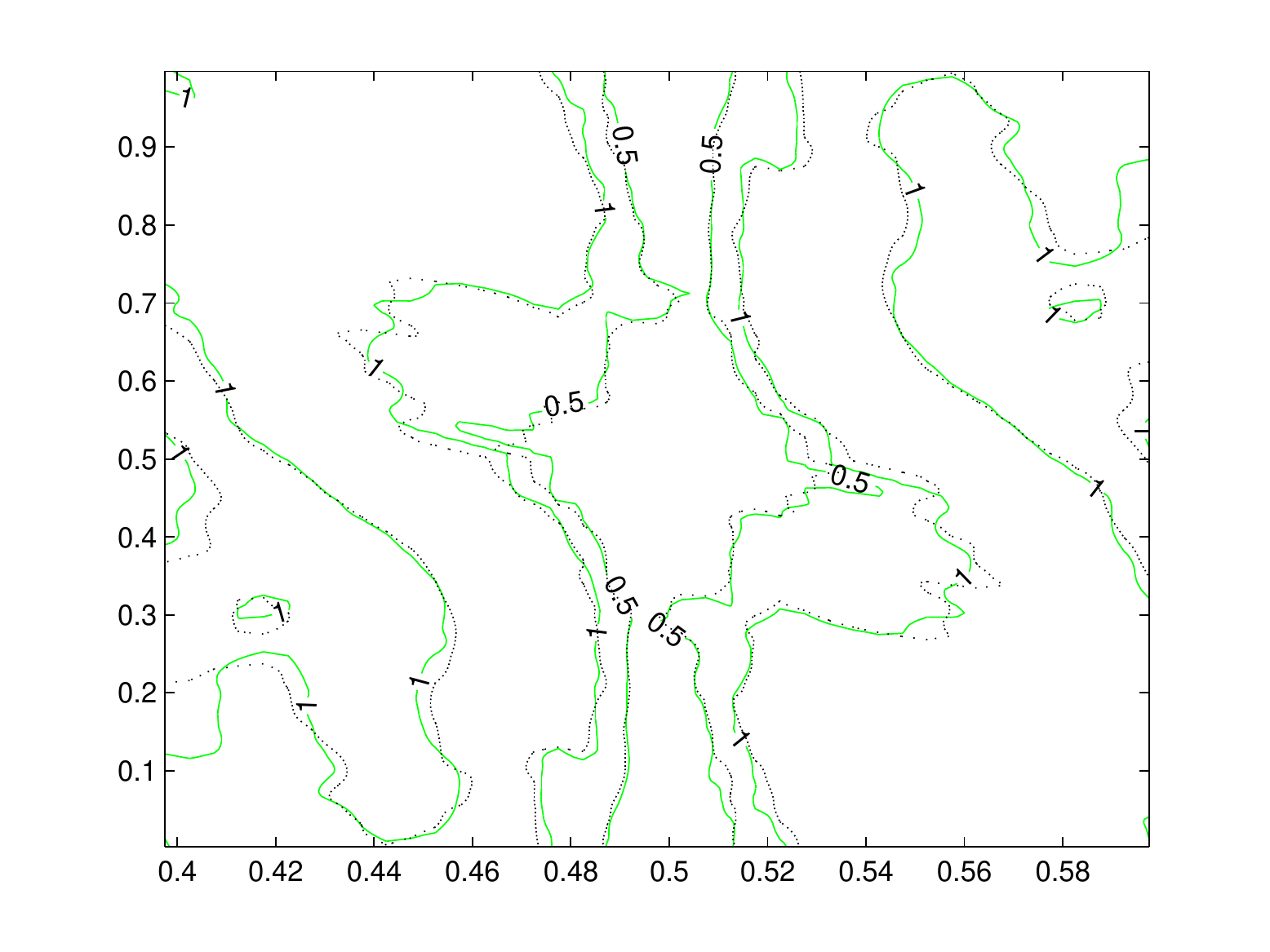}
&
\includegraphics[scale=0.47]{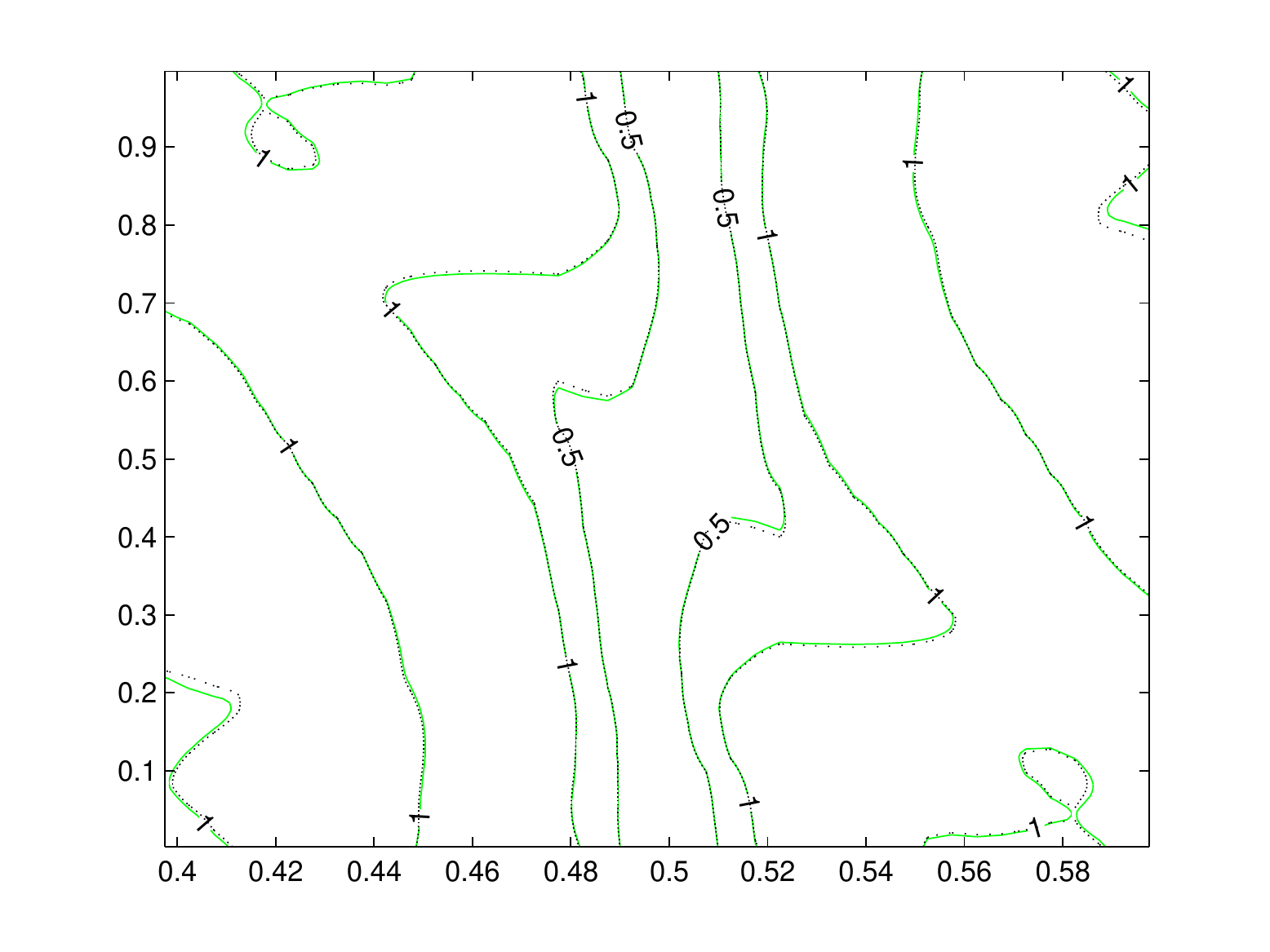}
\end{tabular}
\caption{Superimposed density contours from Kelvin-Helmholtz instability with Mach number
  10 at time $t=10$ at resolution $200^2$ cells. On the left PPM is
  represented by green
  solid lines, and PPM-HLLC with dotted lines. On the right RK-exact with green
  solid lines, and RK-HLLC with dotted lines.RK right.$^*$}
\label{supcont22lateM10}
\end{figure}

We also considered $p=\frac{1}{100\gamma}$, giving a relative Mach number of
$10$. Figure \ref{growthM10} shows the growth of the average transversal
kinetic energy component. The effect of the Riemann solver is indiscernible. Resolution had less influence
here than with relative Mach number $1$, so we only show the data from the
$200^2$-simulation. At Mach numbers this high, the Kelvin-Helmholtz modes are linearly stable, and instead of Kelvin-Helmholtz rolls, kink modes develop, see for example \cite{Woodwardkink}.
For the Mach 10 case differences are also very subtle. In Figure
\ref{ppm22lateM10}, showing filled density contours at time $t=10$, one can
see slightly more small scale structure with PPM than PPM-HLLC by careful
inspection of the plots. The similar plots from the RK-codes in Figure \ref{rk22lateM10}
clearly show more smeared out structures than the PPM-codes. There is no
noticeable difference between RK-HLLC and RK-exact, which means that any effect
of changing the Riemann solver is much less prominent than the smearing due the
RK-algorithm. The superimposed density contours in Figure
\ref{supcont22lateM10} also illustrate the increased numerical diffusion of the RK-algorithm, and that this suppresses the effect of changing the Riemann solver.
\footnotetext{$^*$In the plots $x$ is on the horizontal axis, and $y$ on the vertical axis.}

\subsection{Richtmeyer-Meshkov instability}
\begin{figure}\centering
\begin{tabular}{cc}
\includegraphics[height=5cm, width=7cm]{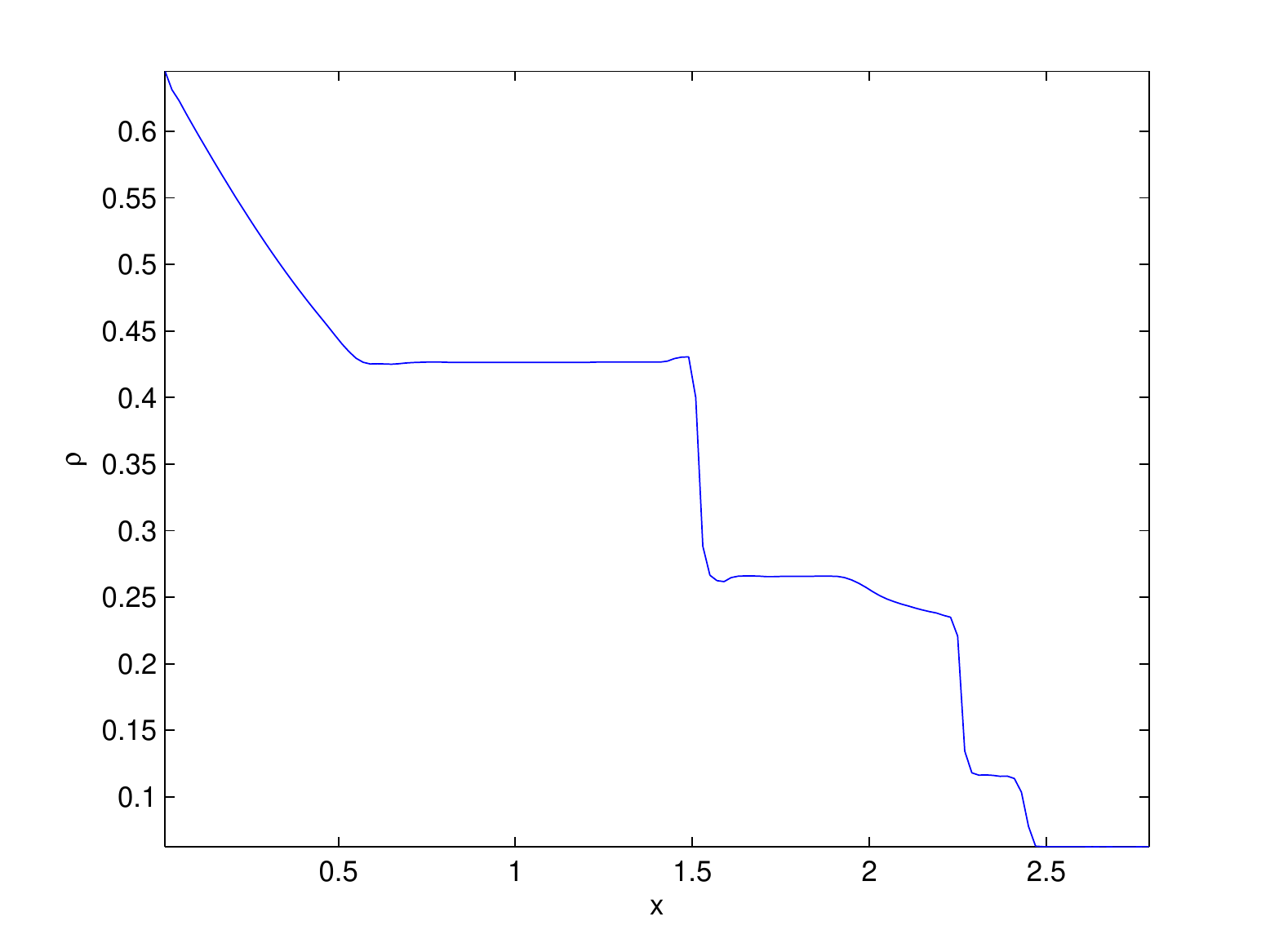}
\end{tabular}
\caption{Slice of Richtmeyer-Meshkov instability at $t=1$, $y=0.49$,
  Computed with PPM at resolution $\Delta x=\Delta y=0.02$.}
\label{rm1}
\end{figure}

\begin{figure}\centering
\begin{tabular}{cc}
\includegraphics[scale=0.47]{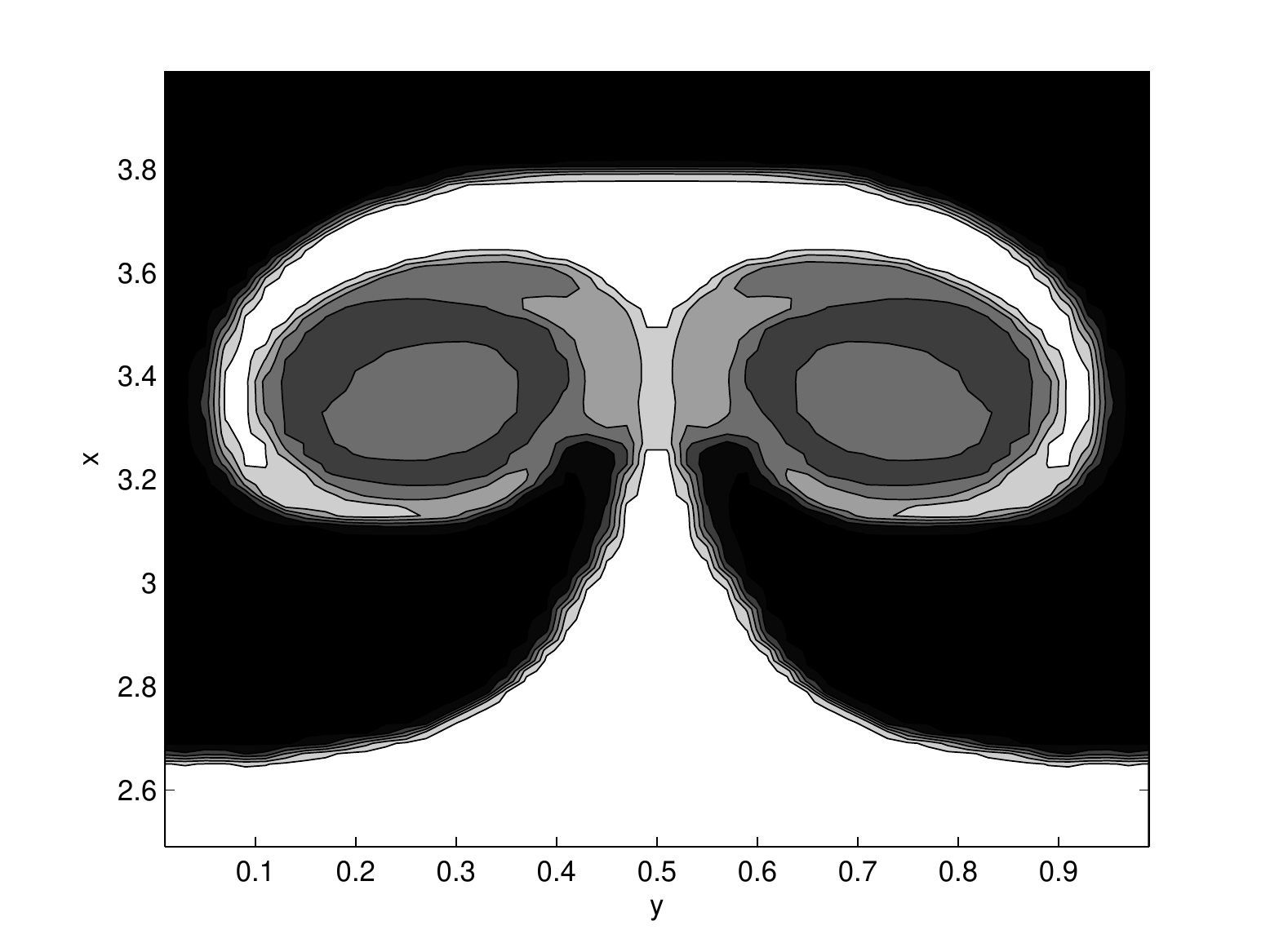}
&
\includegraphics[scale=0.47]{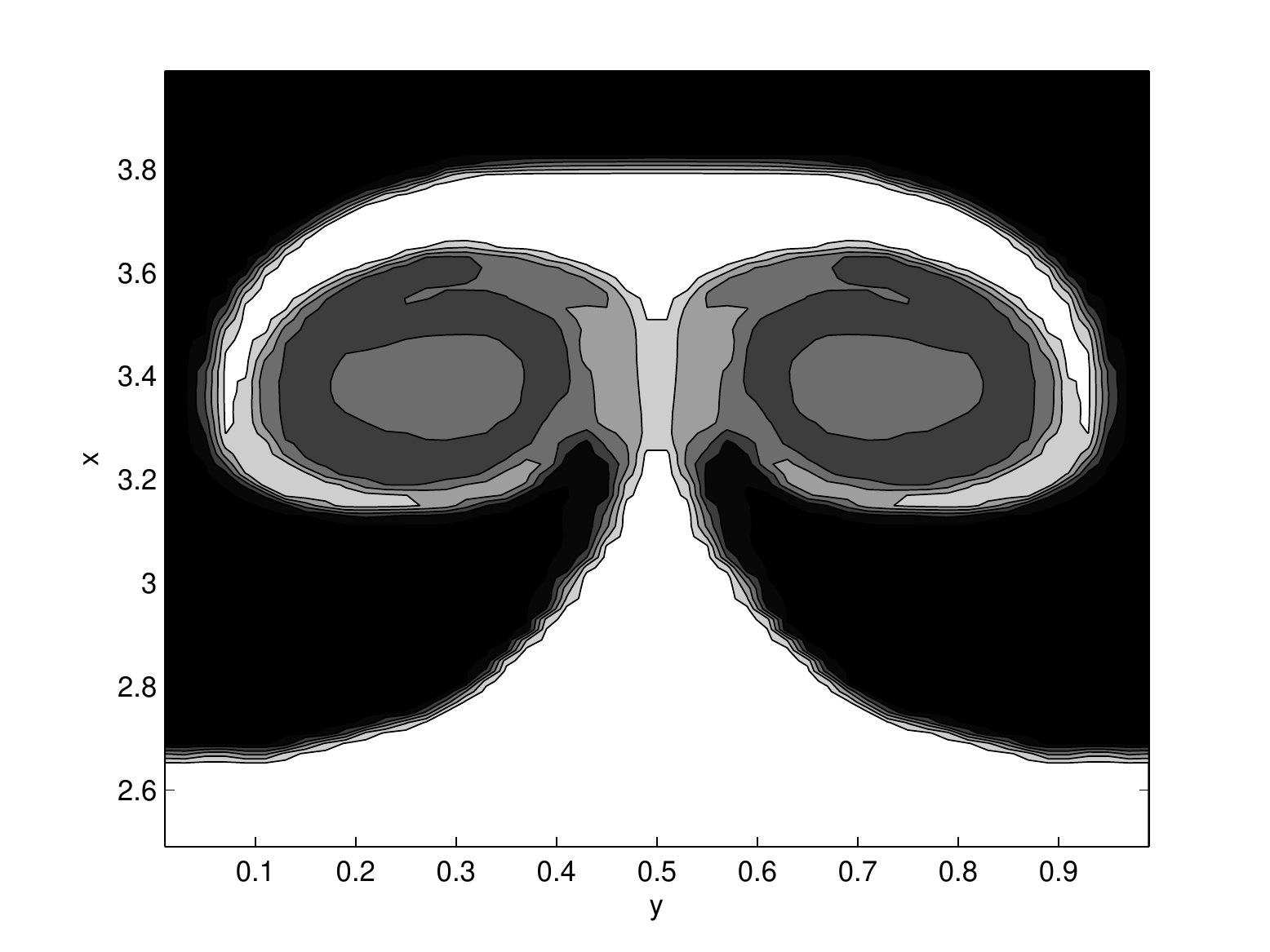}
\end{tabular}
\caption{Density contours from Richtmeyer-Meshkov instability at time $t=14$. We show data
  from PPM to the left, and  PPM-HLLC to the right. Density contours
  range linearly from 0.1 to 0.22. Resolution $ \Delta x=\Delta y=0.02$.}
\label{rm14}
\end{figure}

The Richtmeyer-Meshkov instability occurs when a planar shock hits a parallel,
slightly perturbed density jump. We used the following setup to simulate
this. The domain was $(x,y)\in(0,16)\times(0,1)$ with periodic boundary conditions in $y$, and
Neumann boundary conditions in $x$. We set up a shock tube problem in the $x$-direction
at $x=1.6$ with density and pressure as in the first Toro test and constant
velocity $u=-1$ in the $x$-direction. We considered adiabatic gas $\gamma=1.4$
so as to coincide with the three-dimensional runs rather than the shock tube tests. At the line $x=f(y)=3.2 +0.2\cos(2\pi y)$, the density fell by a factor of
$2$. When the shock goes through the initial density
jump, the boundary $f(y)$ evolves into a mushroom-like structure. In Figure
\ref{rm1} we see a slice in $x$-direction of the density profile at time
$t=1.0$. 
The shock has just hit the density jump, and we see a weaker shock going through,
and a reflected wave moving back towards the contact from the shock tube problem. This
last wave might cause some minor reflected waves, but otherwise the instability
is not influenced by other features. The
boundary at $x=0$ is transparent to the supersonic rarefaction, and we stop
before the shock reflection at $x=16$ affects the instability. The CFL-number was $0.8$. Again it is hard to observe
differences, see Figure \ref{rm14}, but it seems the original PPM resolves the
'extremities' of the high density region a bit sharper, at least they extend
more. 

\begin{figure}\centering
\begin{tabular}{cc}
\includegraphics[scale=0.6]{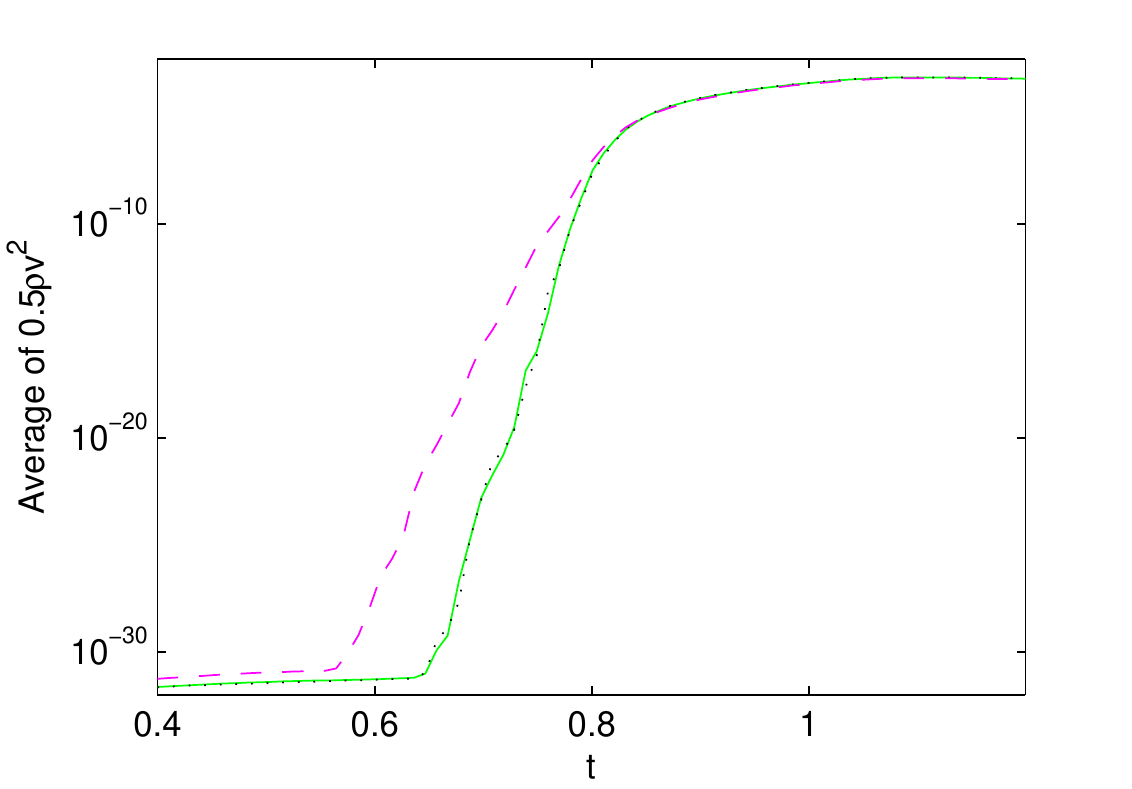}
\end{tabular}
\caption{Logarithm of average $\half\rho v^2$ vs. time $t$. We see data from
  PPM as a dotted line, PPM-HLLC solid blue and RK-HLLC dashed magenta. The
  resolution was  $\Delta x=\Delta y=0.02$.}
\label{rmgrowth}
\end{figure}

We also illustrate the growth of the instability here by the time history of
the transversal component of kinetic energy in Figure \ref{rmgrowth}. With RK-HLLC the density jump is more
smeared out before the shock hits it, which explains the less steep slope.

\begin{figure}\centering
\begin{tabular}{cc}
\includegraphics[scale=0.47]{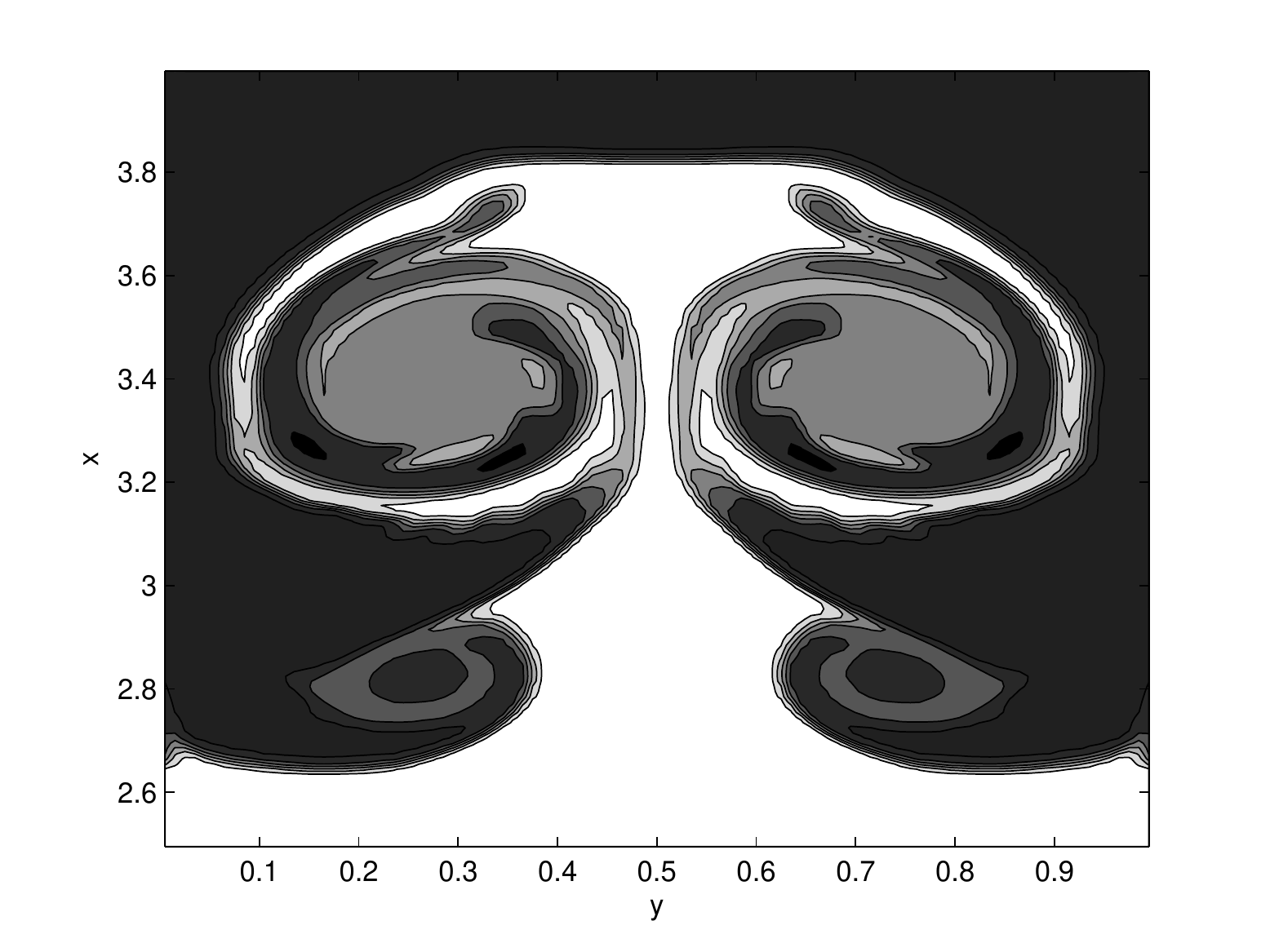}
&
\includegraphics[scale=0.47]{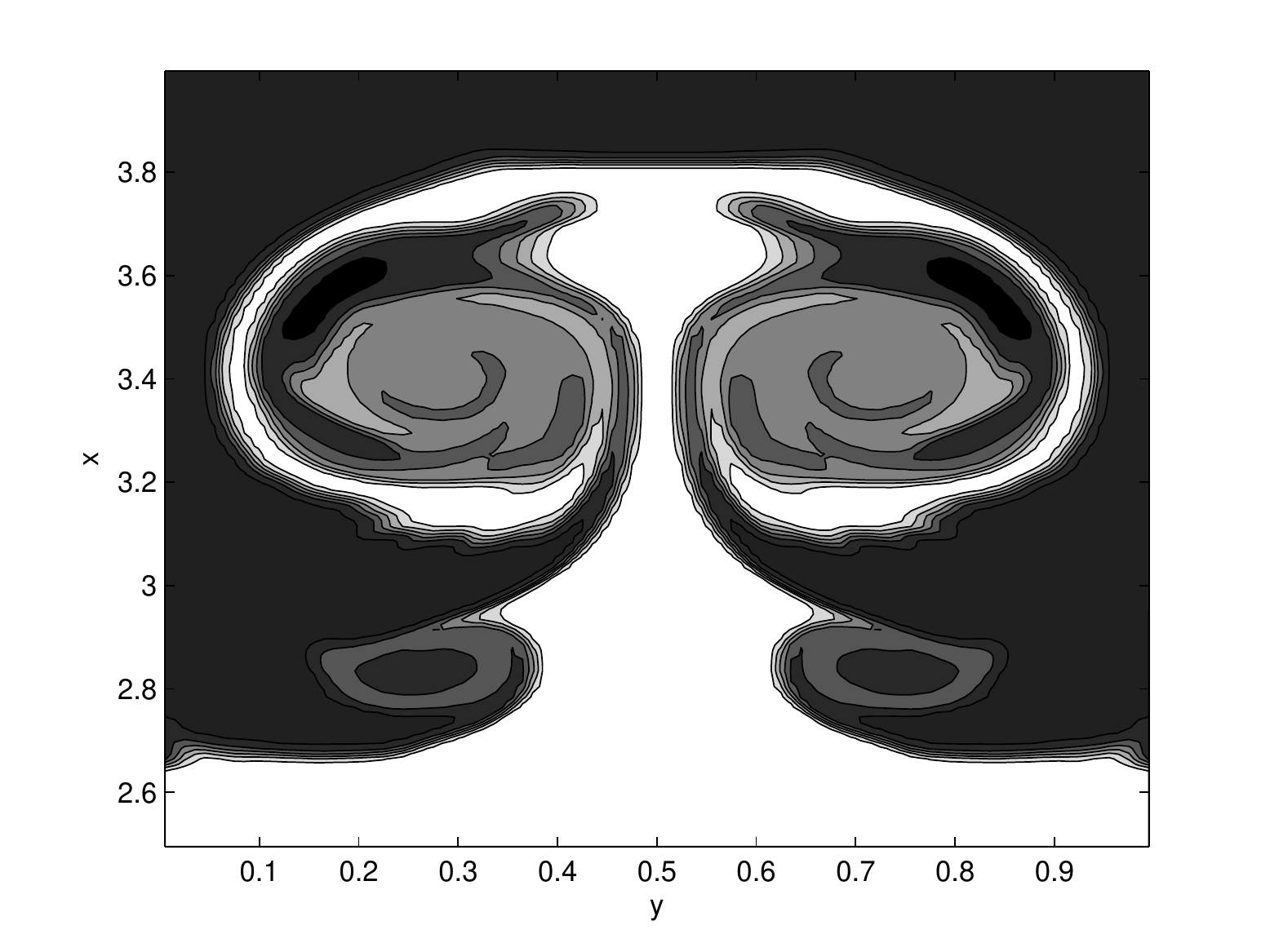}
\end{tabular}
\caption{The same as in Figure \protect\ref{rm14}, but with the
  contact wave steepening turned off in both codes, and resolution $ \Delta x=\Delta y=0.01$. }
\label{rm14nosteephi}
\end{figure}

\begin{figure}\centering
\begin{tabular}{cc}
\includegraphics[scale=0.47]{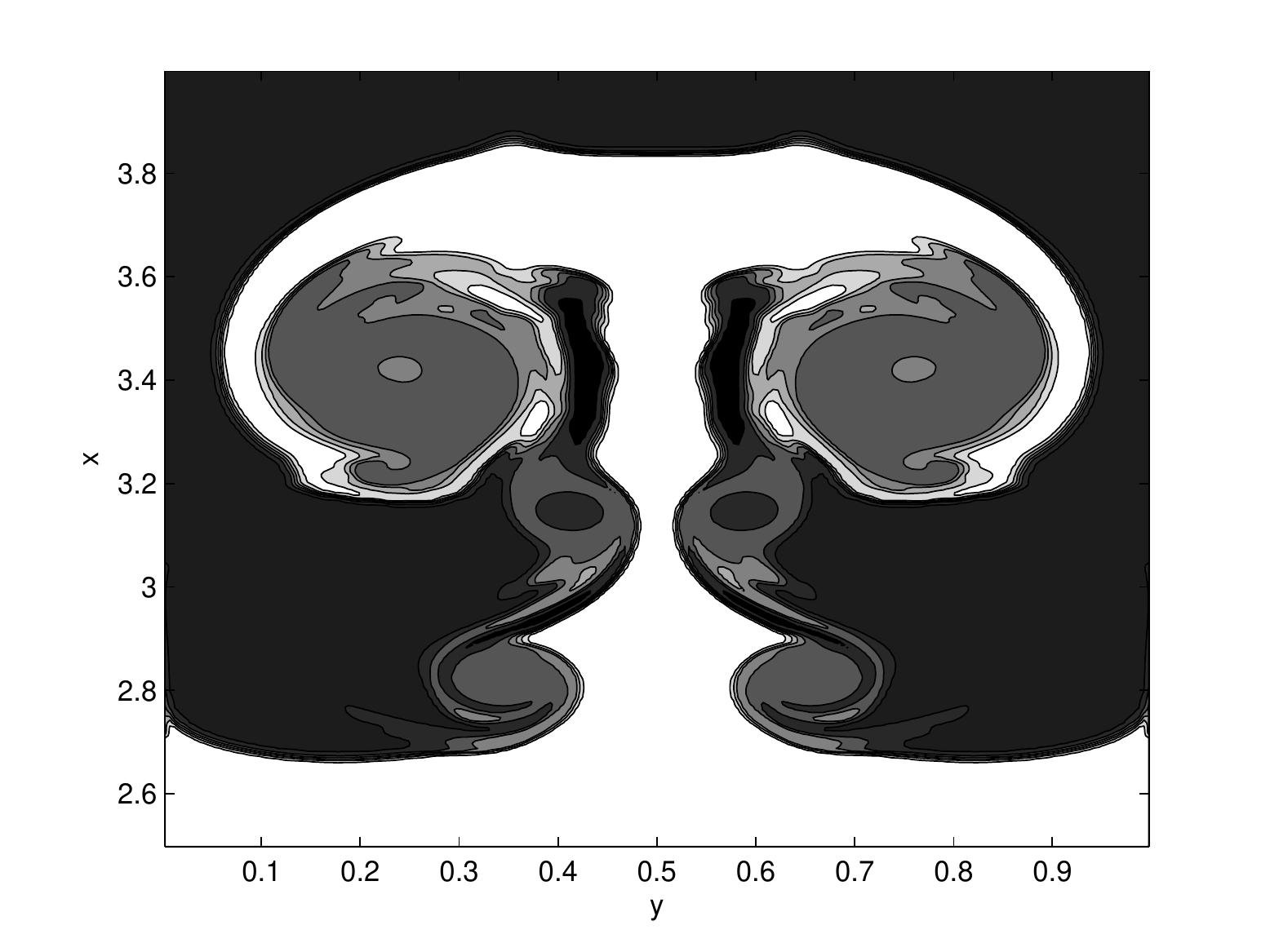}
&
\includegraphics[scale=0.47]{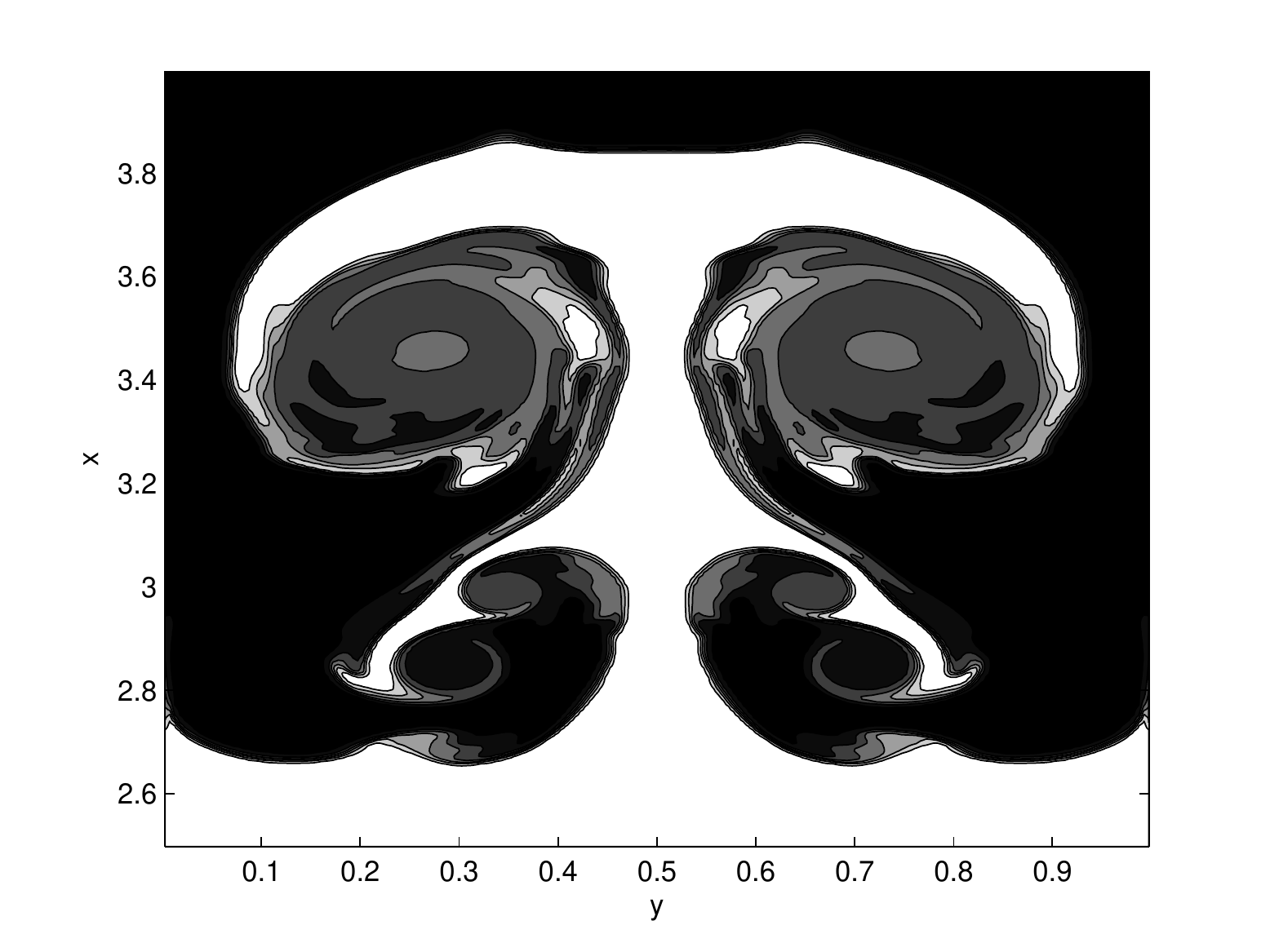}
\end{tabular}
\caption{The same as in Figure \ref{rm14}, but with the
  contact wave steepening turned off in both codes, and resolution $ \Delta x=\Delta y=0.005$. }
\label{rm14nosteepuhi}
\end{figure}

It is known that the contact wave steepening of PPM may artificially
induce instabilities in certain cases. For the PPM-codes we observed
small scale structures that we believe to be numerical noise when repeating the simulation on finer grids. However, by switching off
the steepening, we got reasonable results. In Figures
\ref{rm14nosteephi}-\ref{rm14nosteepuhi} we compare versions of PPM and
PPM-HLLC without steepening. With the highest resolution the density profiles
differ strongly, but there is no way to tell which code is better.
 

\begin{figure}\centering
\begin{tabular}{cc}
\includegraphics[scale=0.47]{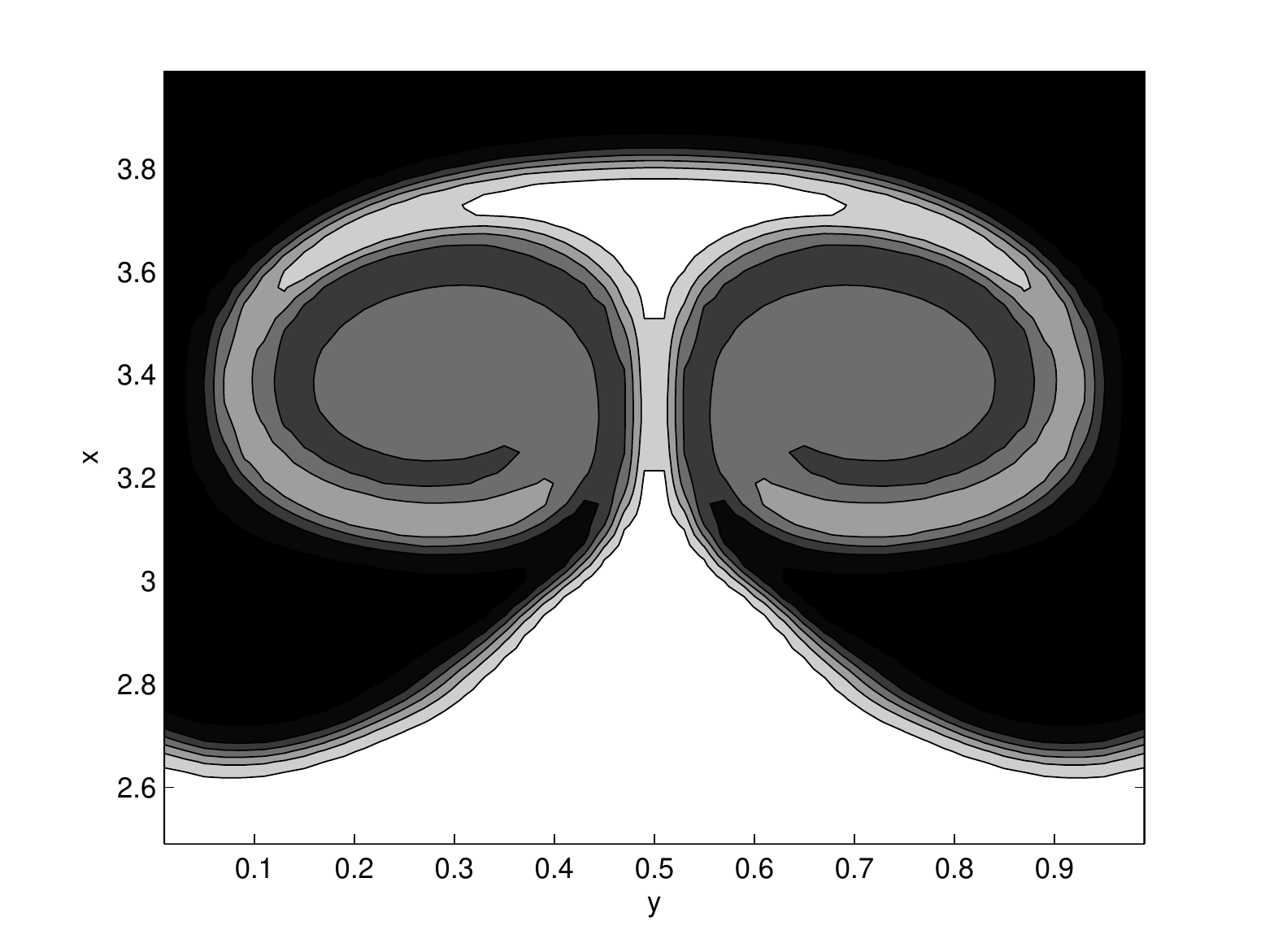}
&
\includegraphics[scale=0.47]{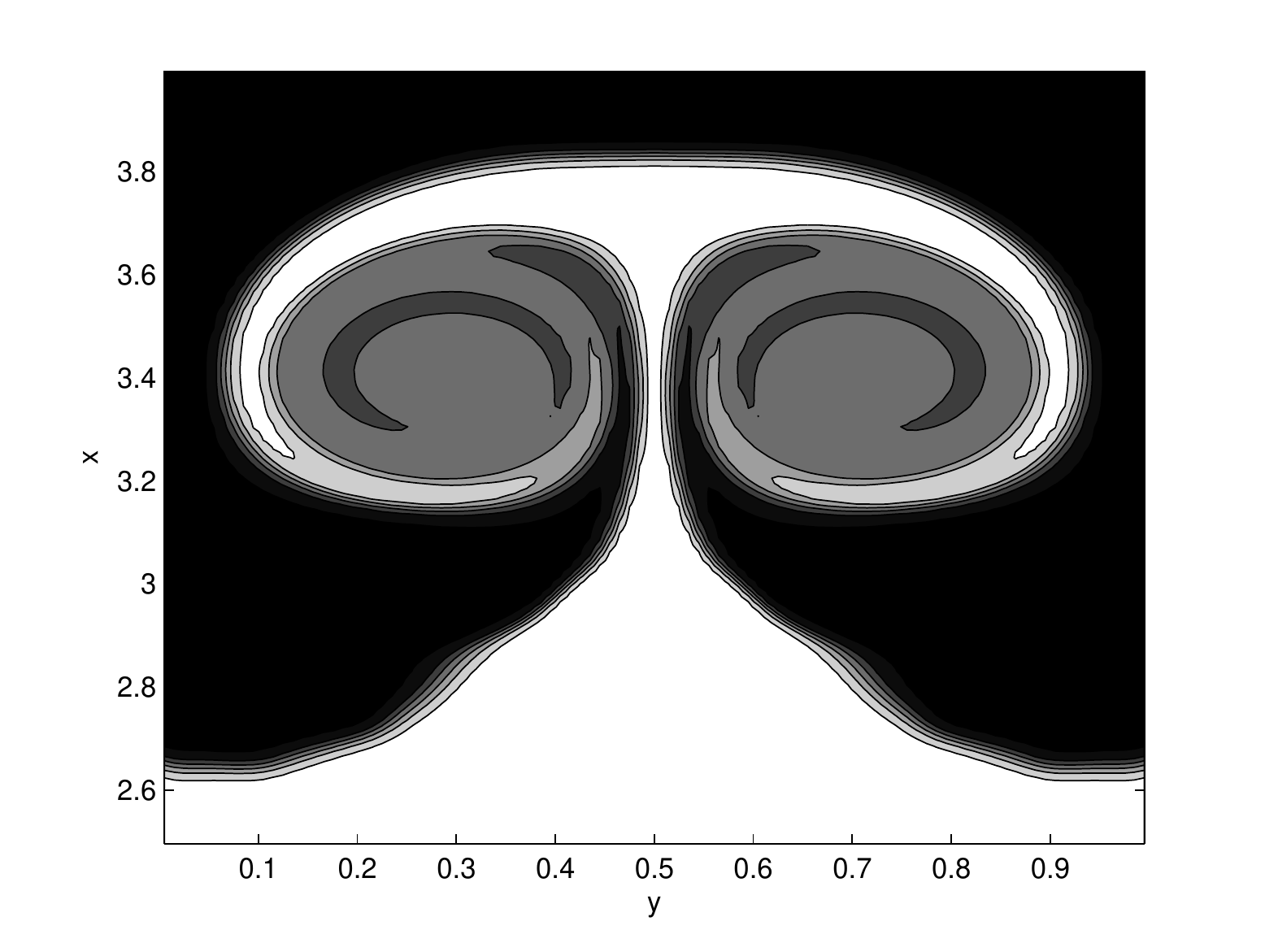}
\end{tabular}
\caption{The same as in Figure \ref{rm14} with RK-HLLC. On
  the left with resolution as in Figure \ref{rm14} and on
  the right with resolution as in Figure \ref{rm14nosteephi}}
\label{rm14rk}
\end{figure}

The RK-HLLC-code produces a more smeared out structure, see Figure \ref{rm14rk}. 

\section{Forced isotropic turbulence}
In many real life flows turbulence is an important feature.
Since we do not know how to infer from simpler test cases how
a numerical method will treat turbulence, we now consider simulations of actual three-dimensional turbulence.  
Because of the three-dimensional nature of turbulence, to get useful results one needs powerful computational resources, and we were able to
perform some parallel simulations on the Hitachi SR8000 at the Leibniz Computing Centre
in Munich. The simulations were part of a larger study on parameters in supersonic turbulence, see \cite{schmidtparamstudy}. We considered the
same type of forced isotropic turbulence experiments described in
\cite{schmidtnum}. The resolution here was $256^3$ equilateral grid cells, and
the boundary conditions periodic as in \cite{schmidtnum}. We refer to
\cite{schmidtphd} and \cite{schmidtnum} for details of the experiments and the
analysis tools. In addition to \cite{schmidtnum},
compressible turbulence simulations with PPM have been investigated by Sytine
et al. in \cite{ppmturb}.

The tests consisted of a constant, zero velocity initial state continuously subjected to a
stochastically varying force field ${\bf f }$. The forcing was given by evolving its Fourier transform by a so called Ornstein-Uhlenbeck process, which is a statistically stationary stochastic process, with parameters such that the resulting force was statistically isotropic. Only the larger wavelengths were given a nonzero contribution. Note that the Fourier transform of a periodic function can be understood as a generalized function given by the coefficients in its Fourier series. By varying the magnitude of the
forcing, the characteristic velocity of the flow was varied
correspondingly. The forcing also had a free parameter $\zeta$ corresponding to a projection operator regulating the solenoidality of the force field. For $\zeta=1$ the force field is divergence free, and for lower values we have progressively
stronger compressive force components. We will not study the influence of this parameter here, just note that all flows considered were highly compressible. How a
gas responds to this injection of energy depends a lot on the equation of
state, as we will show.

Since these flows are highly sensitive to perturbations, it makes no sense to compare the actual solutions. Instead we will compare statistical
properties of the simulated flows, since the statistical approach has been
relatively successful in quantitatively describing turbulence, see for example
\cite{frisch}. Note also that each simulation represented a different
realisation of the stochastic forcing process. One way to extract statistical
information is to make a histogram
of the different values assumed by a scalar
quantity at a fixed time. We can call this to make a probability distribution function (PDF). We will consider PDFs
for $\rho$ and the absolute value of the vorticity $\omega$.

As an indicator of numerical dissipation we will look at the energy spectra, that is, we will look at the energy content in each Fourier mode of the velocity field. Parseval's theorem says
that the total specific kinetic energy equals the integral over the square
of the Fourier transformed velocity field ${\bf\hat{u}}({\bf k},t)$,
\beq
\int |{\bf u}({\bf x},t)|^2{\text d}{\bf x} = \sum_{{\bf k}}{\bf\hat{u}}({\bf k},t)\cdot{\bf\hat{u}}({\bf k},t)^*.%{\text d}{\bf k}.
\eeq
where $\cdot^*$ denotes complex conjugation. In other words, it is given by integrating over the energy
spectrum function $E(k,t)$, which is defined as the sum of the squares of the Fourier coefficients
corresponding to each mode where the three-dimensional wave number vector ${\bf k}$ has absolute value $k$,
\beq
E(k,t)=\sum_{|{\bf k}|=k}\half{\bf\hat{u}}({\bf k},t)\cdot{\bf\hat{u}}({\bf k},t)^*
\eeq
times a scaling factor. We refer to \cite{schmidtphd} and \cite{schmidtnum} for how this was done numerically.

It is intuitively clear that if the solution has a
lot of small scale structure, it indicates low numerical diffusion, although
spurious oscillations could also play a role. The energy spectrum function
gives a way to quantify this idea for these highly complex flows, but it is also connected to deeper ideas about turbulence, in particular Kolmogorov's theory, see for example \cite{frisch}.

Typically a plot of $k\mapsto E(k,t)$ will show three different ranges. For the
lowest wave numbers the stochastic injection of mechanical energy
dominates. Then comes what is called the inertial range, where Kolmogorov's
theory predicts that $E(k,t)$ drops off as $k^{-\frac{5}{3}}$, due to the
transfer of energy from vortices of higher to lower length scales. This 'Kolmogorov cascade'
has been observed for low enough Mach numbers both in experiments and numerical simulations. For the highest wave numbers numerical dissipation becomes dominant, and $E(k,t)$ drops off steeply. Between the inertial range and
the dissipation range, one tends to observe a flattening of $E(k,t)$ in numerical
simulations. This is called the bottleneck effect and it is still debated whether it has physical significance,
or whether it is a purely numerical effect, see \cite{Brandenburgbottleneck}, \cite{schmidtnum} and the references in these. With the resolution here of
$256^3$ cells, the injection range goes straight into a bottleneck range. Since the
Kolmogorov theory is derived for incompressible flow, we also define the
transversal energy spectrum $E_{tr}(k,t)$ which only consist of the part of
${\bf\hat{u}}({\bf k},t)$ orthogonal to ${\bf k}$, so that we only take into
account the divergence free part of the velocity.

Some dimensional quantities need to be defined first, but we choose
not to go into detail about the physical scales as they are not relevant to
the code comparisons. Asymptotically the RMS (root mean squared) amplitude of the force $f$ would approach $(1-2\zeta+3\zeta^2)F_0$ for some prescribed value $F_0$. We use this to define the characteristic velocity $V$ by $V=(F_0 L)^{1/2}$, where $L$ is half the length $l$ of the sides of the periodic box. Hence $V$ is close to the RMS velocity in the fully
developed flow. With characteristic Mach number 'Ma' we refer to the ratio of $V$
to the initial sound speed. The simulations were run for five integral time
scales $T=\frac{L}{V}$. The forcing is strongest at
 wave numbers ${\bf k}$ such that $|{\bf k}|=k_0=\frac{2\pi}{L}$, and zero for
 $|{\bf k}|\geq2k_0$. With $\alpha$ we refer to the integer $\frac{k_0
   l}{2\pi}=2$, and the initial density is denoted by $\rho_0$. As
 scaling factor for $E(k,t)$ we take $\frac{\alpha L}{2\pi}$.
                                
The CFL-number was $0.8$ in all simulations.

\subsection{Adiabatic gas, characteristic Mach number 17.9, $\zeta=0.1$}
We first compare PPM
and RK-HLLC on a set-up with an adiabatic equation of state, that is an ideal
gas with $\gamma=1.4$. Most of our statistics reveal no signifant difference
between the codes, but we see some clear trends in the evolution of the energy
spectra. The spectra imply that RK-HLLC is more dissipative than
PPM when the average Mach number is less than about 5. Furthermore the dissipative
effects of RK-HLLC appears to grow as the Mach number decreases, while the
dissipative effects of PPM is unaffected by Mach number.
\begin{figure}\centering
\begin{tabular}{cc}
\includegraphics[scale=0.46]{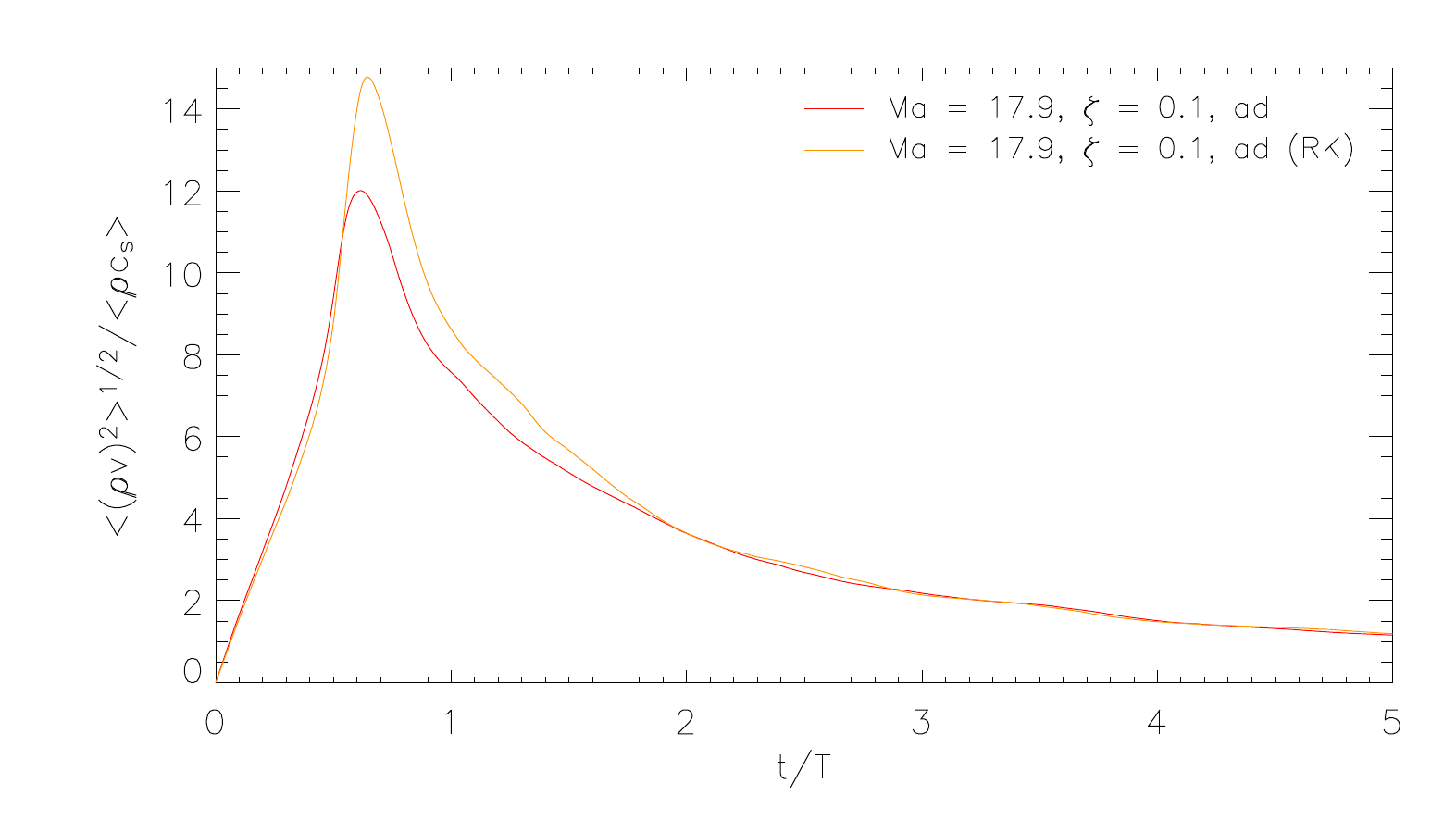}
&
\includegraphics[scale=0.46]{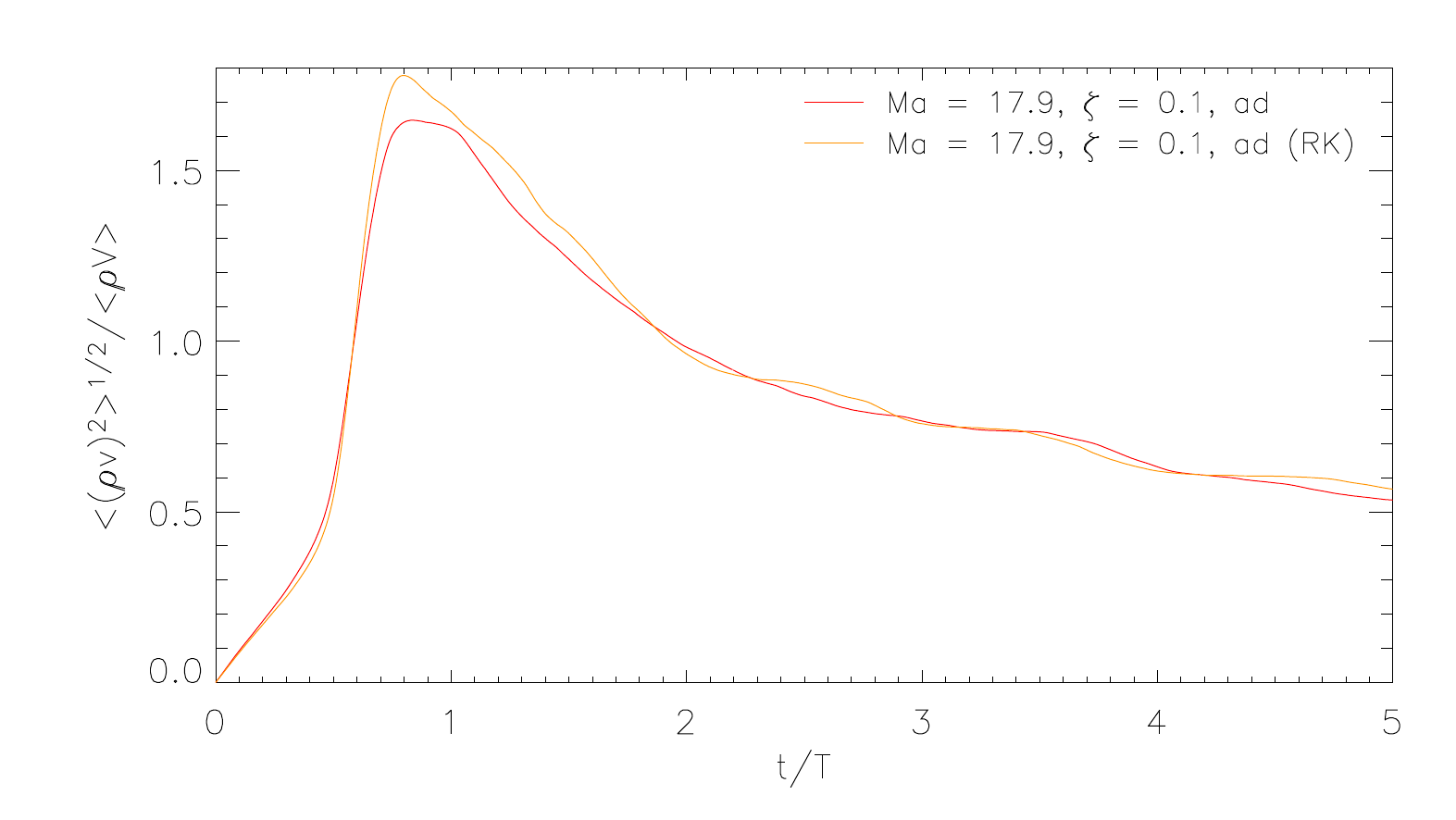}
\end{tabular}
\caption{Time history of RMS (root mean squared) Mach number (left) and
  momentum (right) for
  adiabatic runs. The curve from PPM is labelled '{\it ad}', and the curve from
  RK-HLLC is labelled '{\it ad (RK)}'.}
\label{admachmom}
\end{figure}

\begin{figure}\centering
\begin{tabular}{cc}
\includegraphics[scale=0.9]{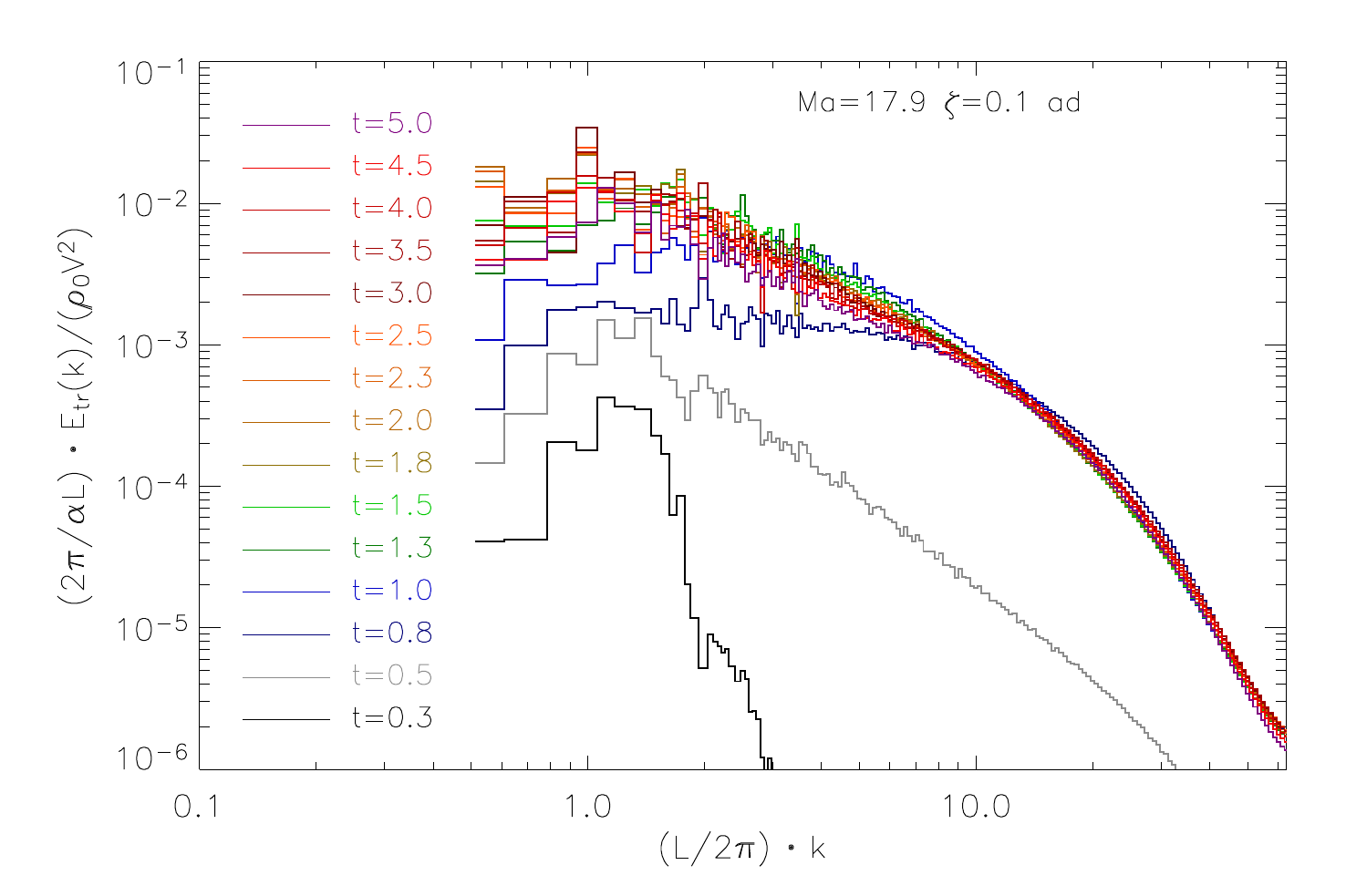}
\\
\includegraphics[scale=0.9]{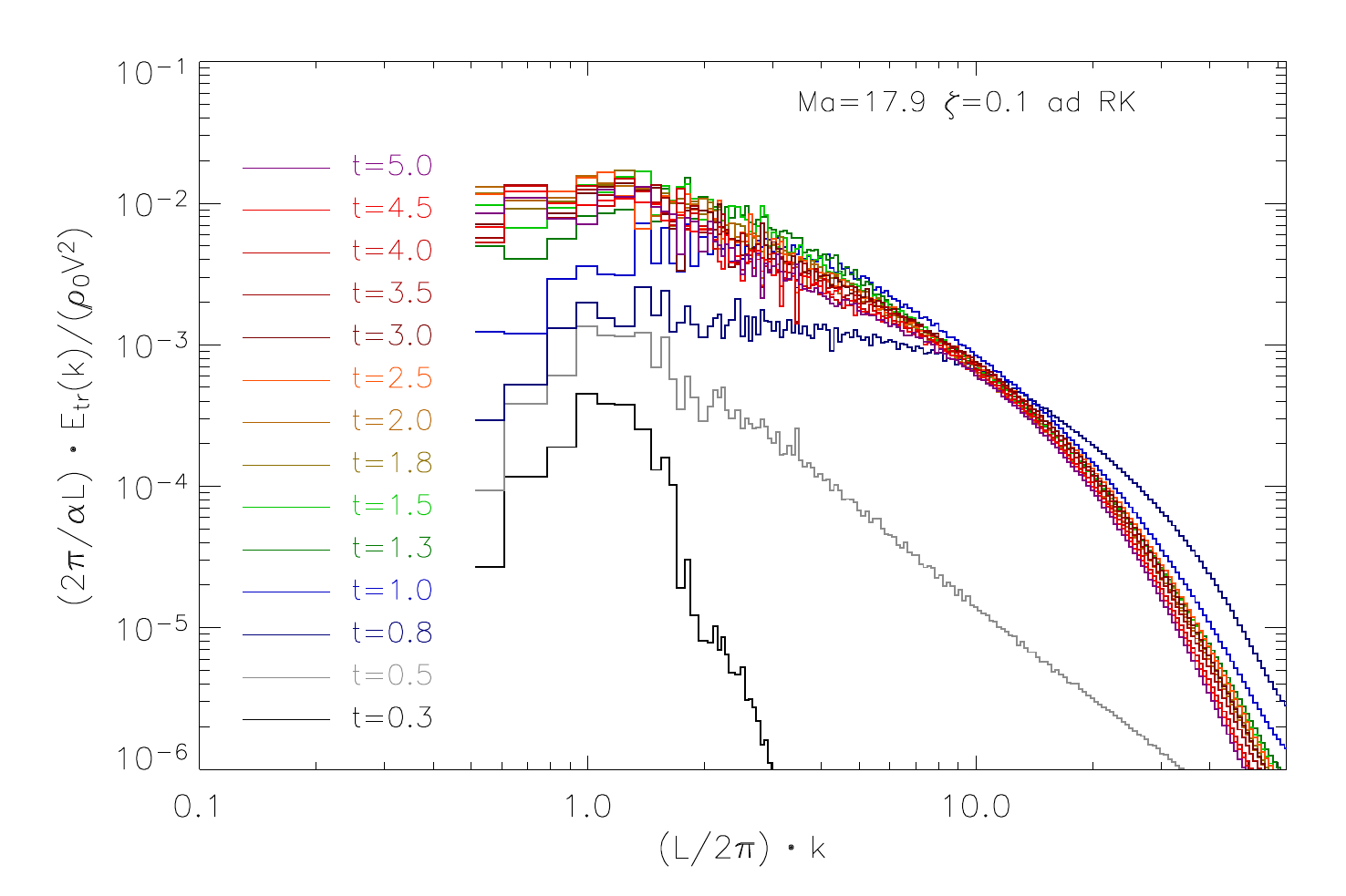}
\end{tabular}
\caption{Time history of transversal energy spectra. Data from PPM are on top, and
  from RK-HLLC underneath. Times are given in units of integral time scale $T$.}
\label{tseriesspectra}
\end{figure}

In the case of an adiabatic gas, the Mach number initially grows sharply, and
then falls off because the injected kinetic energy dissipates into heat, hence
increasing the sound speed, see Figure \ref{admachmom}. The velocity field behaves statistically as
stationary isotropic turbulence after around one integral time scale according to Figure \ref{tseriesspectra}, although
even at the termination point $t=5T$, an equilibrium between
the energy injection and dissipation was not reached, as that would imply a constant RMS momentum in Figure \ref{admachmom}.

\begin{figure}\centering
\begin{tabular}{cc}
\includegraphics[scale=0.48]{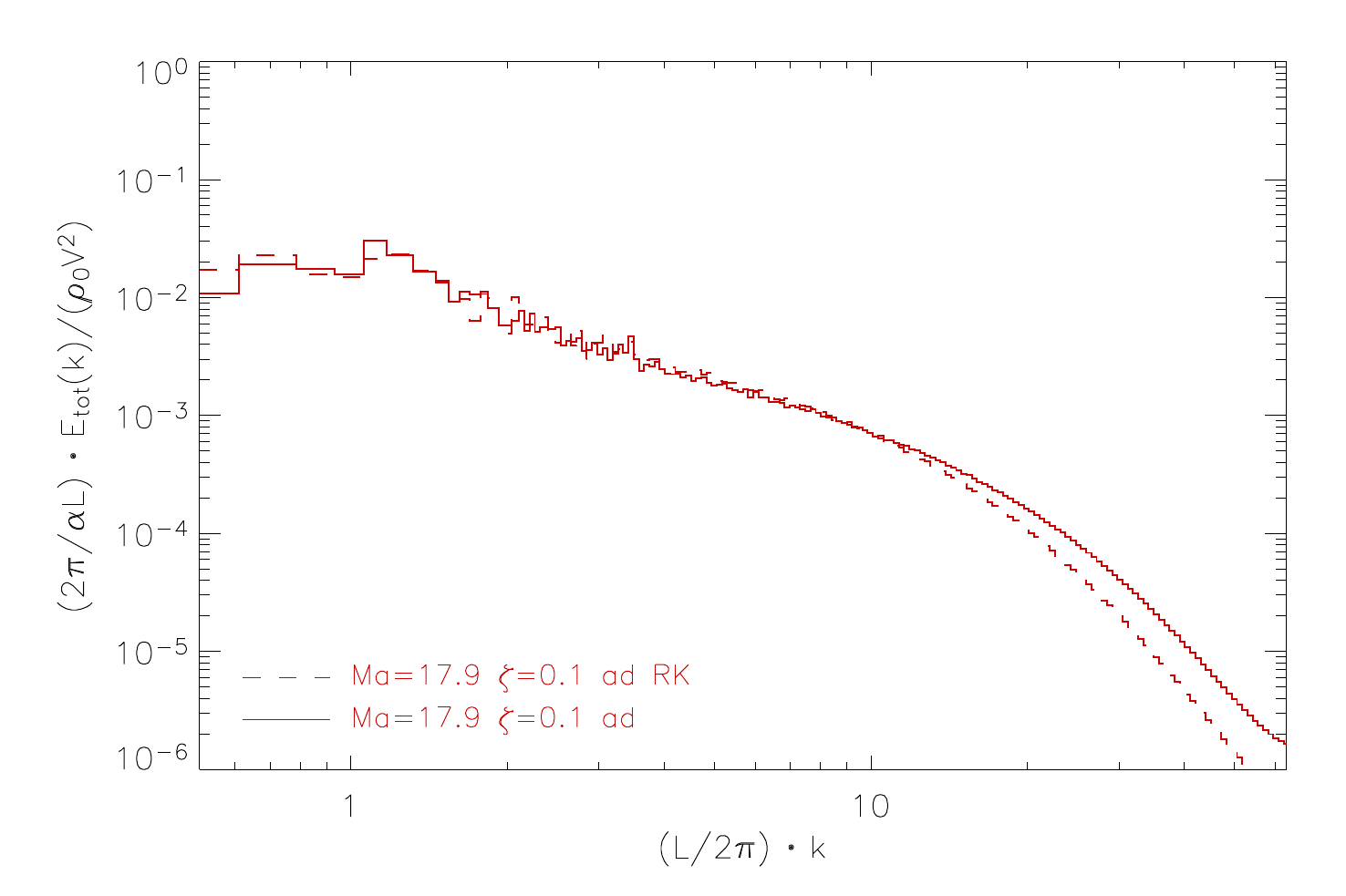}
&
\includegraphics[scale=0.48]{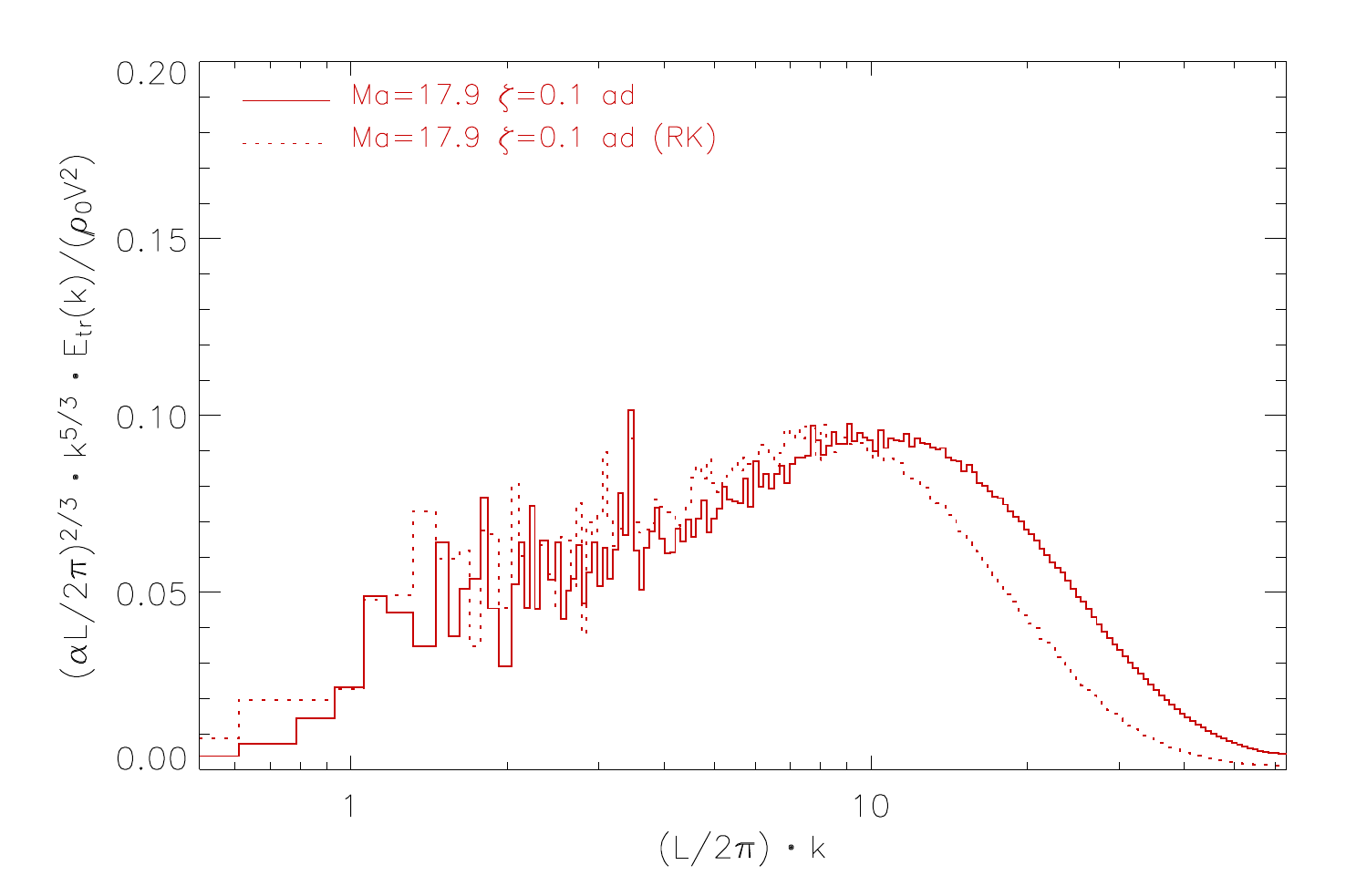}
\end{tabular}
\caption{Energy spectrum at the final time $t=5T$ for the adiabatic runs (left), on
  the right the compensated transversal spectrum. The curves are labelled as
  in Figure \ref{admachmom}.}
\label{adfig}
\end{figure}
 
Figure \ref{adfig} shows energy spectra at the final time. The energy spectrum
function for RK-HLLC drops off significantly
more sharply for the high wave numbers, and this is to be expected due to the less sharp resolution of
the RK-HLLC code. Also note the clear bottleneck
effect, which is best seen in the plot of the 'compensated' transversal energy
spectrum function $\Psi(k,t)$ proportional to $E_{tr}(k,t)k^{\frac{5}{3}}$ in Figure \ref{adfig}. The Kolmogorov theory predicts that $\Psi$ should be constant in the inertial range, and then drop off in the dissipation range.

If we look at other times, however, things become more complicated: In Figure
\ref{tseriesspectra} we see the evolution of the energy spectra.

The fact that in Figure \ref{admachmom}, the curves differ up to time $t=2T$,
we attribute to the different realisations of the stochastic forcing. Hence one
should compare statistics from the different codes only for times $t>2T$. The RK-HLLC-code gave decreasing energy spectra in time (after $t=T$), while
the energy spectra from PPM have no particular time dependency (after $t=T$).
We interpret this as an increase in dissipativity of
RK-HLLC in time, and we associate it with the decrease in Mach number. The observed independence of Mach number for PPM is related to the known fact that this scheme
retains, and even improves, its accuracy as the advective Courant number
decreases. This is caused by the high order upwind advection used in PPM. The typical
advective Courant number in a cell will decrease with the RMS Mach number in these turbulence tests, hence PPM should perform well at the lower Mach numbers.

\begin{figure}\centering
\begin{tabular}{cc}
\includegraphics[scale=0.9]{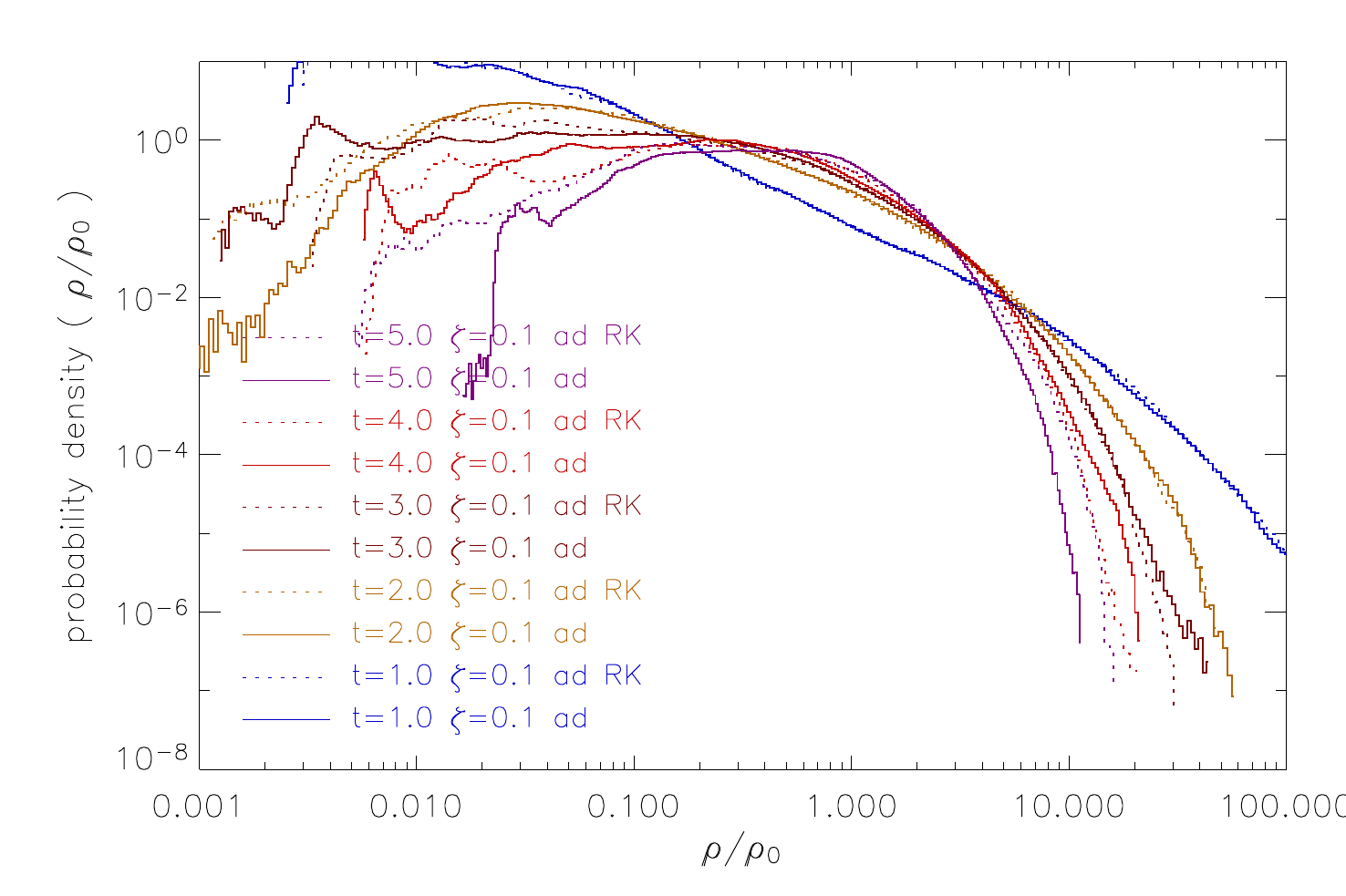}
\end{tabular}
\caption{Time history of mass density PDF for adiabatic runs. The curves are labelled as
  in Figure \ref{admachmom}.  Times are given in units of integral time scale $T$.}
\label{rhopdft}
\end{figure}

\begin{figure}\centering
\hspace{-0.34in}
\begin{tabular}{cc}
\includegraphics[scale=0.9]{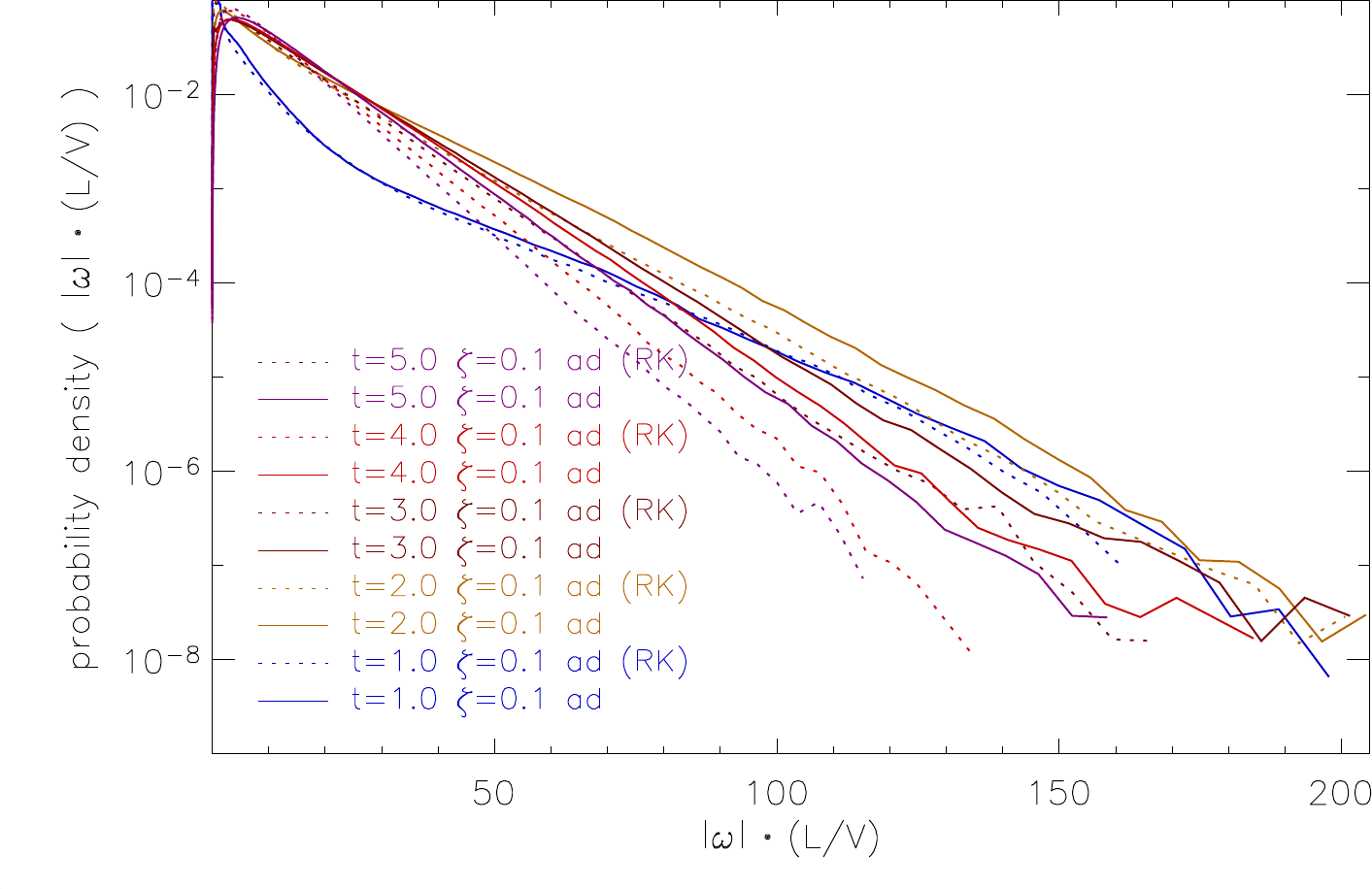}
\end{tabular}
\caption{Time history of $|\omega|$ PDF for adiabatic runs. The curves are labelled as
  in Figure \ref{admachmom}.}
\label{omegapdft}
\end{figure}

Figure \ref{rhopdft} shows mass density PDF at different times. There are clear differences between the simulations in both the high and low density regions, but this seems to be due to fluctuations inherent in the stochastic process behind the forcing, as there is no clear trend. The vorticity PDF's in Figure \ref{omegapdft} show the same tendency as the energy spectra. From time $t=2T$, the tails in the vorticity PDF's from PPM are clearly longer than those of RK-HLLC, meaning that the flow contains more small scale vortices.

\subsection{Isothermal gas }
\begin{figure}\centering
\begin{tabular}{cc}
\includegraphics[scale=0.46]{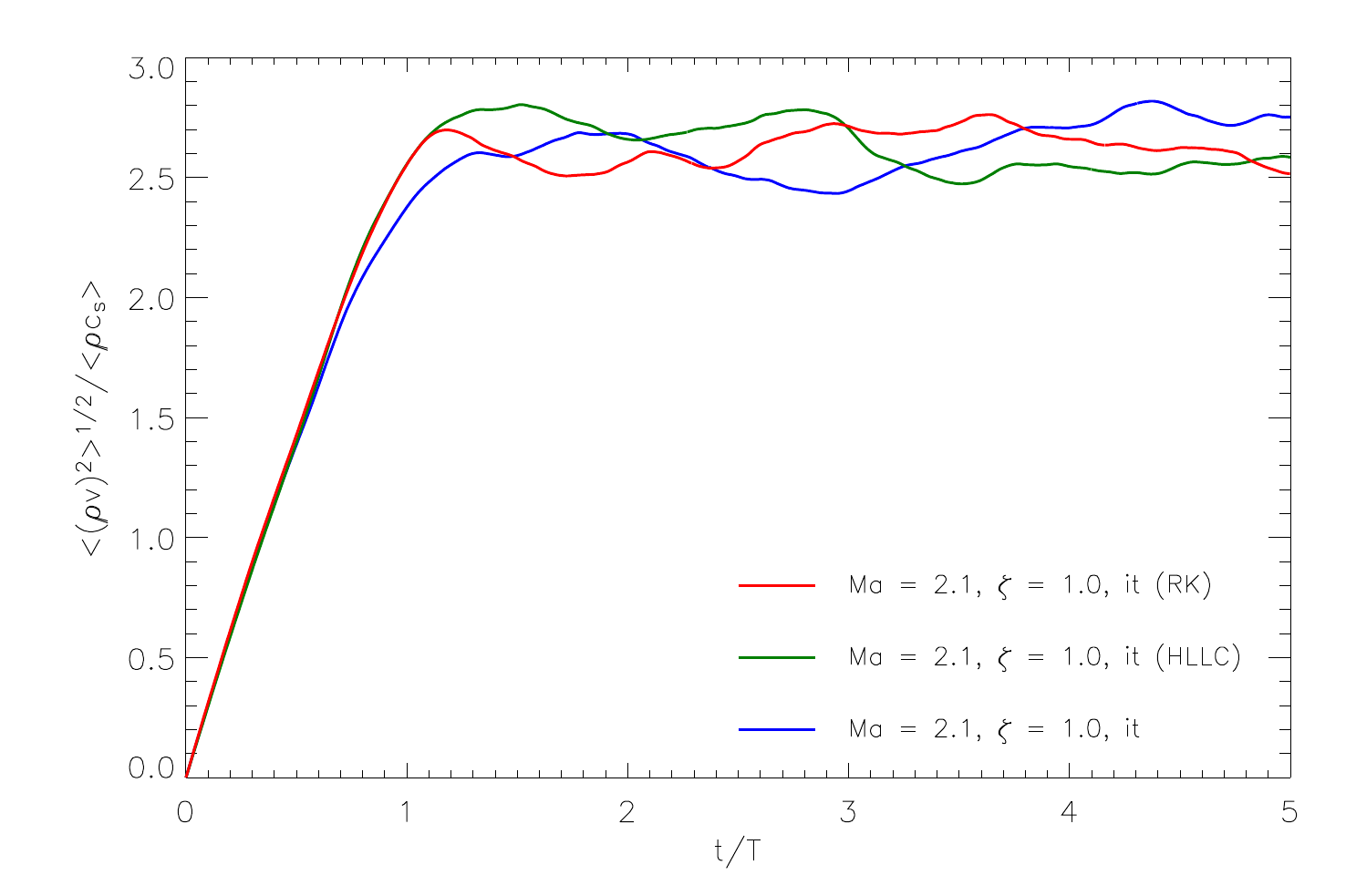}
&
\includegraphics[scale=0.46]{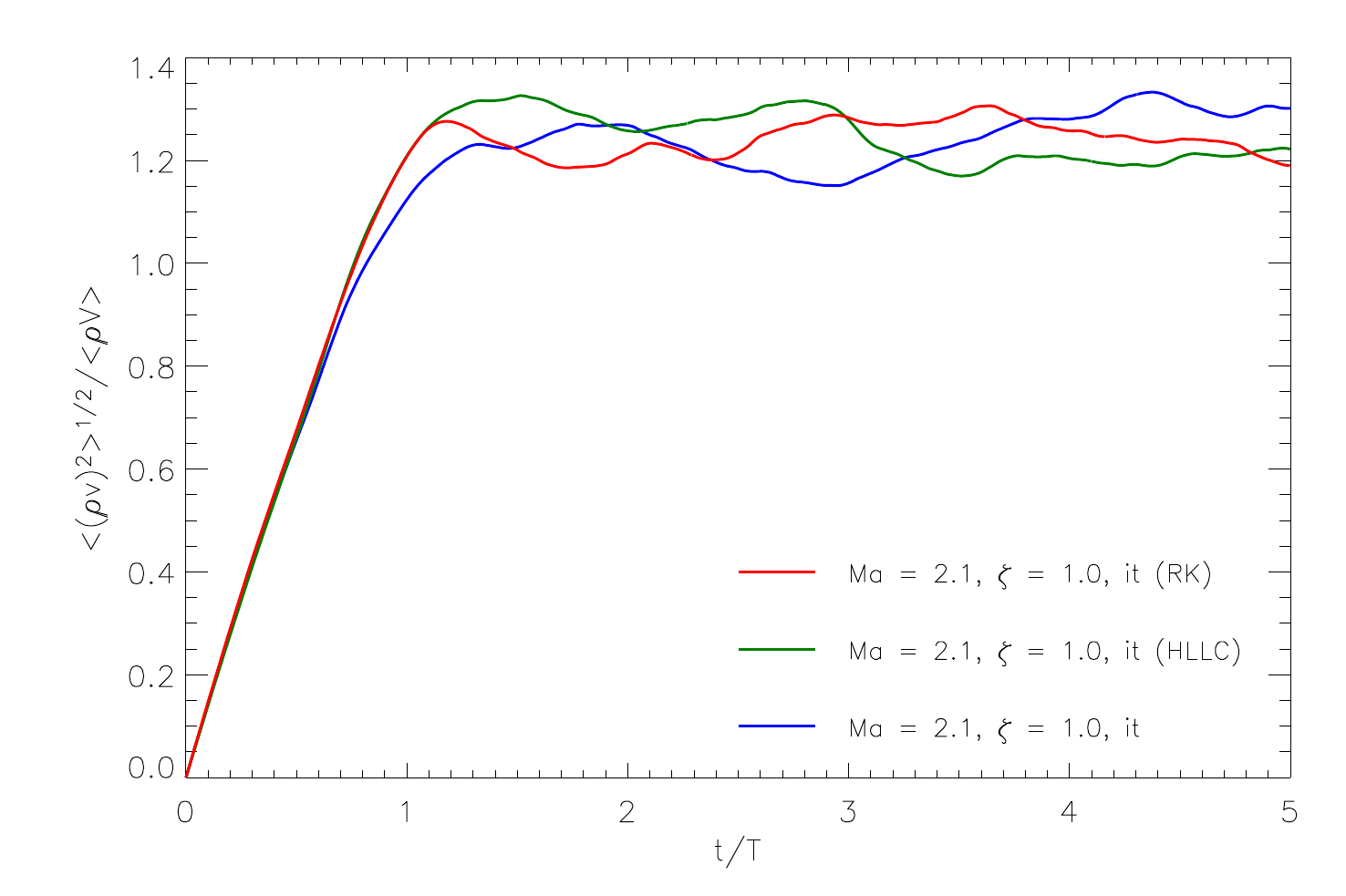}
\end{tabular}
\caption{Time history of RMS (root mean squared) Mach number and momentum for isothermal run
  with characteristic Mach number 2.1. The curve from PPM is labelled '{\it
  it}', from PPM-HLLC '{\it it (HLLC)}', and from RK-HLLC '{\it it (RK)}'.
}
\label{itmachvort021}
\end{figure}

\begin{figure}\centering
\begin{tabular}{cc}
\includegraphics[scale=0.46]{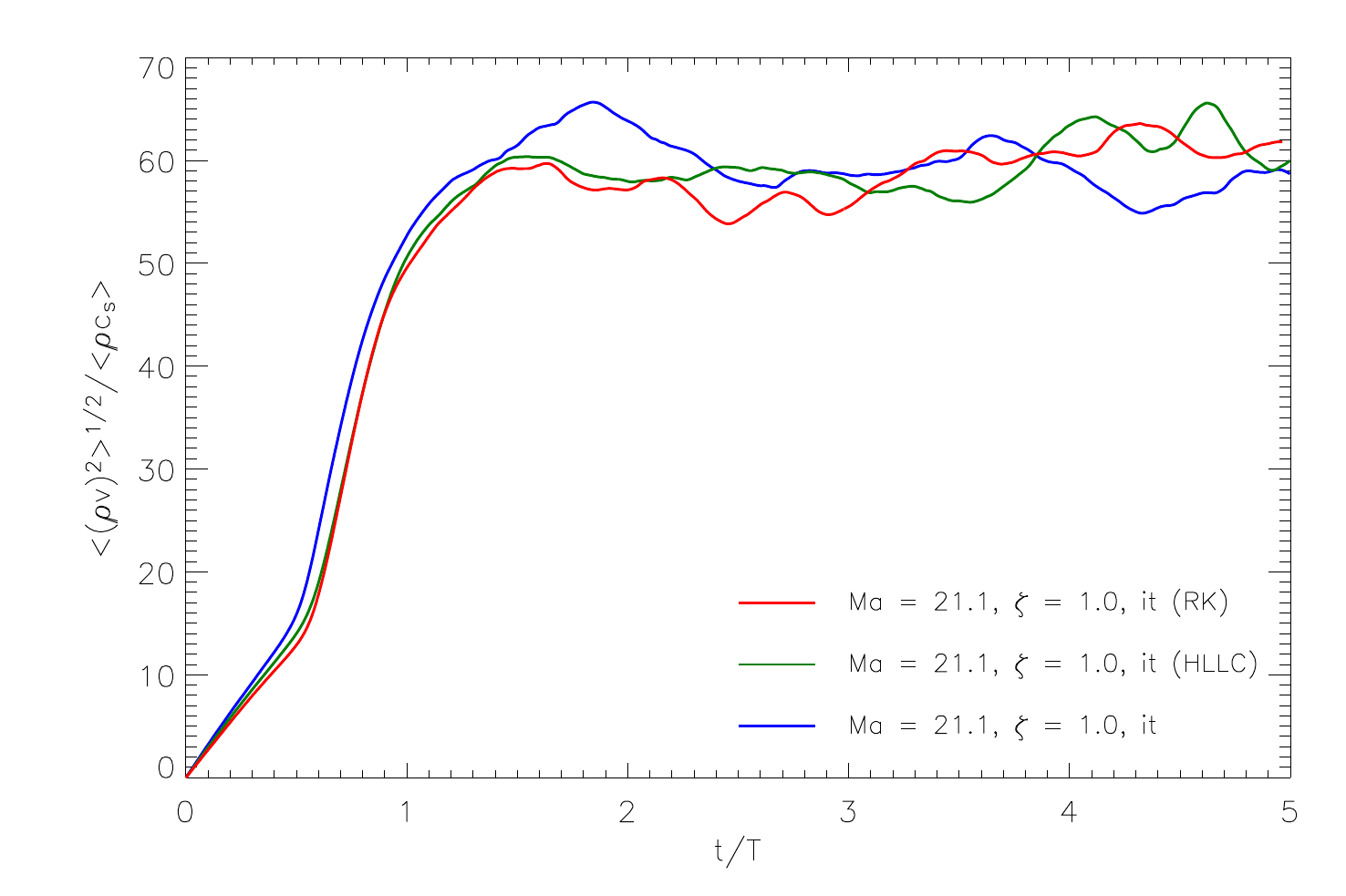}
&
\includegraphics[scale=0.46]{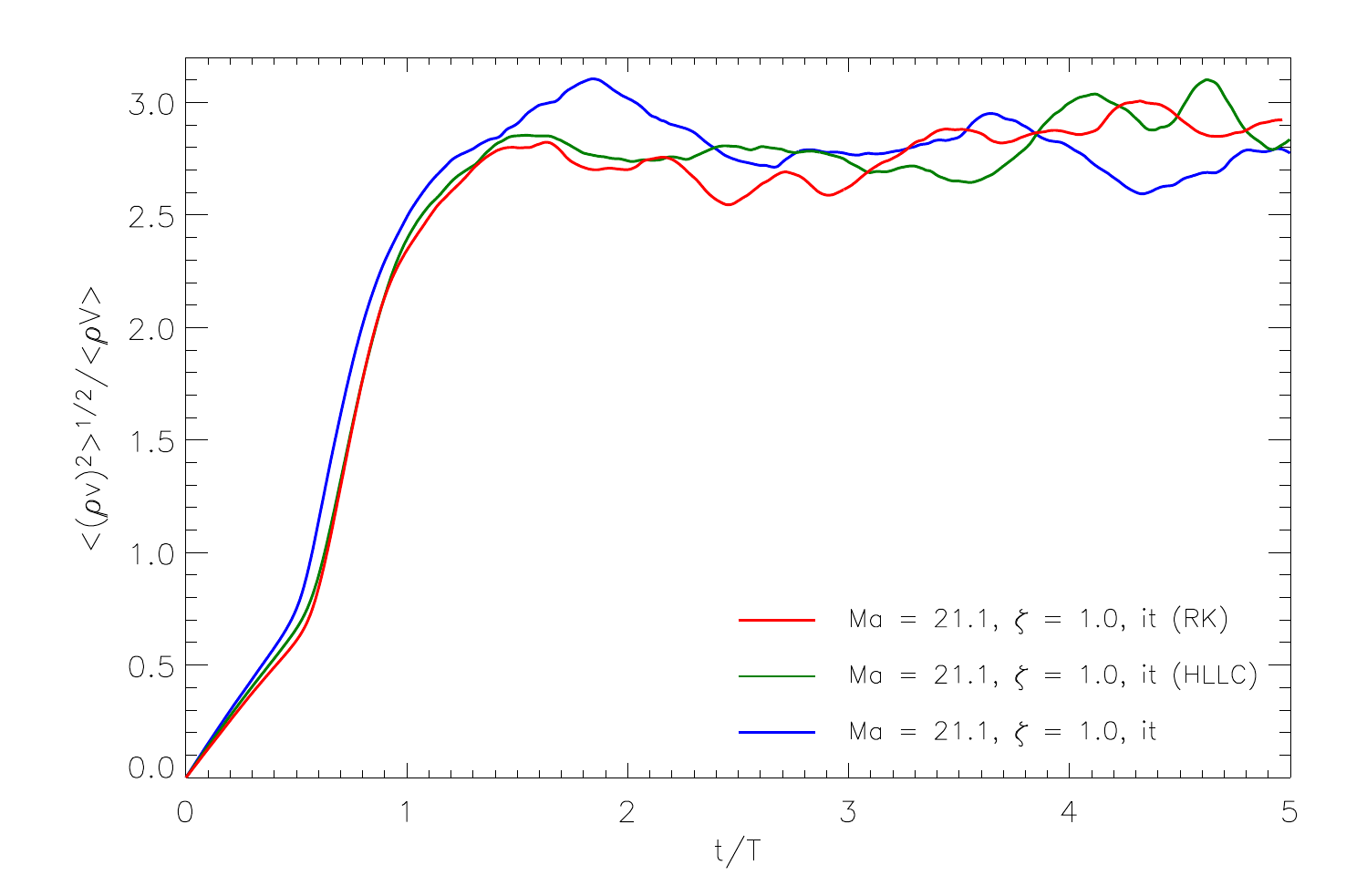}
\end{tabular}
\caption{Time history of RMS (root mean squared) Mach number and momentum for isothermal run
  with characteristic Mach number Ma 21.1. The curves are labelled as in
  Figure \ref{itmachvort021}.}
\label{itmachvort211}
\end{figure}

\begin{figure}\centering
\begin{tabular}{cc}
\includegraphics[scale=0.48]{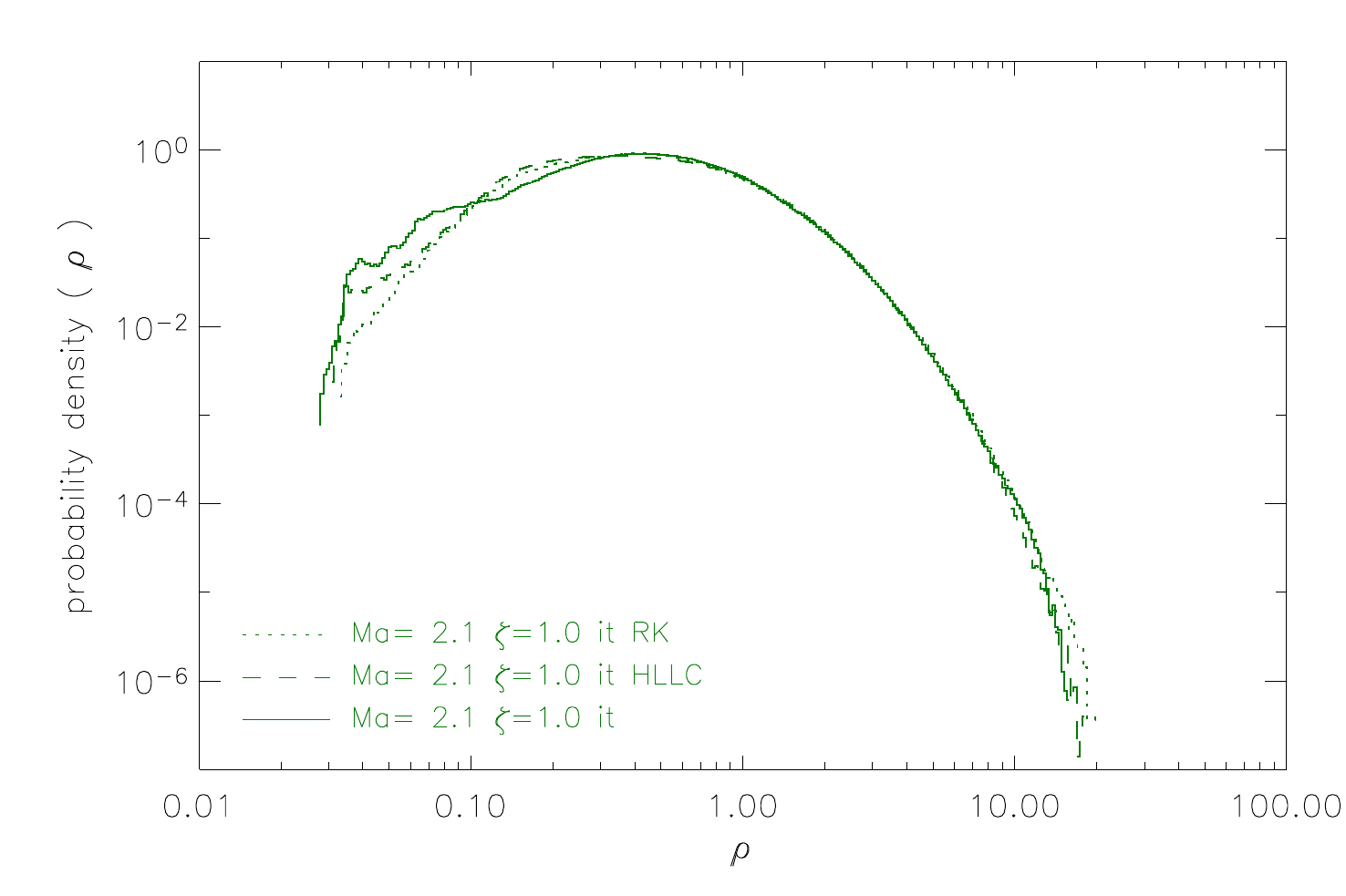}
&
\includegraphics[scale=0.48]{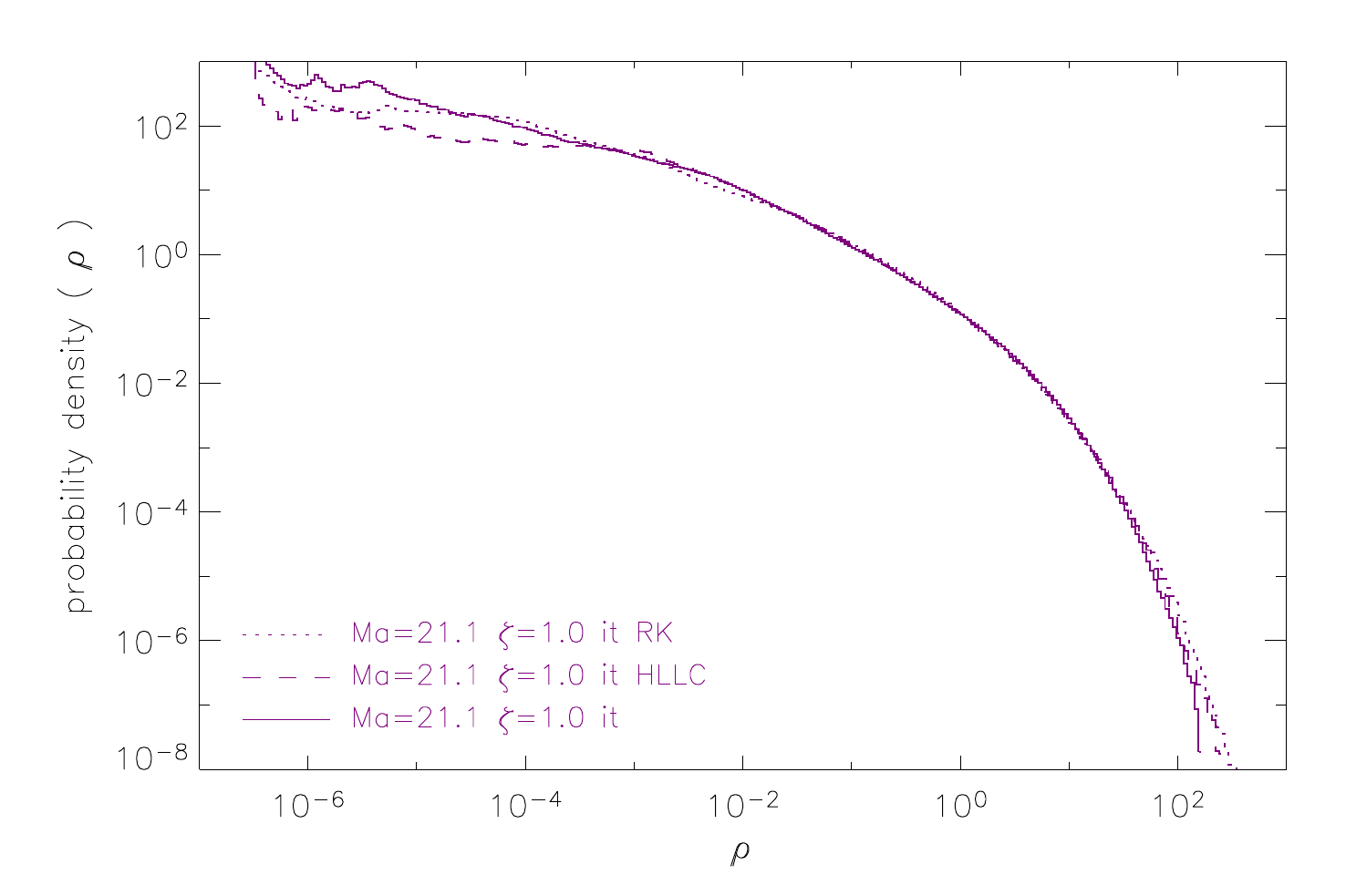}
\end{tabular}
\caption{Density PDFs for isothermal runs from the final time $t=5T$. On the
  left Ma=2.1, and on the right Ma=21.1. The curves are labelled as in
  Figure \ref{itmachvort021}.}
\label{itdens}
\end{figure}

\begin{figure}\centering
\begin{tabular}{cc}
\includegraphics[scale=0.48]{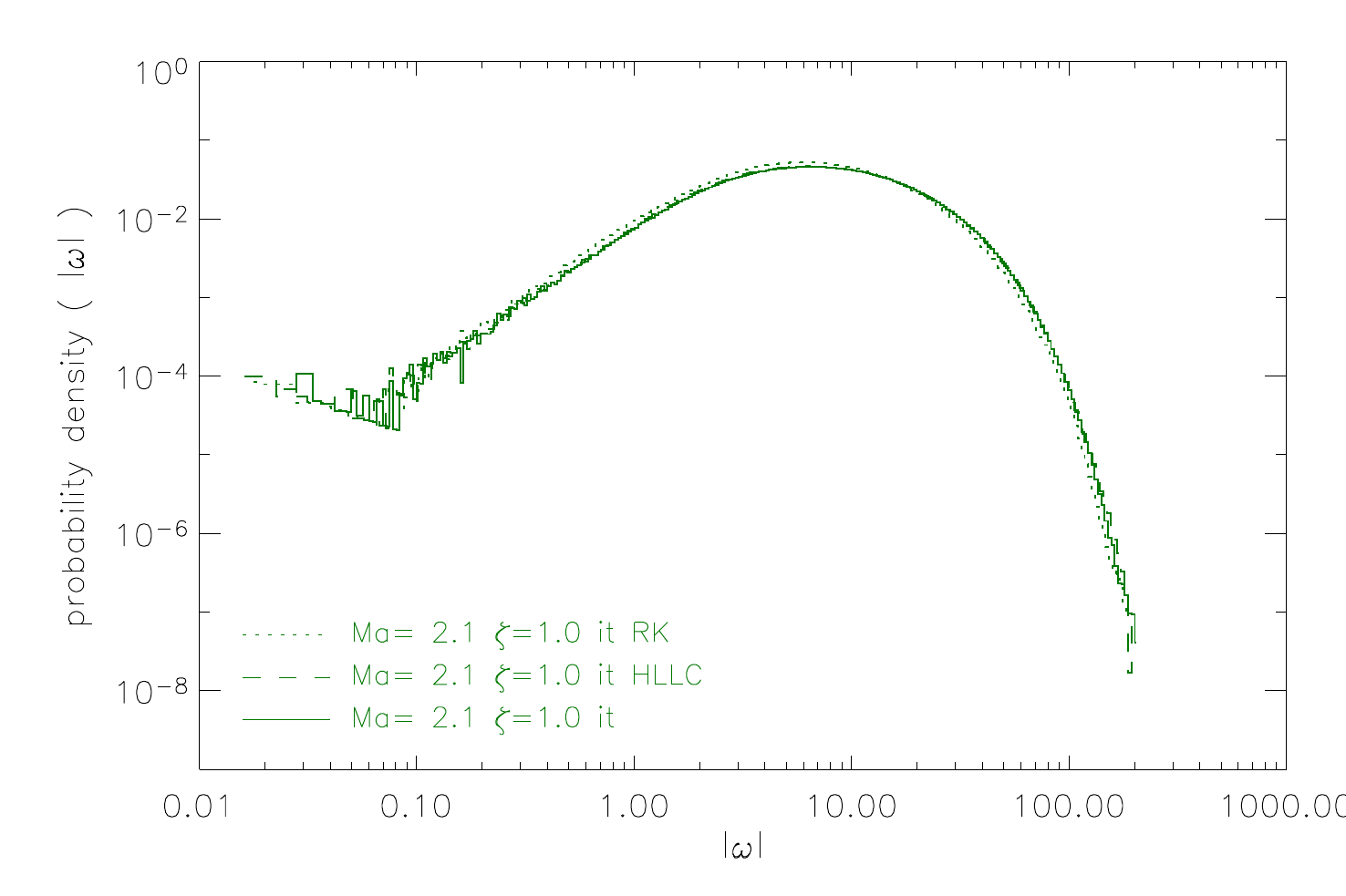}
&
\includegraphics[scale=0.48]{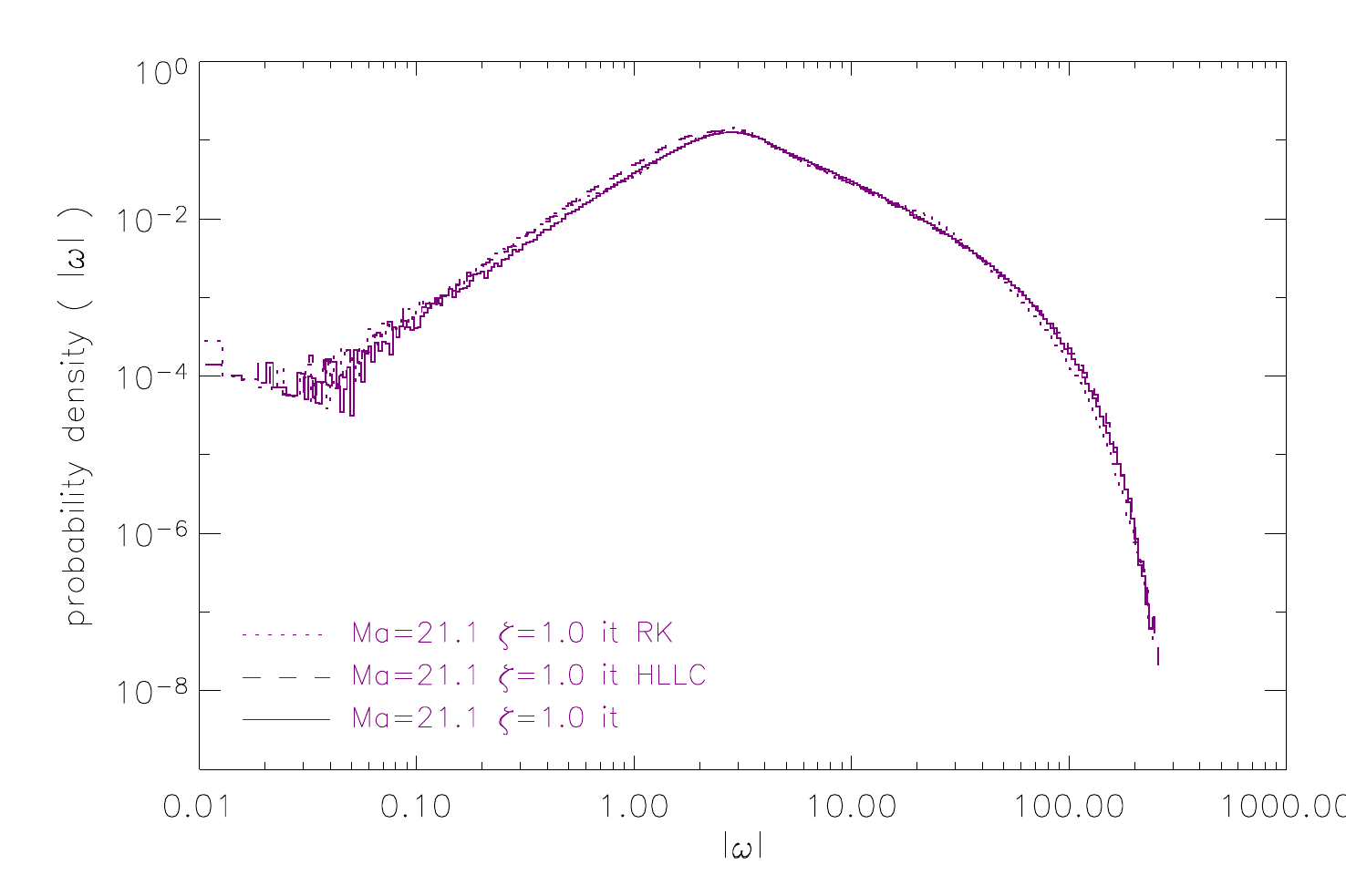}
\end{tabular}
\caption{Vorticity PDFs for isothermal runs from the final time $t=5T$. On the
  left Ma=2.1, and on the right Ma=21.1. The curves are labelled as in
  Figure \ref{itmachvort021}.}
\label{itvort}
\end{figure}

\begin{figure}\centering
\begin{tabular}{cc}
\includegraphics[scale=0.48]{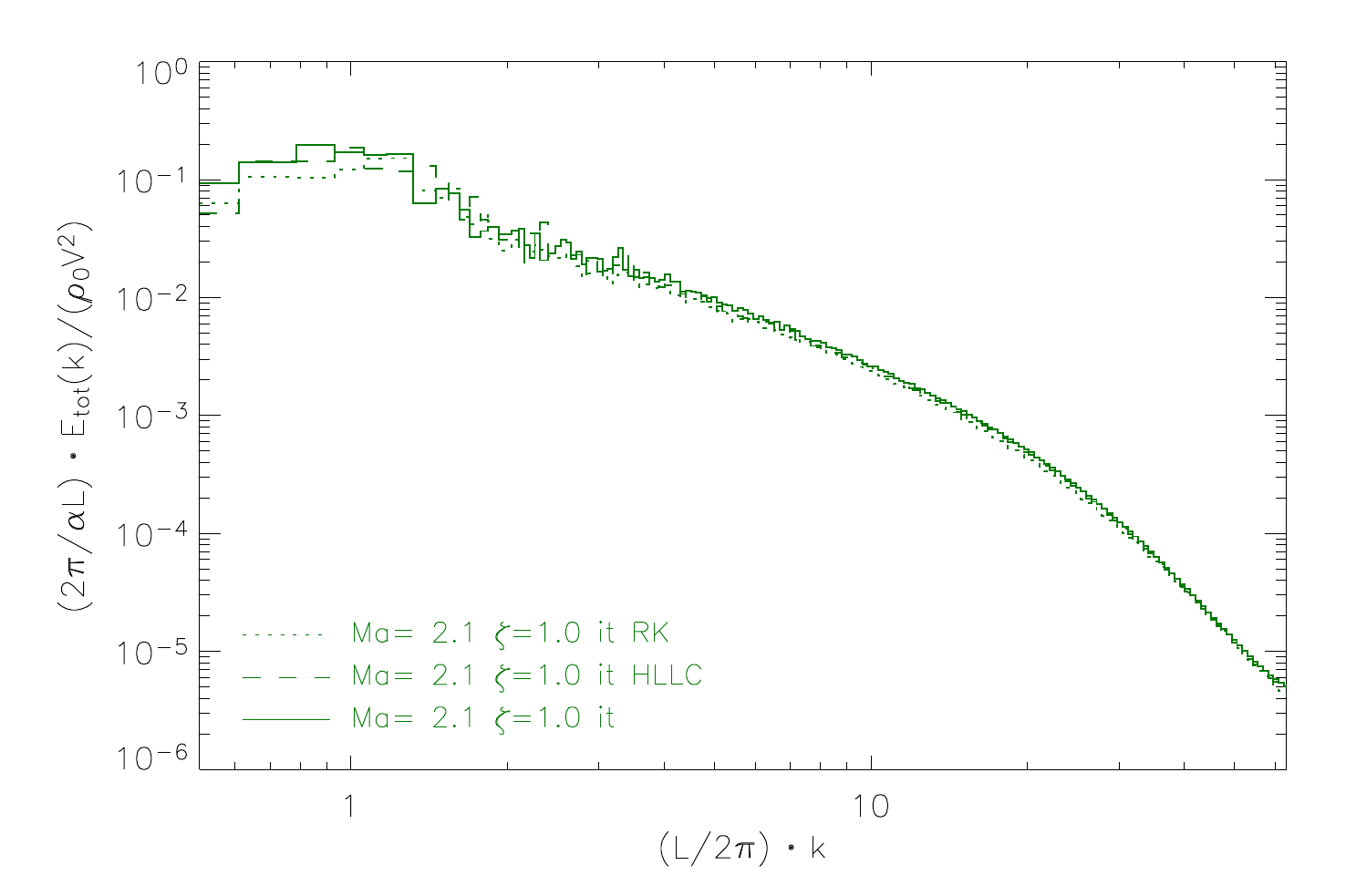}
&
\includegraphics[scale=0.48]{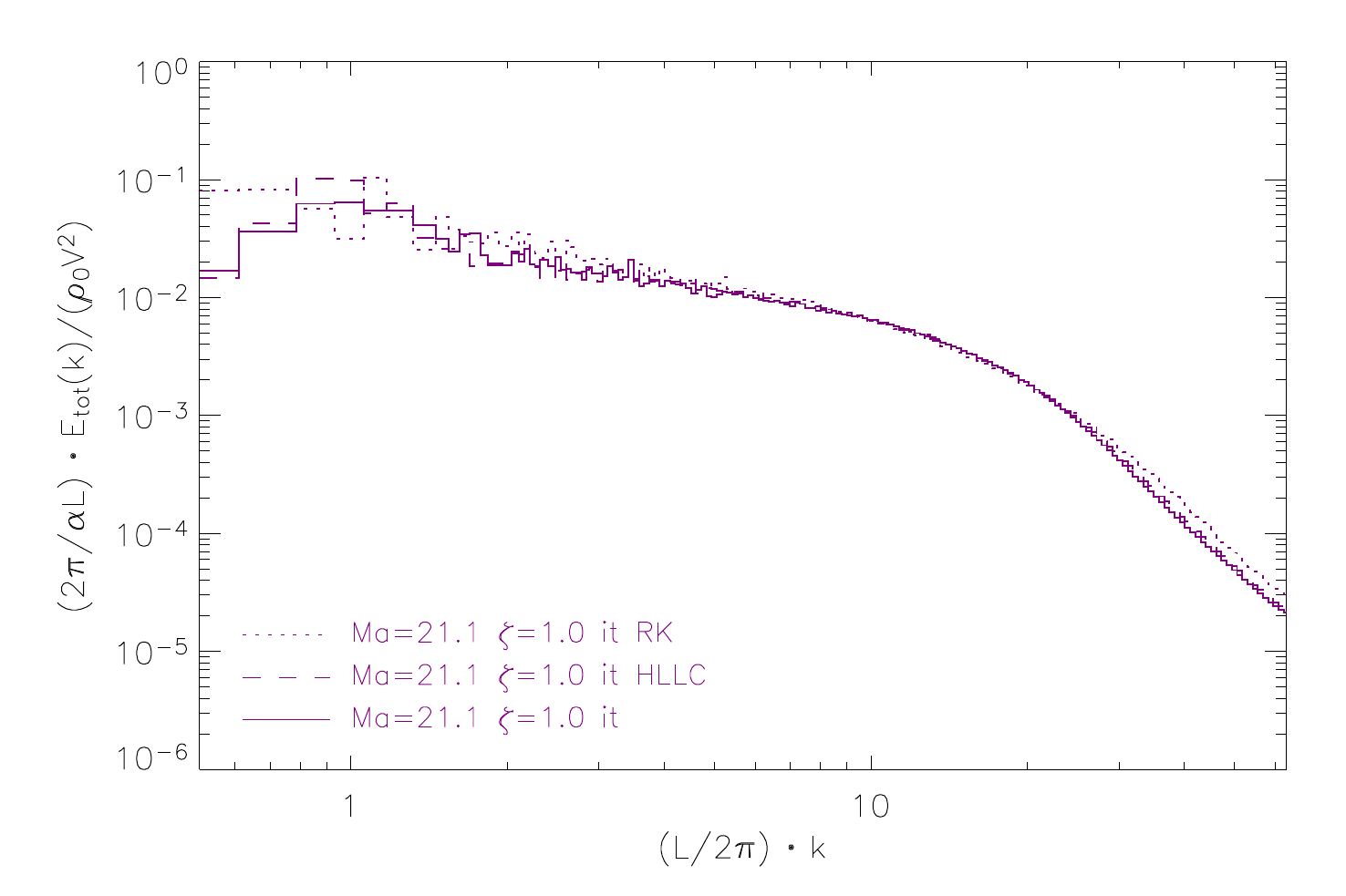}
\end{tabular}
\caption{Energy spectra for isothermal runs from the final time $t=5T$. On the
  left Ma=2.1, and on the right Ma=21.1. The curves are labelled as in
  Figure \ref{itmachvort021}.}
\label{itspect}
\end{figure}

\begin{figure}\centering
\begin{tabular}{cc}
\includegraphics[scale=0.48]{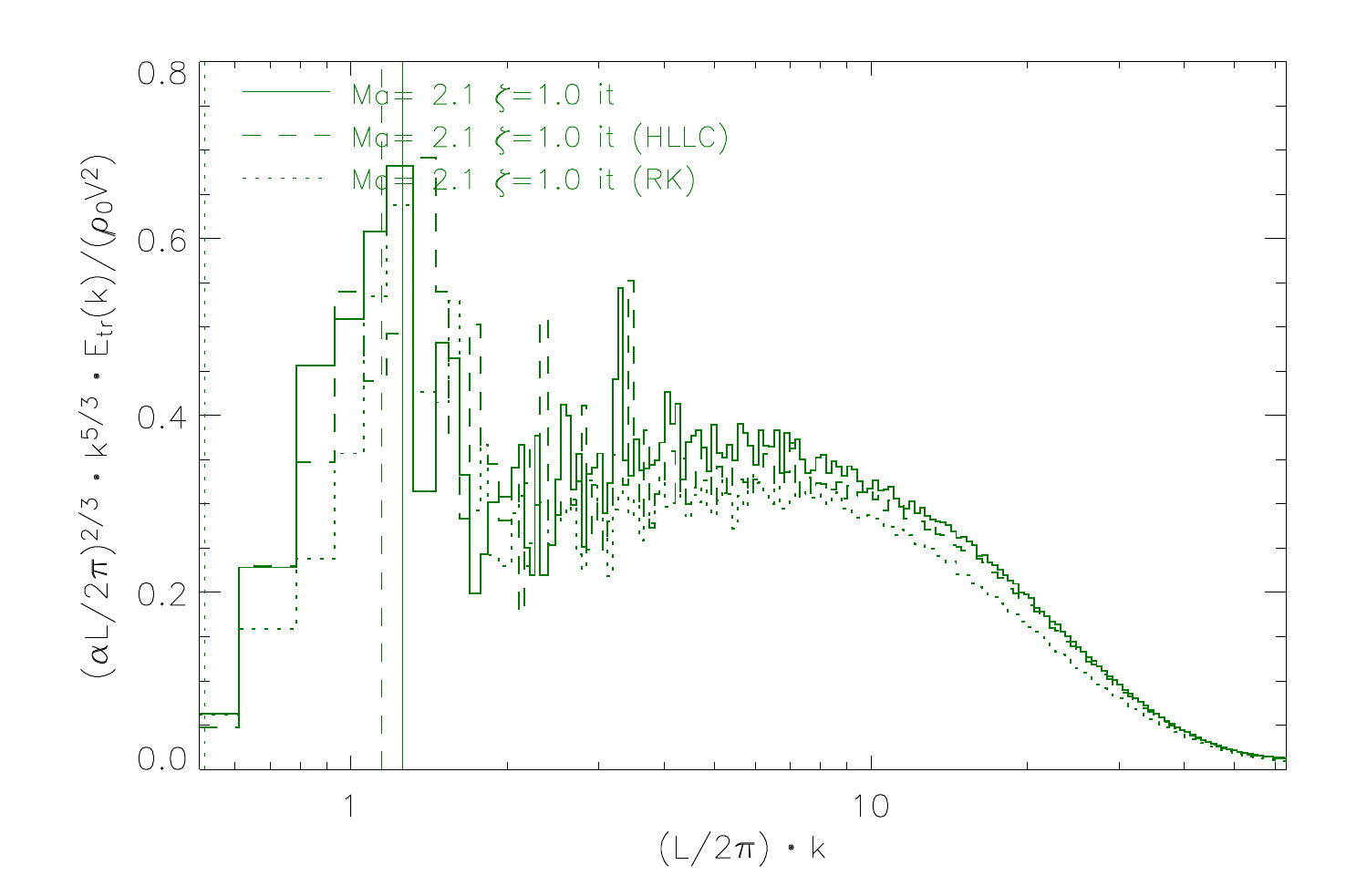}
&
\includegraphics[scale=0.48]{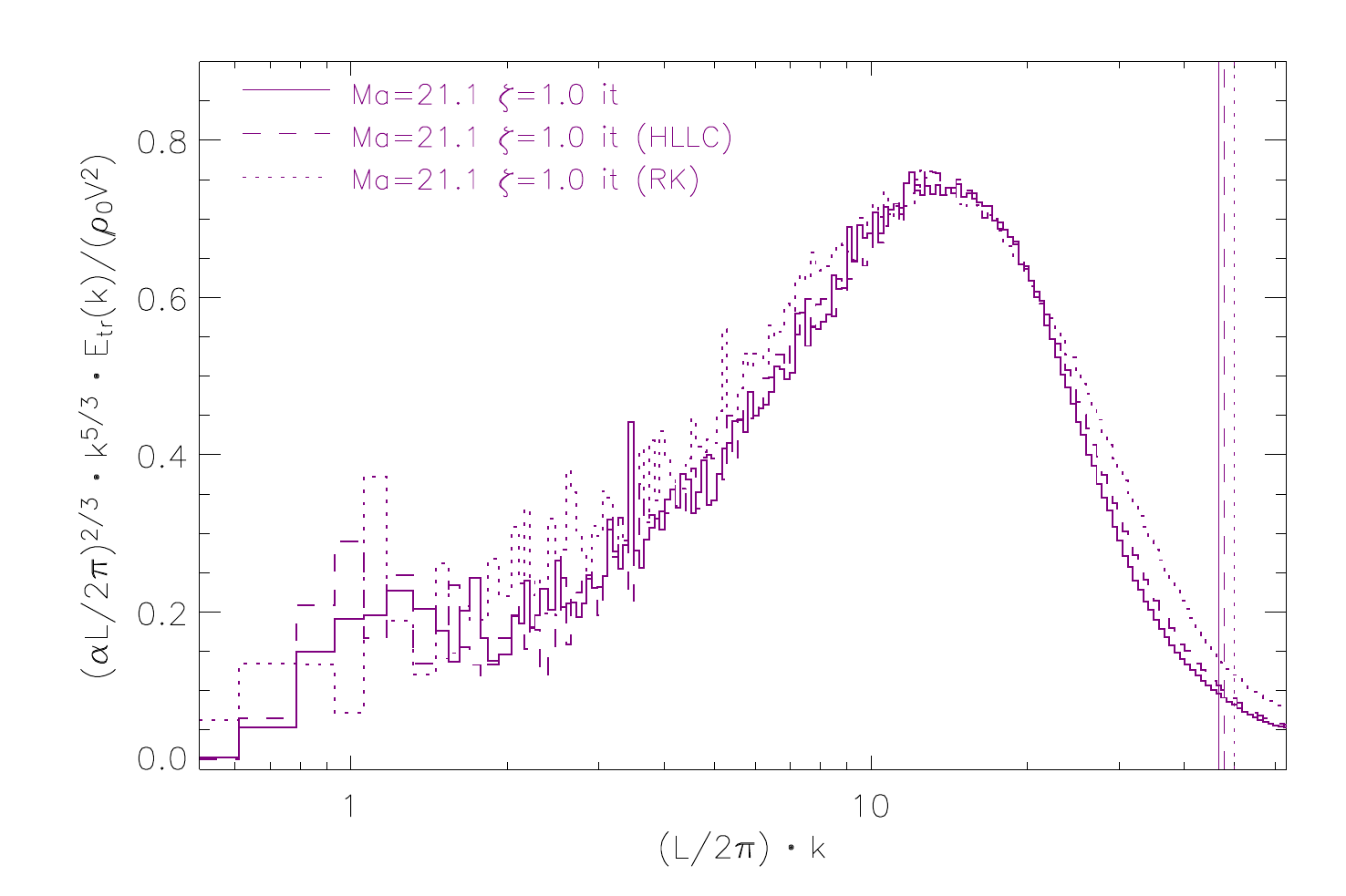}
\end{tabular}
\caption{Compensated energy spectra for isothermal runs from the final time $t=5T$. On the
  left Ma=2.1, and on the right Ma=21.1. The vertical lines
  represent the 'sonic wave number'. Integrating the curve to the left of his
  lin gives $c_s^2$. The curves are labelled as in
  Figure \ref{itmachvort021}.}
\label{itcompspect}
\end{figure}

We also performed simulations with isothermal gas. Here the Mach number
stays near constant after the initial growth phase, as seen in Figures
\ref{itmachvort021}-\ref{itmachvort211}. For this reason, we only analyse the
data from the final time.
We found no significant difference between RK-HLLC and PPM, and there was even less difference between
the two PPM codes. We show PDFs and energy spectra
from simulations with Mach numbers 2.1 and 21.1 in Figures
\ref{itdens}-\ref{itspect}.  Again bottleneck effects are seen in all
simulations, see Figure \ref{itcompspect}.


It is a little surprising that the RK-HLLC code does not appear comparably
more dissipative in these isothermal runs as it did in the adiabatic case, and
in the one- and two-dimensional tests. For the isothermal test with characteristic Mach
number 21.1, it is not so unexpected. This is because the RMS Mach number was much higher than in the adiabatic runs,
and from Figure \ref{tseriesspectra} it seems that the difference is less for
higher Mach numbers.  For the test with characteristic Mach number 2.1
however, the RMS Mach number is comparable to that of the adiabatic tests
at around $t=3T$. Hence, in addition to Mach number, the equation of
state must be taken into account when comparing RK-HLLC and PPM. In section
\ref{1dtests}, we noted that the most significant difference between the RK-
and PPM-algorithms was that both RK-codes smeared out contact discontinuities
more. In the isothermal case contact waves are not present, which might
explain why the codes differ less here than in the adiabatic runs.

\section{Summary}
From our numerical experiments we have made the following observations:
\begin{itemize}
\item
There is slightly more smearing of stationary shocks with HLLC-Bouchut
compared to the exact solver.
\item
The RK-codes smear out most features more than the PPM-codes, and especially contact
discontinuities. %This is due the lower order in space.
\item
RK-HLLC handles small densities better than the other codes.
\item
All codes exhibit spurious oscillations. We see more of them with the
PPM-codes, except at the near stationary shock, where PPM has a specialised
'flattening' procedure.
\item
The growth of Kelvin-Helmholtz and Richtmeyer-Meshkov instabilities appears to be little affected by which of the two Riemann solvers are used.
\item
In turbulence simulations of adiabatic gas, the dissipativity of RK-HLLC seems
to be less for higher than for lower Mach numbers,
while the dissipation with PPM is independent of the Mach number.
\item
For turbulence in an adiabatic gas with an RMS (root mean squared) Mach number less than about 5,
RK-HLLC seems to be more dissipative than PPM.
\item
For turbulence with an RMS (root mean squared) Mach number of 2.5 and higher in an isothermal gas, there
were no significant differences between Riemann solvers or higher algorithms.
\end{itemize}
The widespread use of PPM in the astrophysics community has lead to concern about
how much the results depend on this algorithm. We conclude that with respect
to the Riemann solver their results are accurate. However the efficiency of the
HLLC Riemann solver of Bouchut suggests that it may be used instead.

{\bf Acknowledgements}
The authors would like to thank Prof. Jens C. Niemeyer at the University of
W\"urzburg for valuable comments and suggestions. We also thank Christoph
Federrath for performing the post-processing of the turbulence simulation data.
\bibliography{libturb.bib}
\bibliographystyle{plain}
\end{document}